\documentclass[useAMS,usenatbib]{mnras}
\usepackage{times}
\usepackage{graphicx}
\usepackage{float}
\usepackage{epsfig}
\usepackage{courier}
\usepackage{chngcntr}
\usepackage{caption}
\usepackage{subcaption}
\usepackage{xspace}
\usepackage{epstopdf}
\usepackage{booktabs}
\usepackage{gensymb}
\usepackage{color,soul}
\usepackage{comment}
\usepackage{amssymb}
\usepackage{amsmath}
\usepackage{enumitem}
\usepackage{nicefrac}
\usepackage{rotating} 
\usepackage{tikz}
\usepackage[T1]{fontenc}
\usepackage{times}
\usepackage{graphicx}

\counterwithout{figure}{section}

\graphicspath{ {./plots/} } 

\title[SPH Disc Fragments]{Identifying and Analysing Protostellar Disc Fragments in Smoothed Particle Hydrodynamics Simulations}
\author[Cassandra Hall, Duncan Forgan, Ken Rice]{Cassandra Hall$^{1,2}$\thanks{Email: cassandra.hall@le.ac.uk}, Duncan Forgan$^{3,4}$, Ken Rice$^{1,5}$\\
$^{1}$SUPA\thanks{Scottish Universities Physics Alliance}, Institute for Astronomy, University of Edinburgh, Blackford Hill, Edinburgh, EH9 3HJ, UK\\
$^{2}$Department of Physics \& Astronomy, University of Leicester, Leicester, LE1 7RH, UK \\
$^{3}$SUPA\footnotemark[2], School of Physics \& Astronomy, University of St Andrews, North Haugh, St Andrews, KY16 9SS, UK\\
$^{4}$St Andrews Centre for Exoplanet Science\\
$^{5}$Centre for Exoplanet Science, University of Edinburgh, Edinburgh, UK\\
}
\begin{document}

\date{\today}

\pagerange{\pageref{firstpage}--\pageref{lastpage}} \pubyear{2016}

\maketitle

\label{firstpage}

\begin{abstract}
  We present a new method of identifying protostellar disc fragments in a simulation based on density derivatives, and analyse our data using this and the existing \texttt{CLUMPFIND} method, which is based on an ordered search over all particles in gravitational potential energy. Using smoothed particle hydrodynamics, we carry out 9 simulations of a $0.25$ M$_{\odot}$ disc around a 1 M$_{\odot}$ star, all of which fragment to form at least 2 bound objects. We find that when using all particles ordered in gravitational potential space, only fragments that survive the duration of the simulation are detected. When we use the density derivative method, all fragments are detected, so the two methods are complementary, as using the two methods together allows us to identify all fragments, and to then determine those that are likely to be destroyed. We find a tentative empirical relationship between the dominant azimuthal wavenumber in the disc $m$ and the maximum semi-major axis a fragment may achieve in a simulation, such that $a_{\mathrm{max}}\propto\nicefrac{1}{m}$. We find the fragment destruction rate to be around half that predicted from population synthesis models. This is due to fragment-fragment interactions in the early gas phase of the disc, which can cause scattering and eccentricity pumping on short timescales, and affects the fragment's internal structure. We therefore caution that measurements of eccentricity as a function of semi-major axis may not necessarily constrain the formation mechanism of giant planets and brown dwarfs.
%We compare our results to current gravitational instability population synthesis models, and conclude that fragment-fragment interactions in the early gas phase of the disc should be included in future models, since scattering to large radii reduces the tidal destruction of fragments to $\sim$ 20\%, compared to $\sim$ 40\% suggested by current models. Additionally, these fragment-fragment interactions can produce eccentric orbits on very short timescales, and caution that measurements of eccentricity as a function of semi-major axis may not necessarily constrain the formation mechanism of giant planets and brown dwarfs.
\end{abstract}

%============================================================
%      INTRODUCTION
%============================================================
\section{Introduction}
\label{sec:intro}
There are two distinct modes of planet formation in protostellar discs. The first, and most widely accepted, is the core accretion model (CA) \citep{pollack1996,hubickyj2005}. In this model, growth begins with dust grains of $\sim1\/\micron$ that coagulate rapidly into larger particles, ultimately settling into the disc midplane where there is enough material for them to grow to kilometre-sized planetesimals. These planetesimals can then grow via collisions into planetary cores, and if sufficiently massive, and if the gas disc has not dissipated, will accrete a gaseous envelope, ultimately becoming a gas giant planet \citep{pollack1996,lissauer1993}.

Most observational evidence favours this formation mechanism. For example, gas giant planets are preferentially found around metal-rich stars \citep{santos2004}, with an empirical relationship that quantifies the probability, $\mathcal{P}$, of gas giant planet formation as
\begin{equation}
\mathcal{P} = 0.03\times 10^{2.0[\mathrm{Fe}/\mathrm{H}]},
\end{equation}
where $[\mathrm{Fe}/\mathrm{H}]$ is the metallicity of the host star relative to solar metallicity \citep{fischervalenti2005}. Numerical work \citep{caietal2005} has suggested that this would not be the case if the second mode of planet formation, gravitational instability (GI), were the dominant formation mechanism of these planets, since an increase in metallicity responds to a decrease in cooling rate, resulting in weaker GI activity. This ultimately decreases the likelihood of these systems fragmenting, since weak GI corresponds to smaller stresses in the disc. On the other hand, it has also been shown that metallicity variation makes very little difference to the occurrence of fragmentation \citep{boss2002}.

In the GI scenario, gas giant planets and brown dwarfs form by direct gravitational collapse in the gaseous protostellar disc \citep{kuiper1951,cameron1978,boss1997,boss1998}. This happens rapidly, in a relatively early phase of the disc's life when it is massive enough to be self-gravitating. The advantage of this mechanism is its rapidity; gas giants are able to form on timescales shorter than typical disc dispersion timescales ($\sim$ 5 Myr; \citealt{haischladalada2001}). While CA is certainly the most widely accepted model, there are barriers to grain growth at several length scales which seem to indicate difficulty in forming planetary mass objects within the disc lifetime. The most famous of these is the so-called \textit{metre barrier}; as grains increase in size, so do their relative velocities, which makes grain fragmentation, rather than coagulation, the most likely outcome. 

A promising solution to this problem is the \textit{pebble accretion theory} \citep{lambrechts2012,levison2015}. Pebbles are grouped together due to the \textit{streaming instability} \citep{youdin2005}, whereby solid particles orbit at Keplerian velocity, but the gas is pressure supported from the host stellar radiation, causing the gas to orbit at sub-Keplerian speeds. Feeling a headwind, solids slow, losing angular momentum and migrating inwards. As more solid particles migrate inwards, they cluster together, and if the solid-to-gas ratio is sufficiently large (order unity \citealt{youdin2005}), then the backreaction of the dust on the gas will change the local gas velocity. This, in turn, alters the local drag force, promoting the pile up of solids which may gravitationally collapse if they become sufficiently large \citep{youdin2011}. These groups of solid particles may begin to accrete pebbles until they form giant planet cores \citep{lambrechts2012}. 

%
%reducing local headwind by changing 
 %This encourages further clustering, and when sufficiently large, these groups of solid particles may gravitationally collapse \citep{youdin2011}, and can begin to accrete pebbles until they form giant planet cores \citep{lambrechts2012}. However, the streaming instability requires a sufficiently large dust-to-gas ratio, such that the back reaction of the dust 
%%
%
%and as more pebbles migrate inwards, 
%
%t%hey group together, as the headwind reduces locally. When sufficiently large, these groups of pebbles gravitationally collapse \citep{youdin2011}, and can begin to accrete pebbles until they form giant planet cores \citep{lambrechts2012}.
%
%Doesn't the streaming instability also require a large dust-to-gas ratio so that the backreaction of the dust on the gas also changes the gas velocity and hence changes the resulting drag force on the solids, promoting the pile up of solids?
 %
 At smaller scales, the \textit{bouncing barrier} prevents coagulation of dust grains, as particles of a given size, above a certain velocity, are more likely to bounce off each other than they are to coagulate. This results in growth typically halting at around $\sim$ 1 mm in size. However, there is evidence to suggest that this could be beneficial to planetesimal formation, since the introduction of a few $\sim$ cm sized grains (e.g, through radial drift) can act as catalyst to grain growth, sweeping up grains, while preventing the growth of too many larger objects which would otherwise smash each other apart \citep{windmarketal2012}.

 It is generally accepted that disc fragmentation is very unlikely in the inner regions (< 50 au) of a protostellar disc \citep{rafikov2005}. However, the outer regions of protostellar discs may well be susceptible to fragmentation, offering a formation mechanism for directly imaged planets such as those in the HR8799 system \citep{marois2008,nerobjorkman8799,kratter2010failedbinary}. Core accretion models struggle to explain objects such as those in HR8799, with four planets orbiting at 14, 24, 34 and 68 au, with masses of $\sim$ 5 M$_{\mathrm{J}}$ \citep{marois2008,marois2010}, since there is not thought to be enough material to form such massive objects at these distances. Additionally, the growth timescales of such objects, through core accretion, typically exceed disc lifetimes by a factor of at least $\sim$ 3, using conservative estimates \citep{pollack1996}. Gravitational instability may, perhaps, offer an explanation as to the formation mechanism of these systems.

 However, it has been suggested that disc fragmentation rarely forms planetary mass objects \citep{riceetal2015}, with some hydrodynamics simulations \citep{stamatelloswhitworth2009} suggesting objects formed by this mechanism quickly grow to brown dwarf masses ($M > $ 13 M$_{\mathrm{Jup}}$), with lower limits placed on the fragment mass of $\sim 3- 5$ M$_{\mathrm{J}}$ \citep{kratter2010failedbinary,forganrice2011}. This is compounded by the recent possible observation, for the first time, of disc fragmentation in action \citep{tobin2016}, which shows the birth of three protostars that are well above the upper limit of the planetary mass regime.

The recent reformulation of the GI scenario in to what is now known as "tidal downsizing" (\citealt{boleyetal2010,boleyetal2011,nayakshin2010,nayakshin2011a,nayakshin2011b}) does, however, have positive implications for producing low-mass planets at low semi-major axes. The key is to consider the subsequent evolution of fragments into planetary embryos, through dust growth, radial migration and tidal disruption. \citet{forganricepopsynth2013} combined the physical processes of tidal downsizing with semi-analytic models of disc evolution \citep{ricearmitage2009} to produce the first population synthesis model for planets formed through GI. Given the similarities between these fragments and ``first cores'' (see, e.g. \citealt{masunaga1998}), they were modelled as polytropic spheres, with polytropic index $n=1.5$.

They found that $\sim40\%$ of fragments that formed are ultimately tidally destroyed by the central star, and of those that survive $\sim 40\%$ are gas-giant planets with solid cores of $5\sim 10$ earth masses, and the rest are brown dwarfs with no solid core. They also found that low mass embryos tend to remain at larger semi-major axes due to the tidal downsizing process. Out of over 1 million fragments, there was only one terrestrial type planet (core with no gaseous envelope). These results are inconsistent with GI being the dominant planet formation mechanism, but they are certainly consistent with GI forming brown dwarfs and gas giant planets at large radii.

Population synthesis models, by necessity, make simplifying assumptions about the physics that governs the evolution of each planetary system. In particular, interactions between forming protoplanets, and the interaction of the disc with these protoplanets, are not included in the population synthesis models of \citet{forganricepopsynth2013} that we discuss here. In fact, at the time of writing, these effects are not included in any GI population synthesis models. How important these interactions are in determining the final orbital configuration of a system is something that should be carefully considered before further developments are made to such a model. Quantifying the importance of these interactions is difficult, but some headway can be made by performing SPH simulations of fragmenting protostellar discs, and carefully tracking the evolution of fragments' orbital and physical properties throughout the duration of the simulation.
%To understand the influence of a fragment's environment on the subsequent evolution of the fragment requires careful tracking of the fragment's properties at a variety of stages in its lifetime.

In this work, we analyse a suite of SPH simulations of fragmenting protostellar discs, identifying fragments using two different methods. The first, based on the \texttt{CLUMPFIND} algorithm \citep{clumpfind,smith2008}, is done using the gravitational potential, and the second is a new method that uses density derivatives. We do not use sink particles \citep{bateetal1995} in our simulations, as by using only SPH particles, we are able to determine the fragment internal structure as it migrates through the disc, and better understand the orbital evolution of the fragment, which is sensitive to its radial mass distribution. We discuss the relative merits of our different detection methods for our simulations, which show a variety of fragmentation scenarios. We consider the implications of our results for current GI population synthesis models, and finally we consider the orbital and physical properties of the fragments in the simulations, comparing them to the orbital and physical properties of the population synthesis models of \citet{forganricepopsynth2013}. 

The paper is ordered as follows: In section \ref{sec:method}, we describe our overall method, outlining our chosen formalism of SPH in section \ref{subsec:SPH} and detailing the simulation setup in section \ref{subsec:setup}. We present our algorithms in section \ref{subsec:algorithms}, describing our new approach in section \ref{subsubsec:derivative} and our adaptation of an existing approach in section \ref{subsubsec:clumpfind}. We describe our results in section \ref{sec:results}, detailing the relative merits of the different approaches in section \ref{subsec:relative}. We compare our results to current gravitational instability population synthesis models in section \ref{subsec:compareGI}. We outline orbital and spin properties of our fragments in section \ref{subsec:orbital} and \ref{subsec:spin} respectively. We describe density and temperature profiles of fragments with the most interesting histories in section \ref{subsec:profiles}, and discuss our findings and conclude in section \ref{sec:conclusion}.
%We present the orbital and physical properties of these fragments, show a variety of fragmentation scenarios, and compare our results with current GI population synthesis models.
%
%We do this with a novel adaptation of the CLUMPFIND \citep{clumpfind} algorithm, which allows us to track particle populations during the simulations without using sink particles. By purely using hydrodynamics, we are able to calculate fragment internal structure as it migrates through the disc, and better understand the orbital evolution of the fragment, which is sensitive to the radial mass distribution of the fragment.

%=======================================================
%          MODEL
%=======================================================
\section{Method}
\label{sec:method}
%=================================================
% Column density plots all simulations
%================================================
\begin{figure*}
    \centering
    \begin{subfigure}[b]{0.3\textwidth}
      \centering
      Simulation 1\par\smallskip
      \begin{tikzpicture}
        \node (img) {\includegraphics[width=\textwidth]{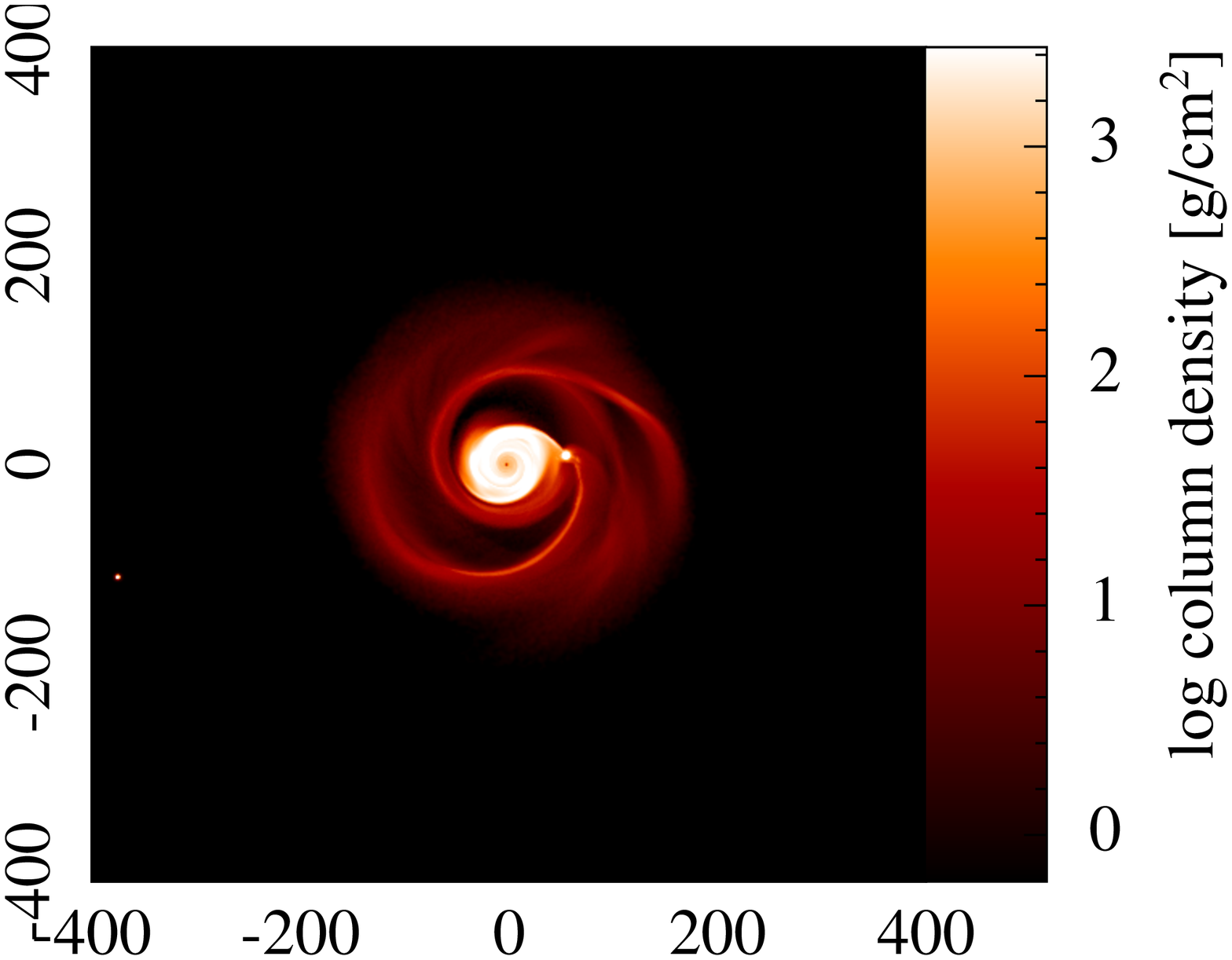}};
        \node[below left,text width=4.4cm,align=center,white] at (img.north east)
             {\\~\\ T=4238 yrs};
      \end{tikzpicture}
        %\includegraphics[width=\textwidth]{sim1.eps}%{output}
       % \caption{$\dot{M}= 2.8\times 10^{-7}\mdotunits$}
       % \caption{}
        \label{fig:sim1density}
    \end{subfigure}
    \begin{subfigure}[b]{0.3\textwidth}
      \centering
      Simulation 2\par\smallskip
      \begin{tikzpicture}
        \node (img) {\includegraphics[width=\textwidth]{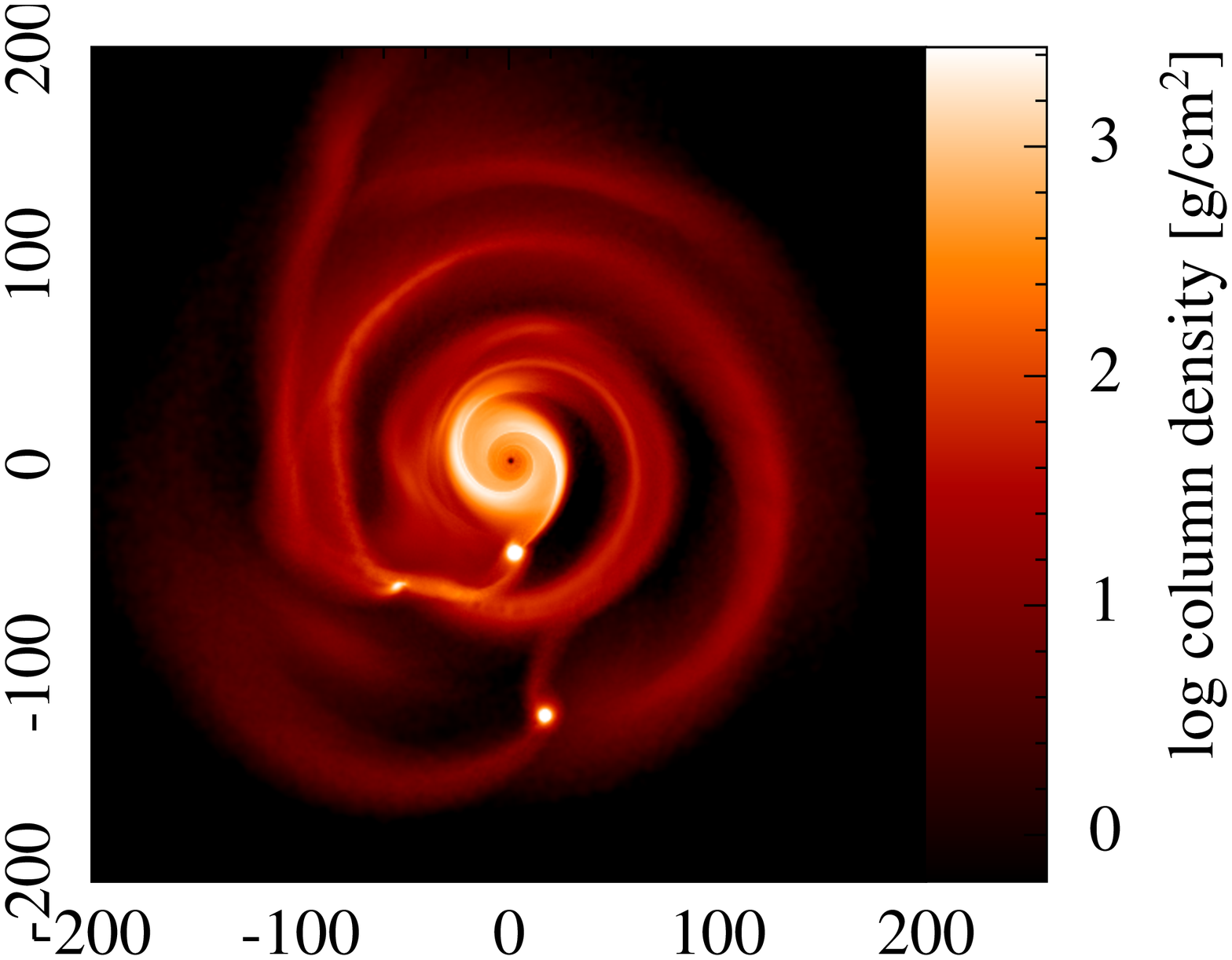}};
        \node[below left,text width=4.4cm,align=center,white] at (img.north east)
             {\\~\\ T=4341 yrs};
      \end{tikzpicture}     
        %\includegraphics[width=1.0\textwidth]{sim2.eps}
        %\caption{$\dot{M}= 3.2\times 10^{-7}\mdotunits$}
        %\caption{}
        \label{fig:sim2density}
    \end{subfigure}
    \begin{subfigure}[b]{0.3\textwidth}
       \centering
       Simulation 3\par\smallskip
      \begin{tikzpicture}
        \node (img) {\includegraphics[width=\textwidth]{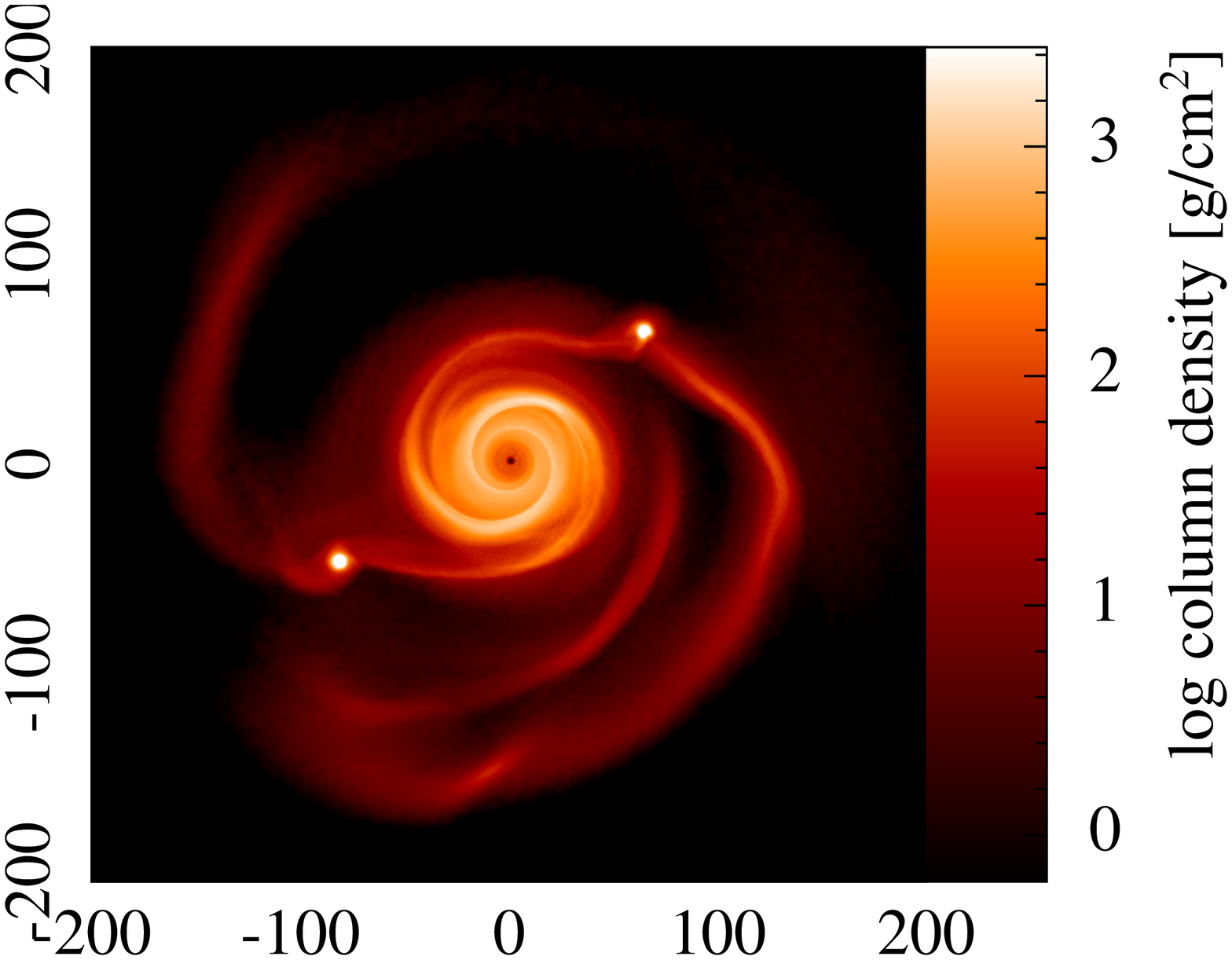}};
        \node[below left,text width=4.4cm,align=center,white] at (img.north east)
             {\\~\\ T=3082 yrs};
      \end{tikzpicture}
        %\includegraphics[width=1.\textwidth]{sim3.eps}
        %\caption{$\dot{M}= 3.6\times 10^{-7}\mdotunits$}
       % \caption{}
        \label{fig:sim3density}
    \end{subfigure}
    \begin{subfigure}[b]{0.3\textwidth}
       \centering
       Simulation 4\par\smallskip
       \begin{tikzpicture}
        \node (img) {\includegraphics[width=\textwidth]{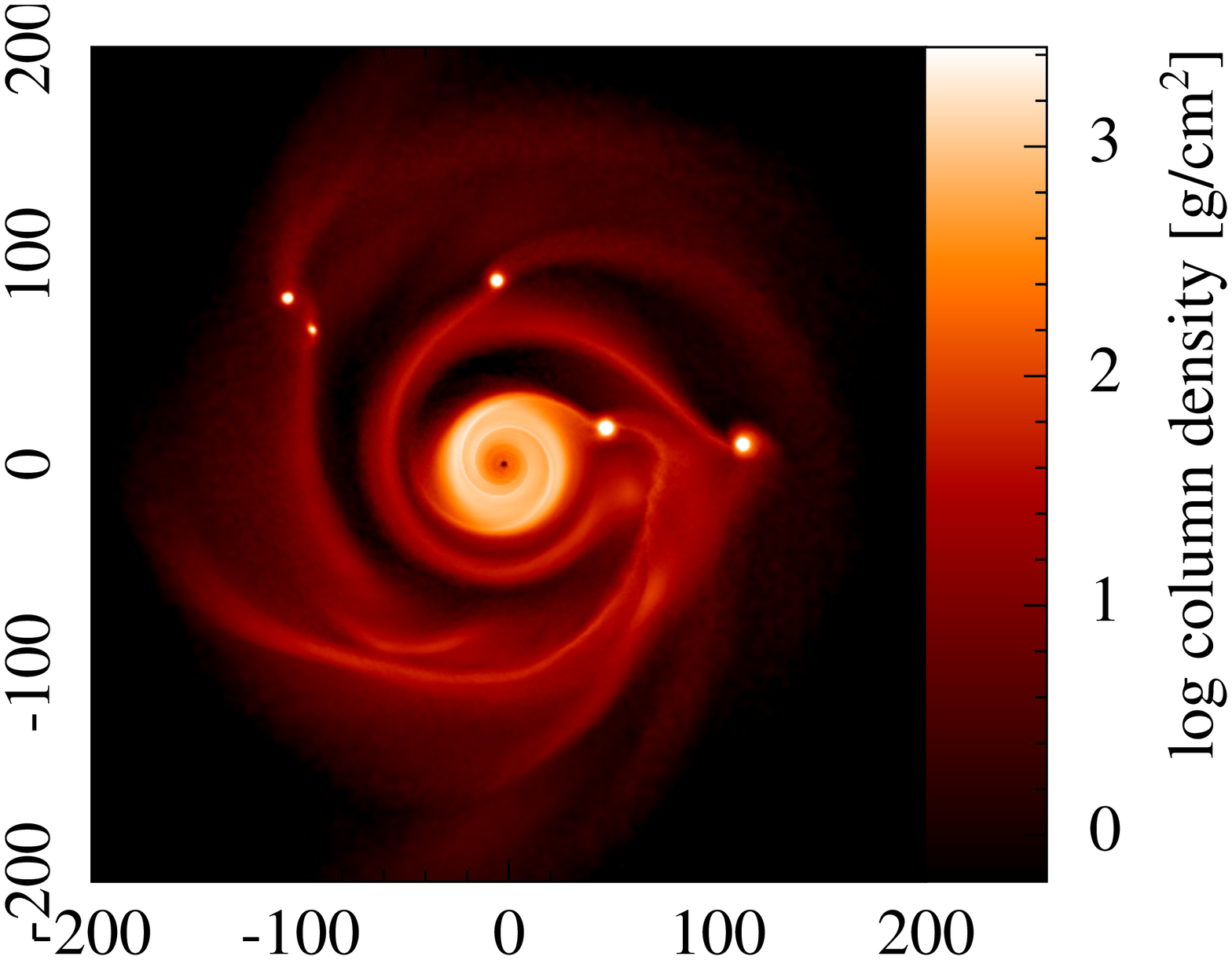}};
        \node[below left,text width=4.4cm,align=center,white] at (img.north east)
             {\\~\\ T=4059 yrs};
      \end{tikzpicture}
        %\includegraphics[width=\textwidth]{sim4.eps}
      %  \caption{$\dot{M}= 2.8\times 10^{-7}\mdotunits$, T=10K, q=0.35}
        %\caption{}
        \label{fig:sim4density}
    \end{subfigure}
    \begin{subfigure}[b]{0.3\textwidth}
       \centering
       Simulation 5\par\smallskip
       \begin{tikzpicture}
        \node (img) {\includegraphics[width=\textwidth]{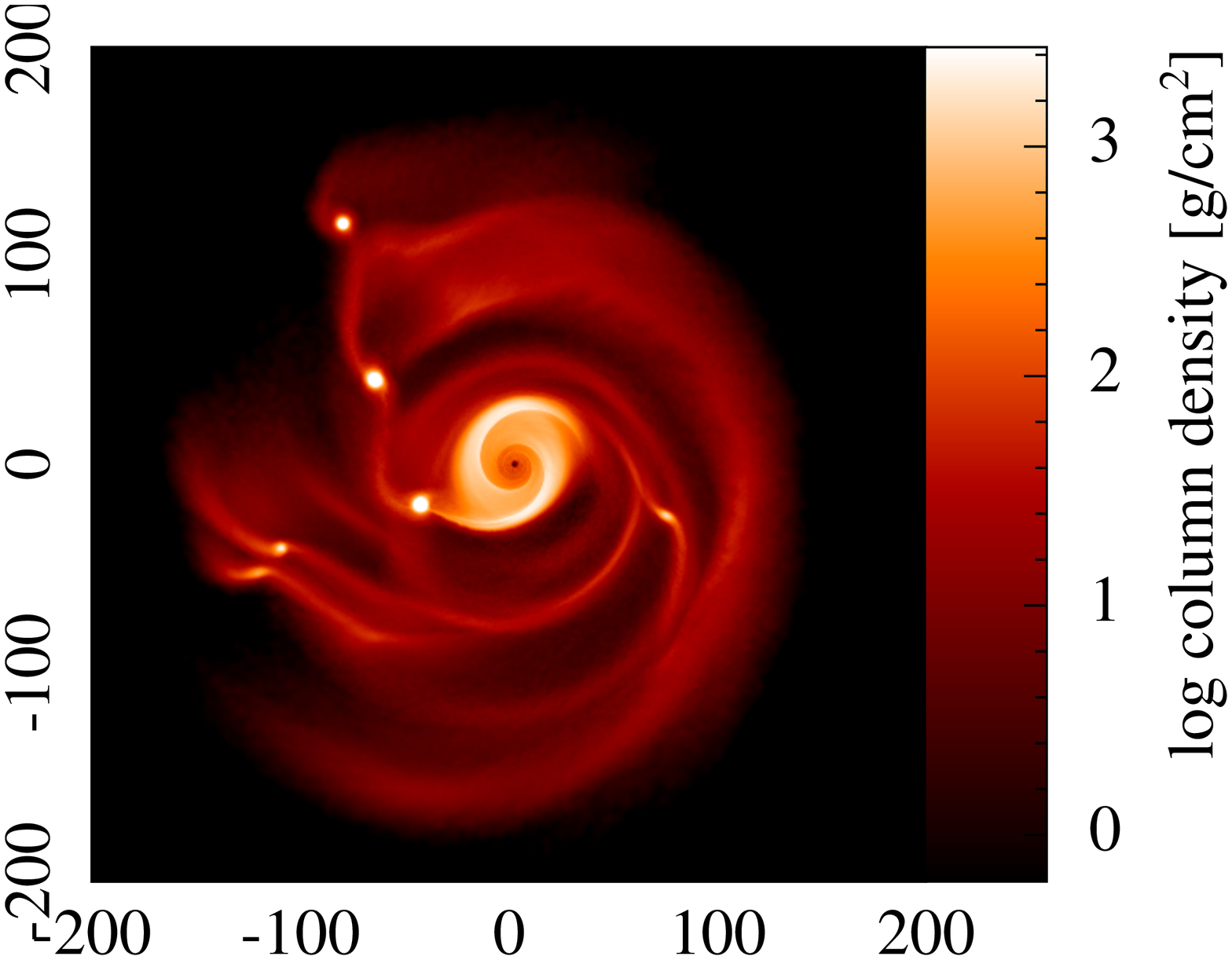}};
        \node[below left,text width=4.4cm,align=center,white] at (img.north east)
             {\\~\\ T=3159 yrs};
      \end{tikzpicture}
        %\includegraphics[width=\textwidth]{sim5.eps}
       % \caption{$\dot{M}= 3.2\times 10^{-7}\mdotunits$, T=10K, q=0.37}
      %  \caption{}
        \label{fig:sim5density}
    \end{subfigure}
    \begin{subfigure}[b]{0.3\textwidth}
       \centering
       Simulation 6\par\smallskip
       \begin{tikzpicture}
        \node (img) {\includegraphics[width=\textwidth]{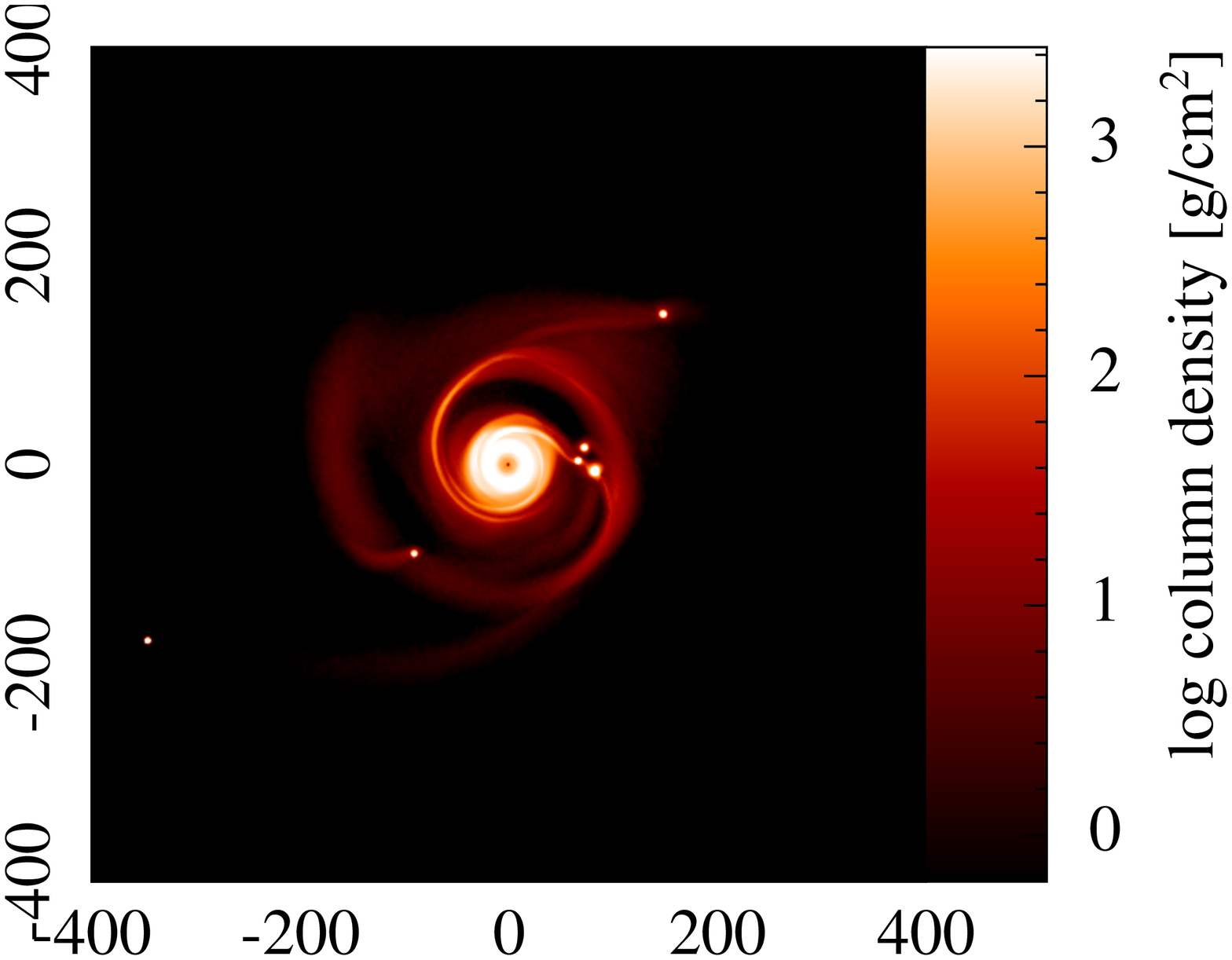}};
        \node[below left,text width=4.4cm,align=center,white] at (img.north east)
             {\\~\\ T=3134 yrs};
      \end{tikzpicture}
        %\includegraphics[width=1.\textwidth]{sim6.eps}
        %\caption{$\dot{M}= 3.6\times 10^{-7}\mdotunits$,T=10K,q=0.35}
       % \caption{}
        \label{fig:sim6density}
    \end{subfigure}
    \begin{subfigure}[b]{0.3\textwidth}
        \centering
        Simulation 7\par\smallskip
        \begin{tikzpicture}
        \node (img) {\includegraphics[width=\textwidth]{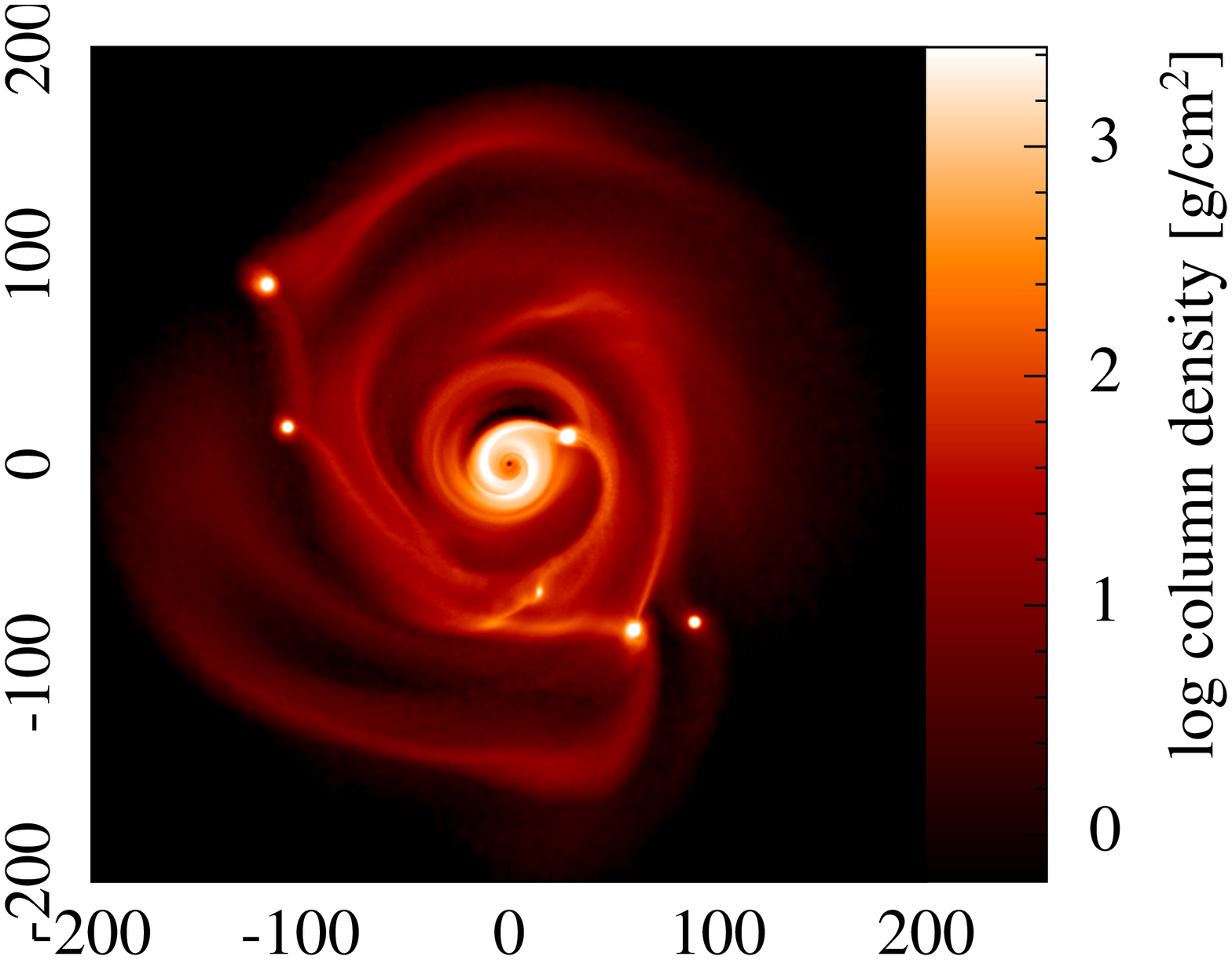}};
        \node[below left,text width=4.4cm,align=center,white] at (img.north east)
             {\\~\\ T=3956 yrs};
      \end{tikzpicture}
        %\includegraphics[width=\textwidth]{sim7.eps}
        %\caption{$\dot{M}= 2.8\times 10^{-7}\mdotunits$, T=30K,q=0.44}
       % \caption{}
        \label{fig:sim7density}
    \end{subfigure}
    \begin{subfigure}[b]{0.3\textwidth}
        \centering
        Simulation 8\par\smallskip
        \begin{tikzpicture}
        \node (img) {\includegraphics[width=\textwidth]{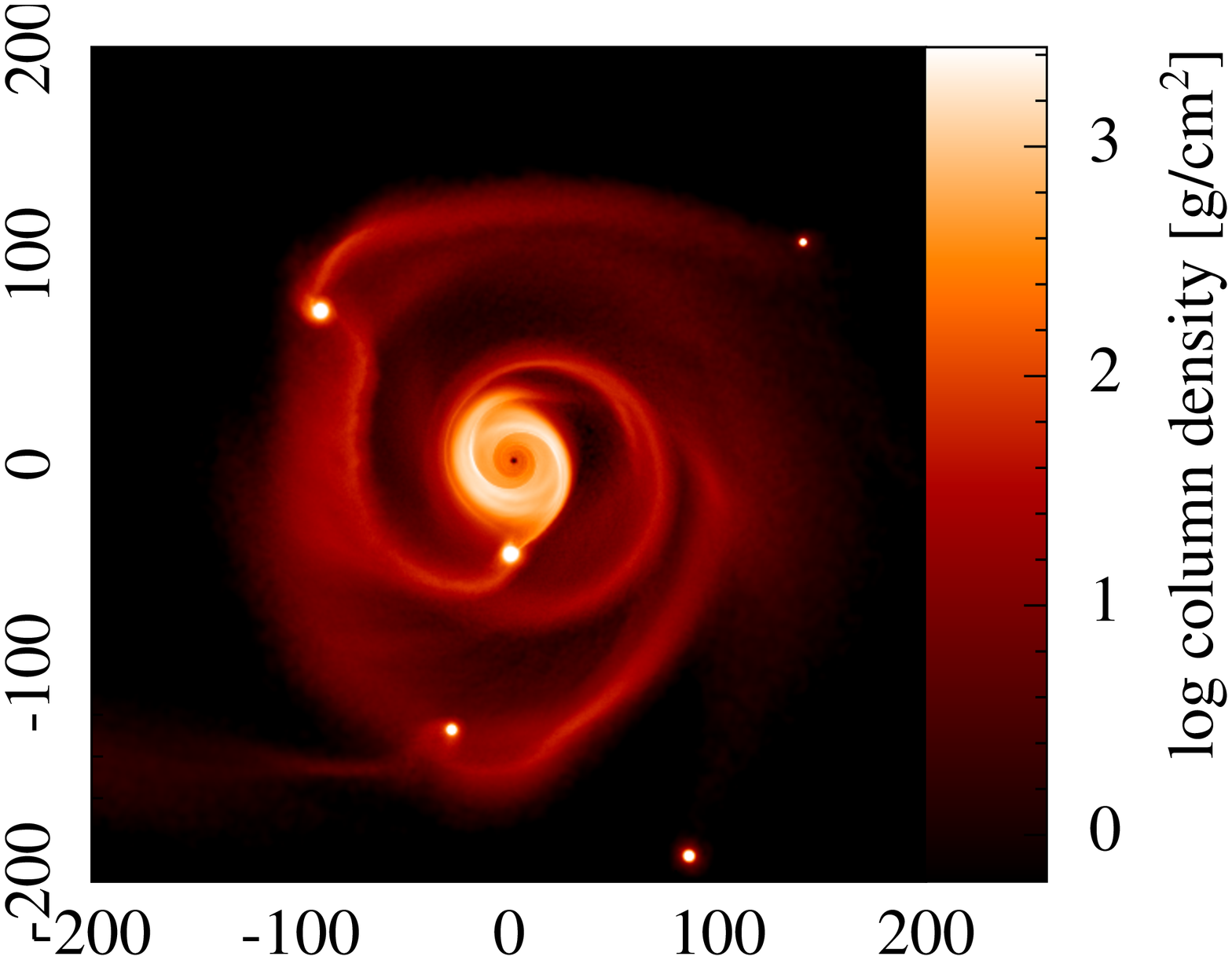}};
        \node[below left,text width=4.4cm,align=center,white] at (img.north east)
             {\\~\\ T=3648 yrs};
      \end{tikzpicture}
        %\includegraphics[width=\textwidth]{sim8.eps}
       % \caption{$\dot{M}= 3.2\times 10^{-7}\mdotunits$, T=30K,q=0.46}
       % \caption{}
        \label{fig:sim8density}
    \end{subfigure}
    \begin{subfigure}[b]{0.3\textwidth}
        \centering
        Simulation 9\par\smallskip
        \begin{tikzpicture}
        \node (img) {\includegraphics[width=\textwidth]{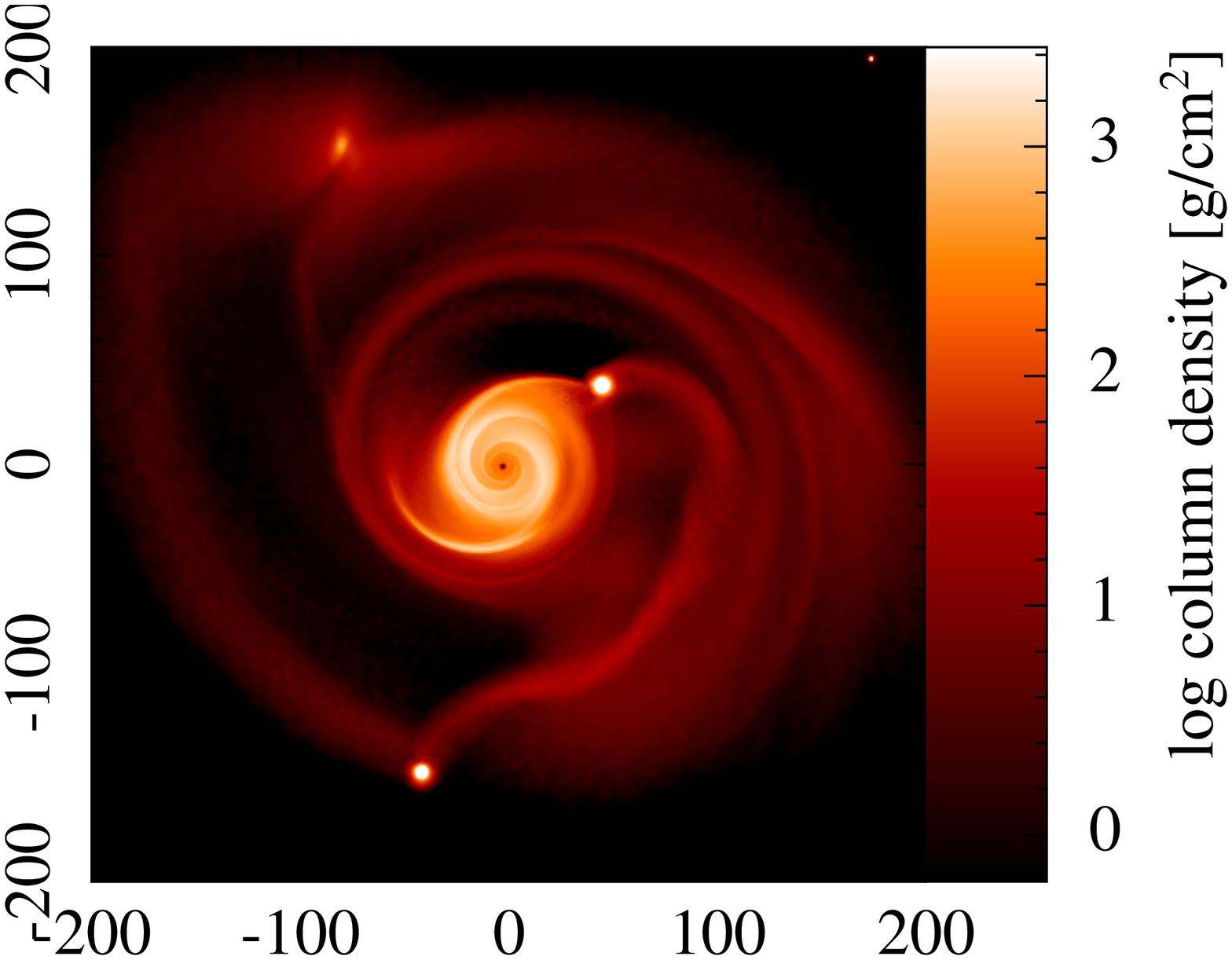}};
        \node[below left,text width=4.4cm,align=center,white] at (img.north east)
             {\\~\\ T=4881 yrs};
      \end{tikzpicture}
       % \includegraphics[width=1.\textwidth]{sim9.eps}
        %\caption{$\dot{M}= 3.6\times 10^{-7}\mdotunits$,T=30K,q=0.49}
        %\caption{}
        \label{fig:sim9density}
    \end{subfigure}
    \caption{Column density plots of the final fragment configuration for all 9 simulations. Despite almost identical initial conditions, there are a variety of ultimate configurations, and a variety of times at which it becomes computationally unfeasible to continue the simulation using only hydrodynamics. Simulations 1 and 6 show an ejected clump at large radial separation, and the top left hand corner of simulation 9 shows a fragment forming just below the threshold detection of our algorithm.}\label{fig:columndensity}
%\end{minipage}
\end{figure*}
%==End column density plots==========================================
We run a suite of 9 smoothed particle hydrodynamics (SPH) simulations, with the radiative cooling method of \cite{forgan2009}. Each disc uses 4 million particles, fragments to form at least two bound objects, and - aside from the random number seed used to initialise each disc - has identical initial conditions. Since our aim is to take advantage of running a fully hydrodynamic simulation by tracing the evolution of fragment radial profiles and mass distribution, we run the simulations for as long as it is computationally feasible using hydrodynamics only (without switching to sink particles). In practice, this means running each simulation until the most dense clump becomes too dense to calculate the next timestep. We give a brief description of SPH, and our selected radiative transfer formalism, in section \ref{subsec:SPH}. We discuss our methods of finding fragments in section \ref{subsec:algorithms}. 
%Once a fragment has been detected, we refer to it as a ``clump'', and adopt the nomenclature S1C2 to mean simulation 1, clump 2. Clump 1 in each simulation is the central star and the majority of the disc.
%Every simulation fragments to form one bound object, which from now on we refer to as ``clumps''. These clumps are identified by our modified CLUMPFIND algorithm, detailed in section \ref{subsubsec:clumpfind}. These clumps are then linked throughout the simulation using a merger tree algorithm, detailed in section \ref{subsubsec:mergertree}.
\subsection{Smoothed Particle Hydrodynamics}
\label{subsec:SPH}
Smoothed Particle Hydrodynamics (SPH) is a Lagrangian hydrodynamics formalism, that evolves a fluid by means of a distribution of pseudo-particles \citep{lucy1977,gingoldmonaghan1977}. There are many review articles about SPH (see e.g. \citealt{monaghan1992,monaghan2005,rosswog2009}), but the basic idea is that each particle has a position, mass, internal energy and velocity, and these parameters can be interpolated over to give fluid quantities at any position. Density is calculated by interpolation over the mass distribution, and pressure is determined using an equation of state with internal energy. Gravitational forces are usually computed using a TREE algorithm \citep{barneshut1989}, and then the discretised energy and momentum equations are solved. Particle velocities are updated using pressure and gravitational forces, and positions are then updated using these velocities. Internal energy changes are computed by calculating $P\mathrm{d}V$ work, viscous dissipation and radiative cooling and heat conduction.

Cooling calculations in SPH are no simple task. Accounting for polychromatic radiative transfer within a hydrodynamics simulation is not possible with current computational resources, and even post-processing a single snapshot with radiative transfer is computationally expensive \citep{stamatellos2005}. Historically, approximations to individual features of radiative transfer were used, such as the cooling time formalism:  $\dot{u} = -u/t_{\mathrm{cool}}$ \citep{rice2003}. Although this parameterisation is useful, allowing us to probe the effects of different cooling timescales in protoplanetary discs, it is somewhat limited, as it only allows us to model energy lost from an SPH particle. Realistically, if energy is lost from one SPH particle, at least some of that energy will be gained by its surrounding neighbours - this is known as radiative transfer.

Since our aim here is to trace the orbital and profile evolution of fragments within protostellar discs, we wish to capture the effects of radiative transfer as far as is feasible. Therefore, the cooling we implement is the hybrid method of \cite{forgan2009}. The details of the algorithm are given in \cite{forgan2009}, however the basic ideas merge the polytropic cooling method of \cite{stamatellos2007} with the flux-limited diffusion method of \cite{mayeretal2007}, which builds on conduction modelling work by \citet{clearymonaghan1999} and the flux-limiter described in \cite{bodenheimeretal1990}. The biggest advantage is the complementary nature of these two methods, energy loss is handled by polytropic cooling (which flux-limited diffusion cannot do), and positive energy exchange between neighbouring particles is handled by flux-limited diffusion (which polytropic cooling cannot do). Since each method handles a separate process, there is no ``double counting'' in any part of the system's overall energy, and these separate parts can simply be summed to calculate the total energy change.
\subsection{Simulation setup}
\label{subsec:setup}
We run a total of 9 simulations of 0.25 M$_\odot$ discs, with a 1 M$_\odot$ central star, an inner radius of 10 au, an outer radius of 100 au, and a radial density profile of $\Sigma\propto r^{-1}$. All discs are initially identical in global properties, varying only the random number seed (the integer used to set the starting point for a sequence of random numbers) used to initialise the disc. The SPH particles are randomly distributed in $\phi$, where $\phi$ is azimuthal angle, and the $r$ position of each particle is determined through the iterative use of an accept-reject algorithm, accepted so long as the position of the particle maintains the desired surface density profile.The $z$ position of the particle is similarly determined, accepted so long as the position maintains hydrostatic equilibrium. The velocity of each particle is exactly Keplerian. This technique results in discs that are identical in their global properties, differing only through a small amount of noise at the particle-separation level.

All discs are evolved until it is no longer computationally feasible to continue, which in reality means the density of a fragment has become so high that timesteps cannot be computed without switching that mass to a sink particle. However, since here we wish to examine physical and orbital properties of the fragments which are influenced by their radial mass distribution, we do not do this. All of the simulations fragment to form at least two bound objects, and their ultimate configuration is shown in Figure \ref{fig:columndensity}, which shows 9 column density plots, in physical units, and illustrates a variety of fragmentation scenarios as the simulation's final configuration. We discuss this in detail in section \ref{sec:results}, but include images now to make the explanation of our methods in section \ref{subsec:algorithms} clearer.

%==End column density plots==========================================
\subsection{Algorithms}
\label{subsec:algorithms}
We present here two methods of detecting fragments in SPH simulations, and one method of linking them together between timesteps. Once a fragment has been identified, we then refer to it as a "clump". The first method of detection is based on the clump finding approach of \cite{smith2008} (which is in turn based on the publicly available \texttt{CLUMPFIND} algorithm developed by \citealt{clumpfind}). The basic idea is to perform an ordered search on SPH particles from high (physical) density to low density. The highest density particle $i$ forms the center of a clump, and if the next particle in the list is a neighbour (i.e. in close spatial proximity), it is also added to this clump. If it is not a neighbour, it forms the center of a new clump. This process is continued to the next most dense particle, until a minimum density threshold is reached.

%==============================
% DDS Figure
%==============================
\begin{figure}
\centering
\includegraphics[width=\linewidth]{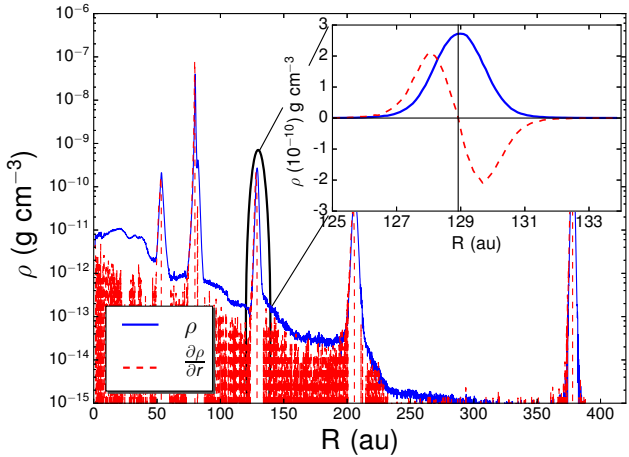}
\caption{Our density derivative search method on the disc shown in Figure \ref{fig:columndensity}, simulation 6. The solid blue line shows the radial density profile of the disc, and the dashed red line shows the derivative of this with respect to $r$. The zoomed region demonstrates how the negative zero crossing of the derivative identifies one of the real density peaks.\label{fig:derivative}}
\end{figure}
%==============================
The search in this manner, from least dense to most dense particle for our SPH simulations of protostellar discs was unsuccessful in identifying clumps in our simulation. We are faced with a different scenario to \citet{smith2008}, who used their algorithm in molecular cloud cores. Once our discs have evolved enough to fragment, the inner disc is so dense that many of the particles inside $\sim 20$ au fulfil the criteria to become the head of their own clumps. This results in neighbours, since this is a friends-of-friends algorithm, belonging to these even if they are in the outer disc, where fragmentation has actually occurred.

This problem was solved, to some extent, by using the gravitational potential of the particles, rather than the density, for the ordered search. We discuss this method in section \ref{subsubsec:clumpfind}. The inability of this method to identify low mass, fluffy fragments, or fragments that are so deep in the potential well of the central star that they are ultimately tidally destroyed, prompted the development of an approach that could correctly identify such fragments. The approach uses a gridded derivative search of the SPH interpolated density of the particles, and is discussed in section \ref{subsubsec:derivative}.

Finally, in both cases, the clumps are linked between timesteps using a merger tree algorithm, typical of dark matter halo tracers in cosmological simulations (see, for example, \citealt{mergertrees}). This process is detailed in section \ref{subsubsec:mergertree}.

%and use a neighbour-finding approach to assign particles
%\cite{smith2008}
%The first method of detection is based on the the CLUMPFIND \cite{clumpfind} algorithm, which we have modified
%We present here a novel method of tracing particles in a hydrodynamics simulation using an adaptation of the CLUMPFIND %algorithm, which is described fully in \cite{clumpfind}. Here we describe how our version, modified for use in SPH simulations, %works. We then describe our method for linking clumps using a merger tree in section \ref{subsubsec:mergertree}.

\subsubsection{Gravitational potential search (\texttt{CLUMPFIND})}
\label{subsubsec:clumpfind}
%=======================================
% Radial migration plot
%=======================================
\begin{figure*}
\begin{center}
\begin{tabular}{cc}
\textbf{Density derivative search} & \textbf{Potential search}\\
\includegraphics[width=0.5\linewidth]{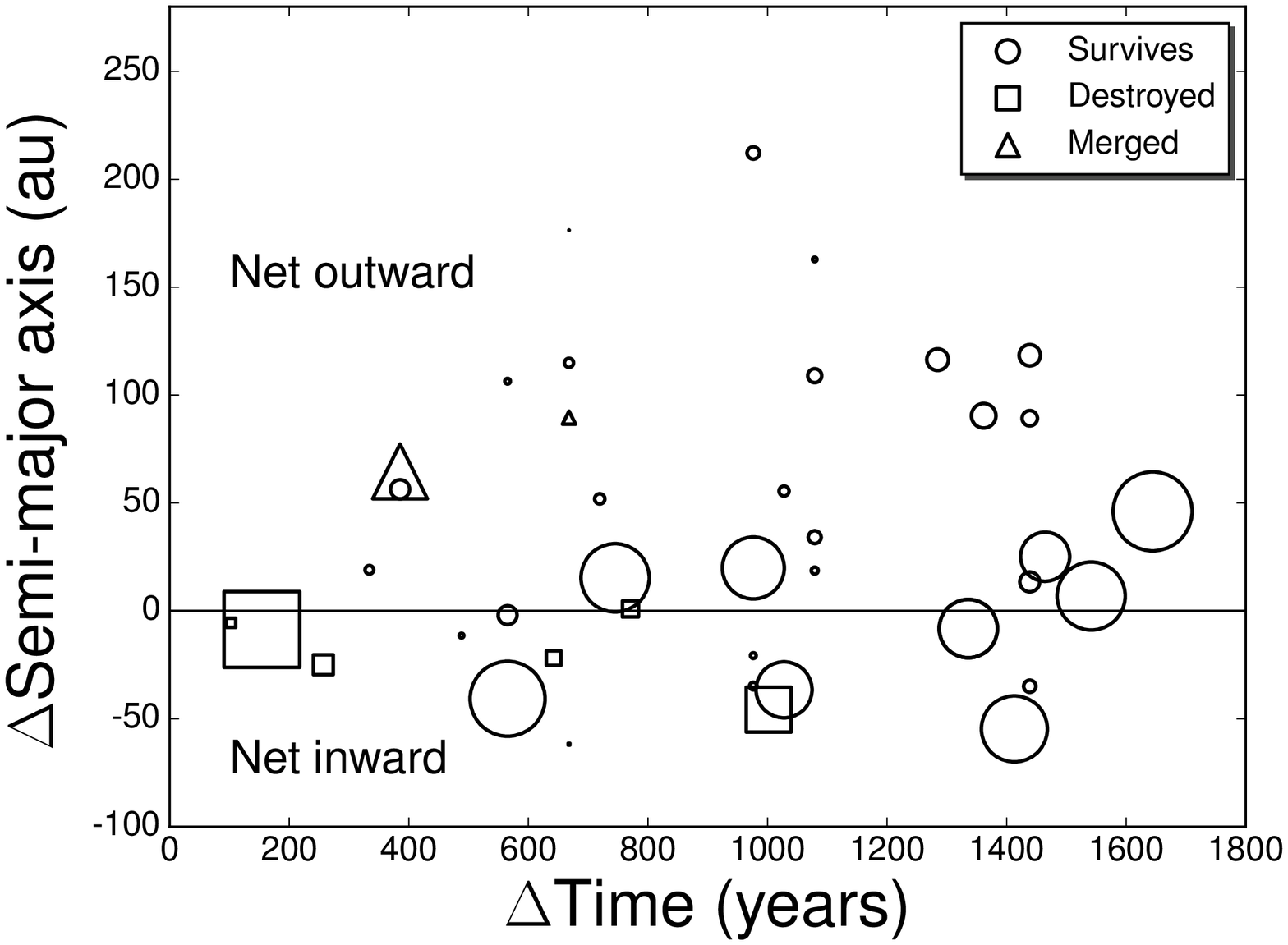}& \includegraphics[width=0.5\linewidth]{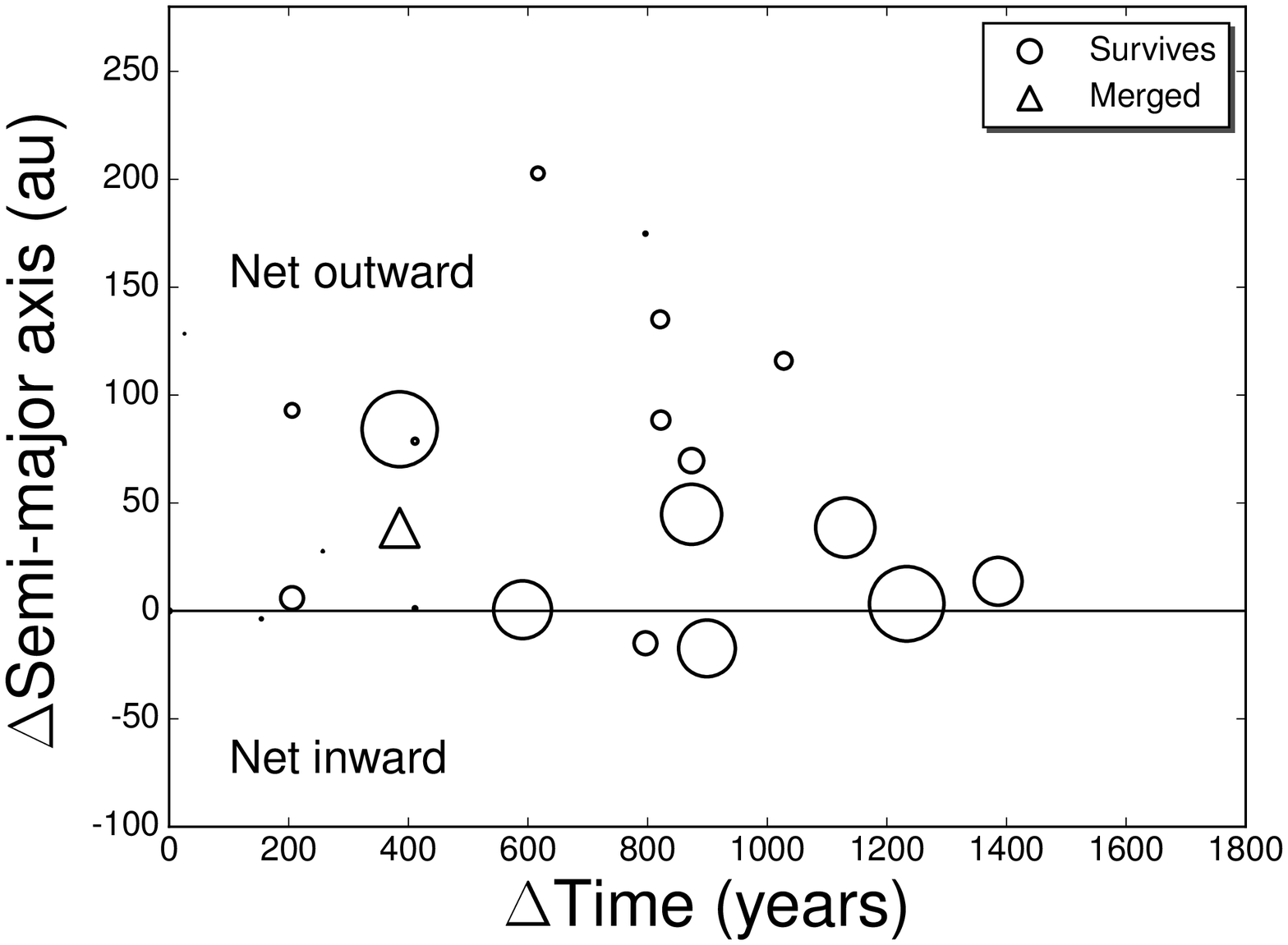}

\end{tabular}
%Plot made by plotradialmigration.py
\caption{ Plots showing total change in semi-major axis, for all clumps in all 9 simulations, plotted against the time between initial identification, and the either the end of the simulation, or the last timestep in which they are identified, if they are destroyed or merged. Larger markers correspond to more massive clumps. The left hand panel shows clumps as they are identified by our density derivative search method, and the right hand panel shows clumps as they are identified using our ordered potential search method. Circular markers indicate the clump survived the duration of the simulation, square markers indicate tidally destroyed clumps, and triangle markers indicate the clump was subsumed by another clump. The right hand panel shows that clumps that are ultimately destroyed are not detected by the ordered potential search. The left hand panel shows that $\sim 20\%$ of fragments are ultimately tidally destroyed. There are no identical markers in both plots because the clumps are detected at different times in the simulation, and thus migrate different distances.\label{fig:migration}}
\end{center}
\end{figure*}
%======================================
Broadly speaking, clumps are created with a unique integer identifier (ID) at the local minima of the gravitational potential, so long as there are at least a minimum number of neighbour particles above some defined ``noise'' level. For our purposes, we define this critical number as $n_{\rm{critical}} = n_{\rm{mean}}-5n_{\sigma}$, where $n_{\rm{mean}}$ is the mean number of neighbours each particle has, and $n_{\rm{sigma}}$ is the standard deviation of the number of neighbour particles. We do this because the neighbour lists of particles at low density can be sparse due to the algorithms used to calculate the smoothing length, $h$.

We begin by creating a clump at the location of the central sink particle (star). All particles are then sorted by their gravitational potential energy, and we loop over the particles in order of most negative to least negative gravitational potential energy (i.e. most bound to least bound). We select the particle, $i$, with the most negative potential energy, and as long as it does not already belong to a clump (in which case, we exit and select the next particle), we iterate over particle $i$'s neighbours, $j$. If the majority of the neighbours $j$ (>50\%) are in a clump $k$, the particle is also in clump $k$. We assign the particle ID of $i$ to ID$_{i}=k$ and exit. If the majority of $i$'s neighbours are not in a clump, then since $i$ is the most bound particle it starts a new clump $l$, provided that $n_{\mathrm{neighbours}}>n_{\rm{critical}}$. We then loop over particle $i$'s neighbours $j$, assigning ID$_j=l$ so long as $j$ is not already in a clump. We then proceed to next most bound particle, $i+1$, and repeat \citep{forganpricebonnel2016}.

\subsubsection{Density Derivative Search}
\label{subsubsec:derivative}
In this method, we compute a 2D grid in cylindrical $r$ and $\phi$ coordinates, and bin all particles into these grid cells. The maximum density of a particle in each cell is then taken to be the peak density in that cell, $\rho_i$, which gives us a 2D sheet describing density maxima. The number of bins required to reasonably identify all clumps varies due to the stochastic nature of the simulations, more fragments with low density at $\sim$ 100 au require a larger number of bins to properly resolve, with our resolution criterion being that the number of clumps ultimately detected by the search is equal to the number of clumps that are determined ``by-eye''. If fewer clumps are detected by the search than ``by-eye'', then the resolution is increased until we detect these low-mass, low-density clumps. A typical resolution is 10,000 radial bins and 7200 azimuthal bins.

We next take the derivative of the peak density in that cell with respect to $r$ and $\phi$. As this is noisy in density space, we smooth these derivative, equivalent to making a new signal, where the element is now the average of $n$ adjacent elements, such that:
\begin{equation}
\frac{\partial\rho_i}{\partial r} = \frac{1}{n}\sum\limits_{{j=-\frac{n-1}{2}}}^{j=\frac{n-1}{2}} \frac{\partial\rho_{i+j}}{\partial r}
\end{equation}
and
\begin{equation}
\frac{\partial\rho_i}{\partial \phi} = \frac{1}{n}\sum\limits_{{j=-\frac{n-1}{2}}}^{j=\frac{n-1}{2}} \frac{\partial\rho_{i+j}}{\partial \phi}
\end{equation}
Despite its simplicity, this approach is best at removing white noise while keeping the sharpest step response. The value of $n$ ($\sim$ 100 is typically sufficient), like the number of bins in which to bin the data, must be optimised by the user to get the best compromise between smooth data, which removes false peaks, and data which is sensitive enough to identify small clumps.

We then use this smoothed derivative to identify clumps, which will be a ``real'' peak in the density. A real peak is identified by a sustained zero-crossing of the derivative with a negative gradient. Peaks due to noise will also have zero-crossings, but they are sustained for fewer bins than real peaks. These false peaks can be eliminated by requiring that the zero-crossing is sustained for $m$ bins, with $m$ optimised by the user to remove most (if not all) false detections while still detecting less dense clumps.

The radial search is shown in Figure \ref{fig:derivative}, which shows the radial density profile (blue solid line) and the derivative of the radial density profile (dashed red line) for the disc shown in simulation 6 of Figure \ref{fig:columndensity}. The zoomed section shows the peak of a clump, with the negative gradient zero-crossing of the derivative identifying the peak.

Once the particle marking the center of a clump ($i$) has been identified, we add all of $i$'s neighbour particles $j$ to that clump. We now loop over all particles which form that clump, adding their neighbours to this clump as well. We repeat this until we reach some density threshold. We found that adding neighbour particles until more than half the particles in the neighbour sphere are less dense than the inner 1 au of the disc produced good results. Once we have identified the bulk of the clump, we then proceed with a potential search described in section \ref{subsubsec:clumpfind}, which determines to which clumps the rest of the unidentified particles in the simulation belong.
\subsubsection{Merger Tree}
\label{subsubsec:mergertree}
%=========================================
% Time plots of mass
%=========================================
\begin{figure*}
  \textbf{\large{Potential search}}\par\smallskip
    \begin{subfigure}[b]{0.3\textwidth}
        \includegraphics[width=1.0\textwidth]{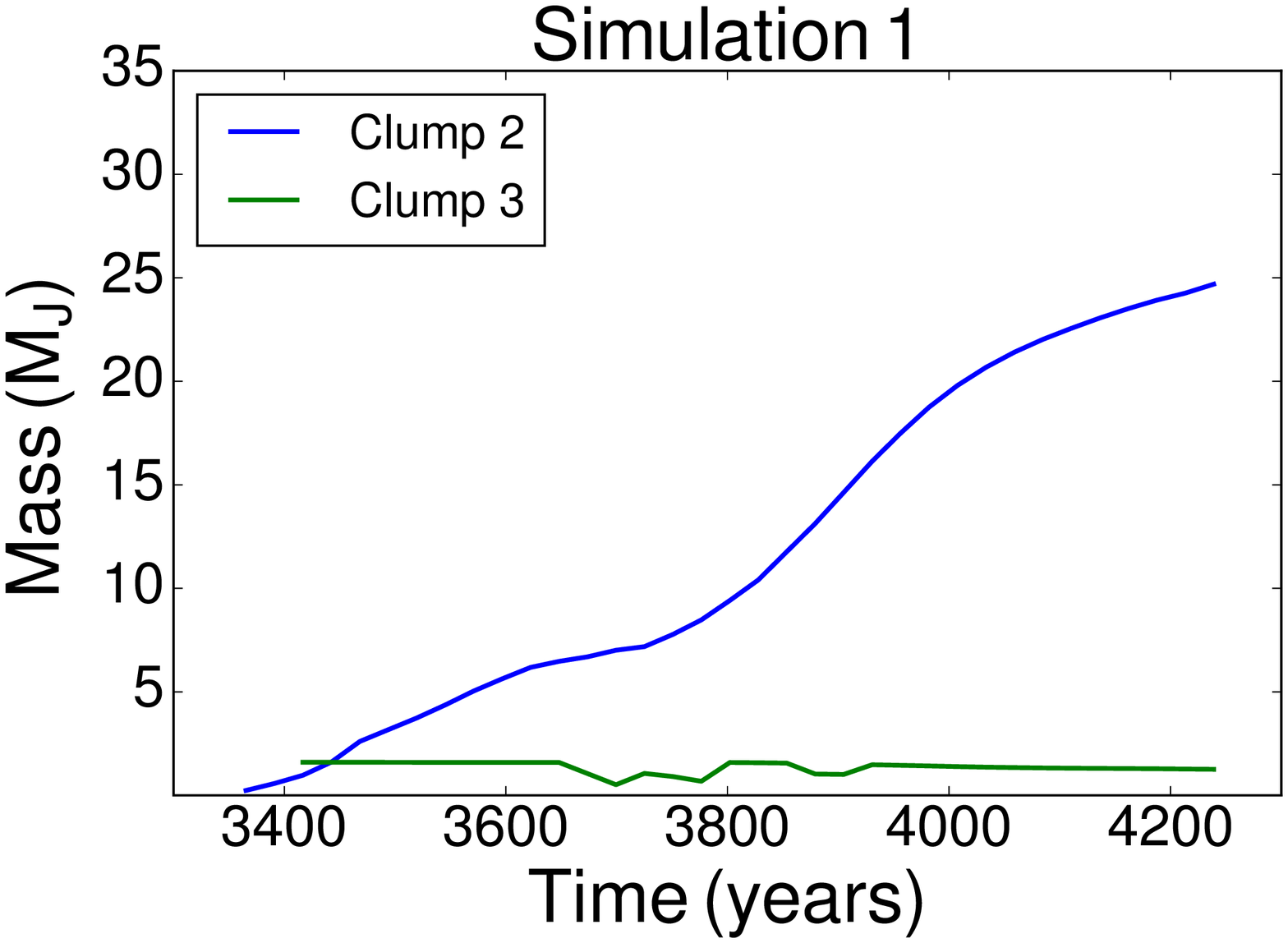}%{output}
       % \caption{$\dot{M}= 2.8\timeUonlys 10^{-7}\mdotunits$}
       % \caption{}
       \vspace{0cm}
        \label{fig:sim1massVtimeUonly}
    \end{subfigure}
    \begin{subfigure}[b]{0.3\textwidth}
        \includegraphics[width=1.0\textwidth]{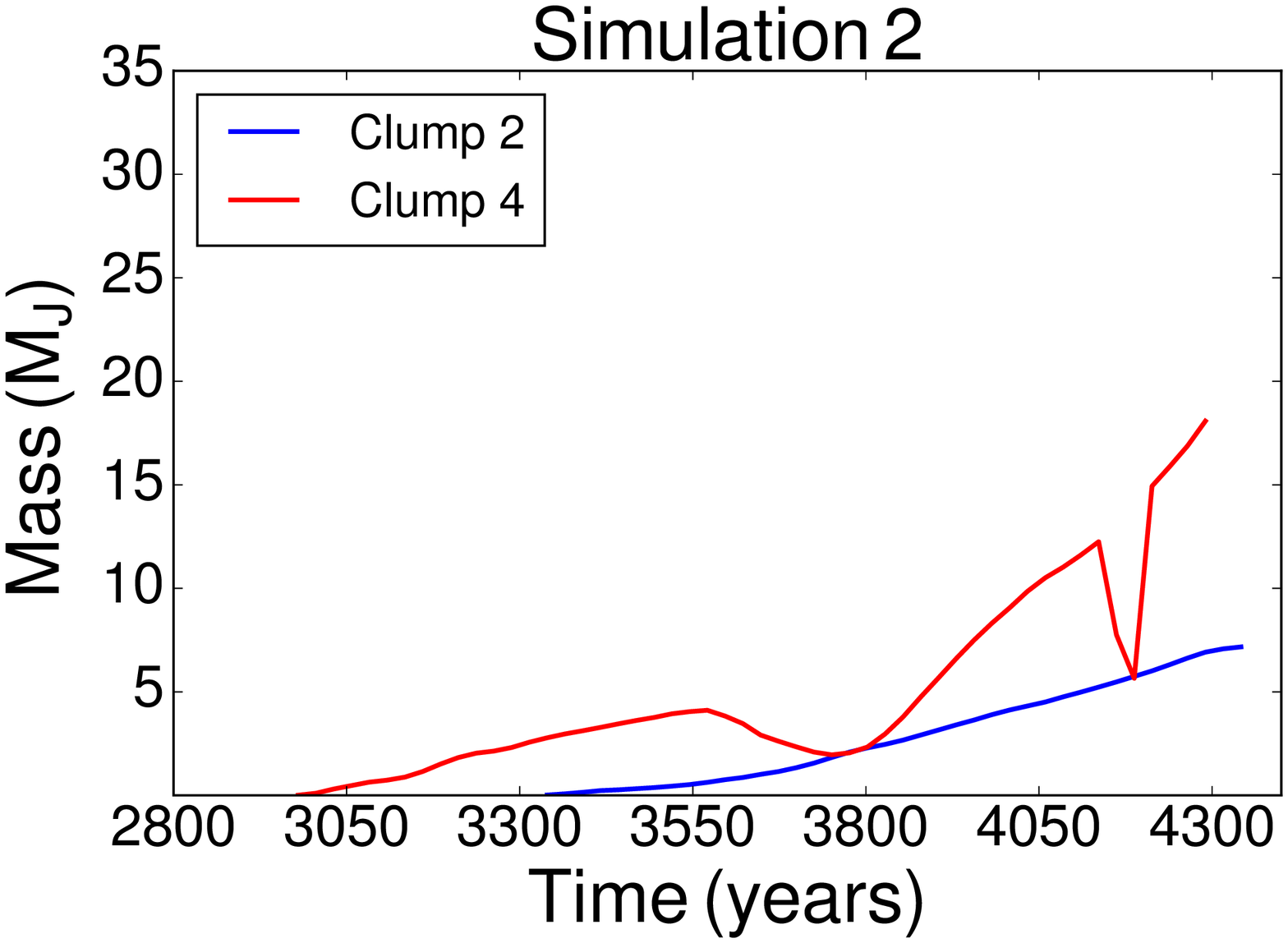}
        %\caption{$\dot{M}= 3.2\timeUonlys 10^{-7}\mdotunits$}
        %\caption{}
        \label{fig:sim2massVtimeUonly}
    \end{subfigure}
    \begin{subfigure}[b]{0.3\textwidth}
        \includegraphics[width=1.\textwidth]{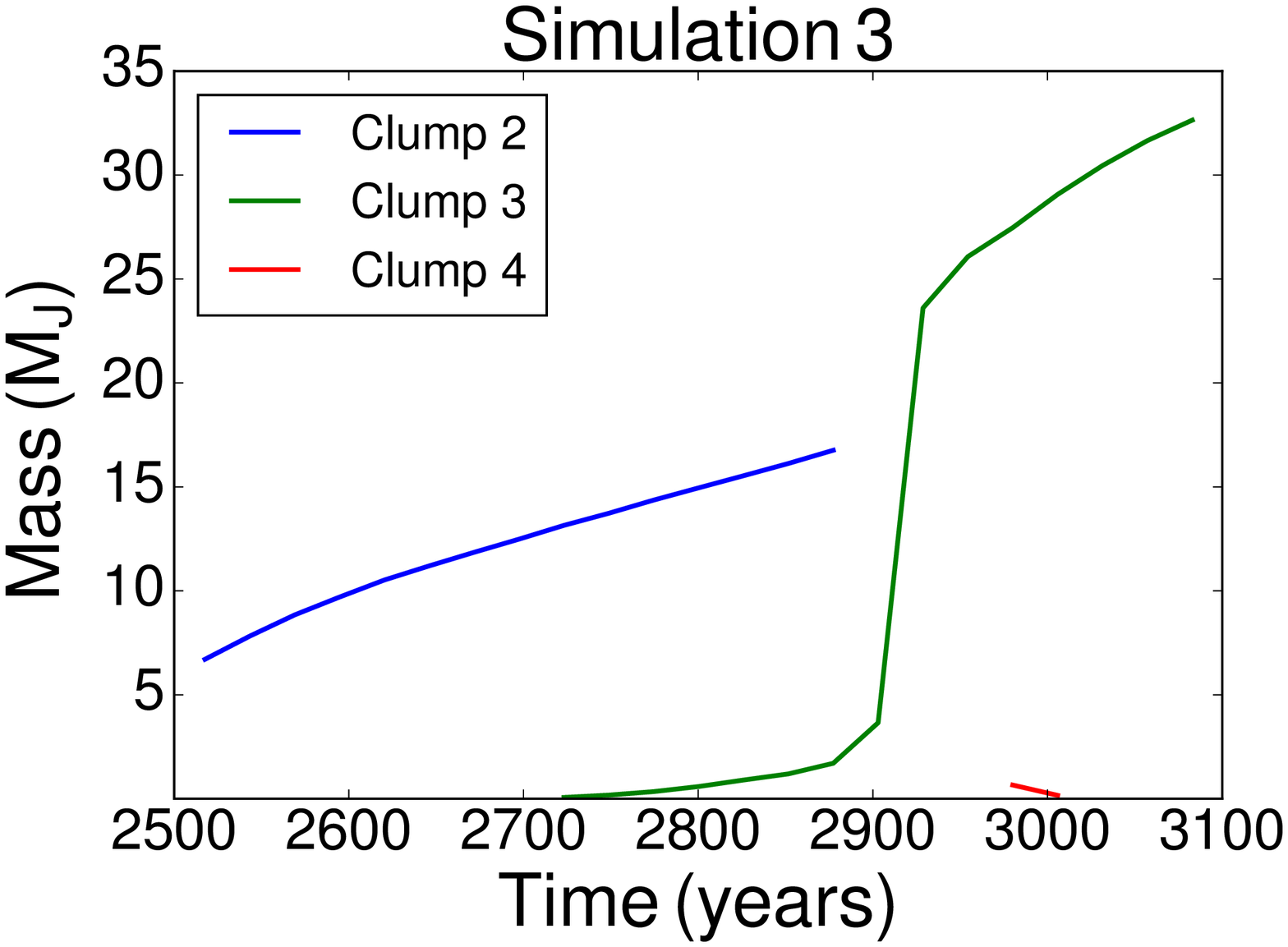}
        %\caption{$\dot{M}= 3.6\timeUonlys 10^{-7}\mdotunits$}
       % \caption{}
        \label{fig:sim3massVtimeUonly}
    \end{subfigure}
    \begin{subfigure}[b]{0.3\textwidth}
        \includegraphics[width=\textwidth]{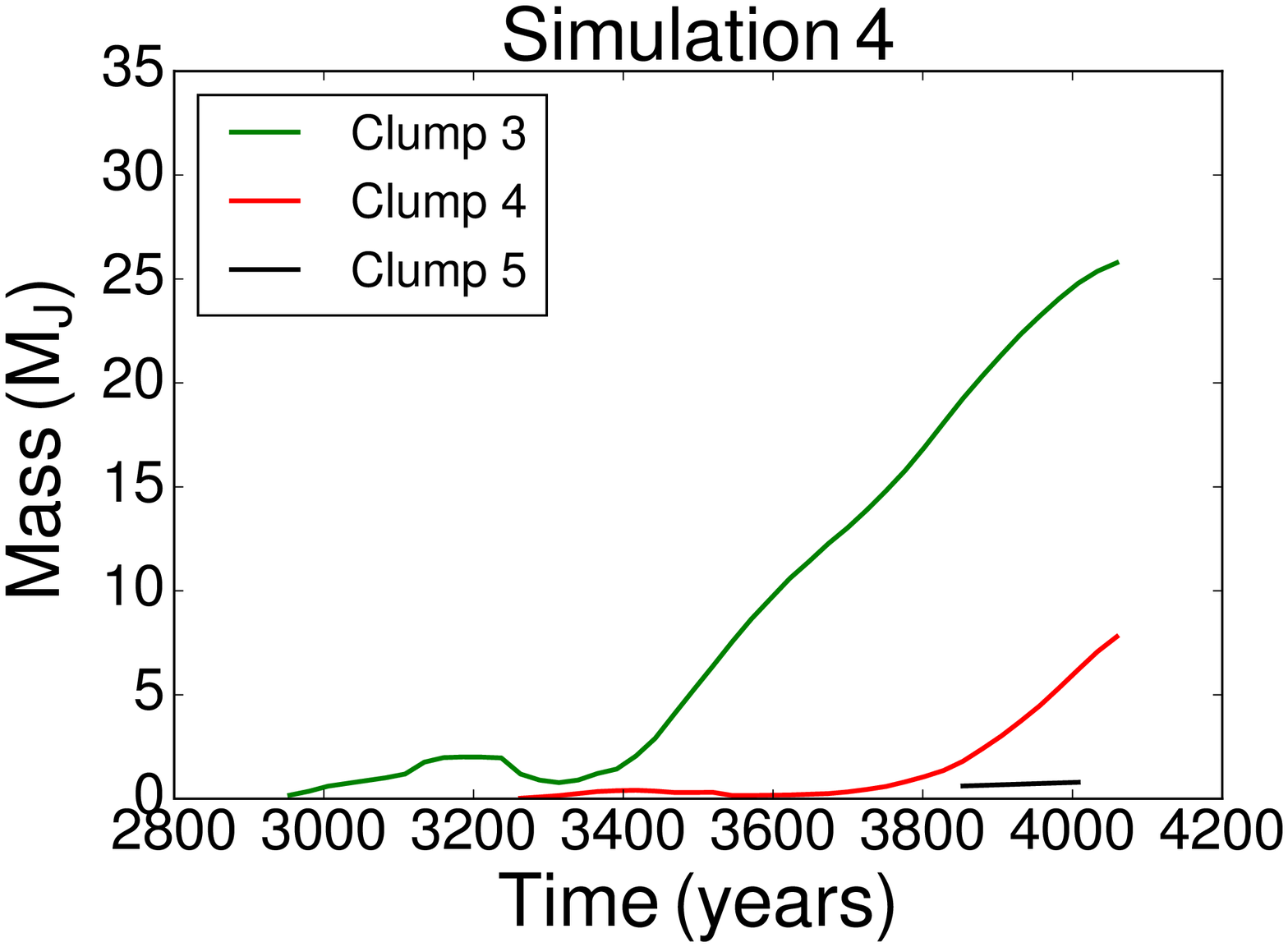}
      %  \caption{$\dot{M}= 2.8\timeUonlys 10^{-7}\mdotunits$, T=10K, q=0.35}
        %\caption{}
        \label{fig:sim4massVtimeUonly}
    \end{subfigure}
    \begin{subfigure}[b]{0.3\textwidth}
        \includegraphics[width=\textwidth]{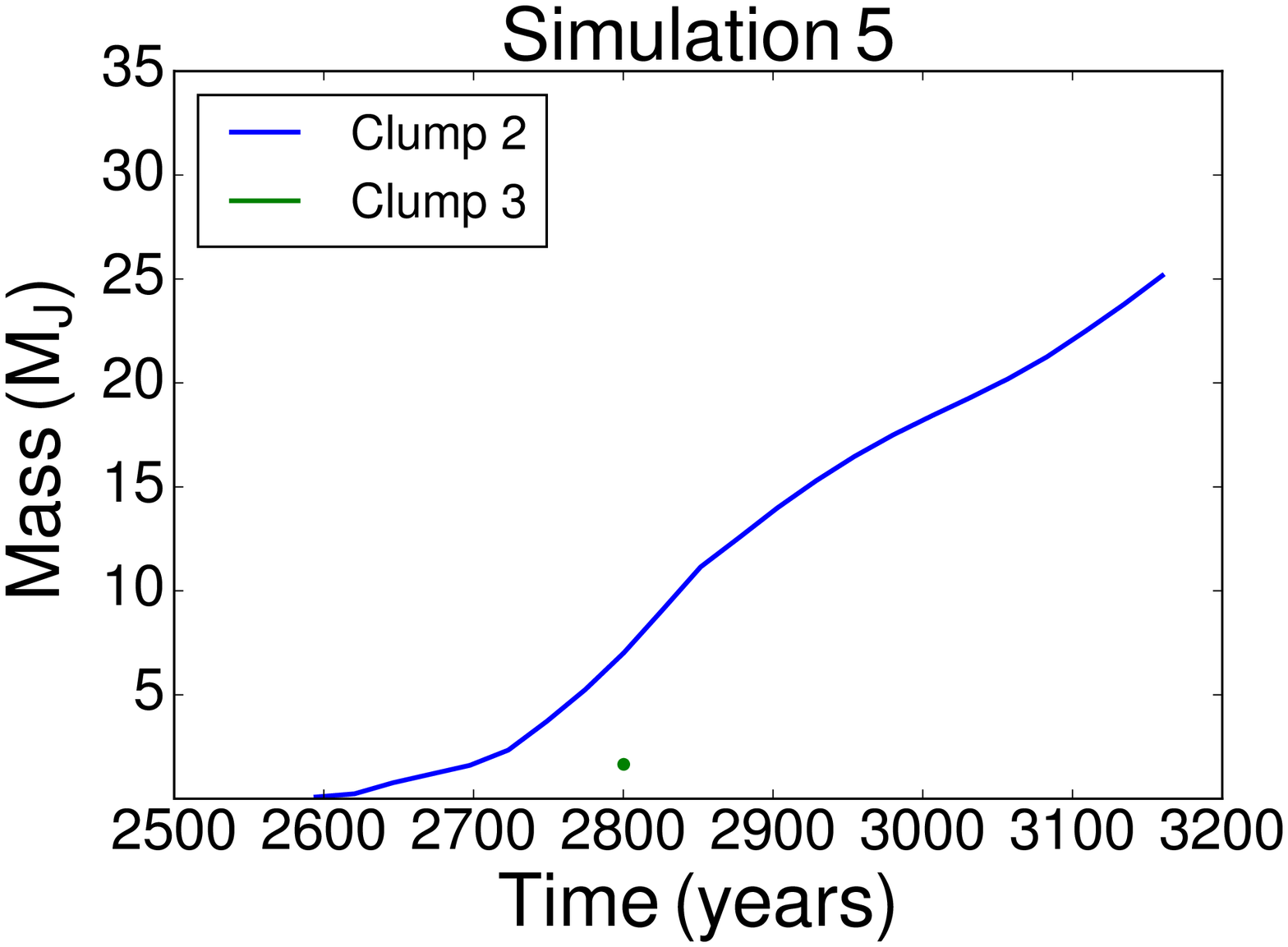}
       % \caption{$\dot{M}= 3.2\timeUonlys 10^{-7}\mdotunits$, T=10K, q=0.37}
      %  \caption{}
        \label{fig:sim5massVtimeUonly}
    \end{subfigure}
    \begin{subfigure}[b]{0.3\textwidth}
        \includegraphics[width=1.\textwidth]{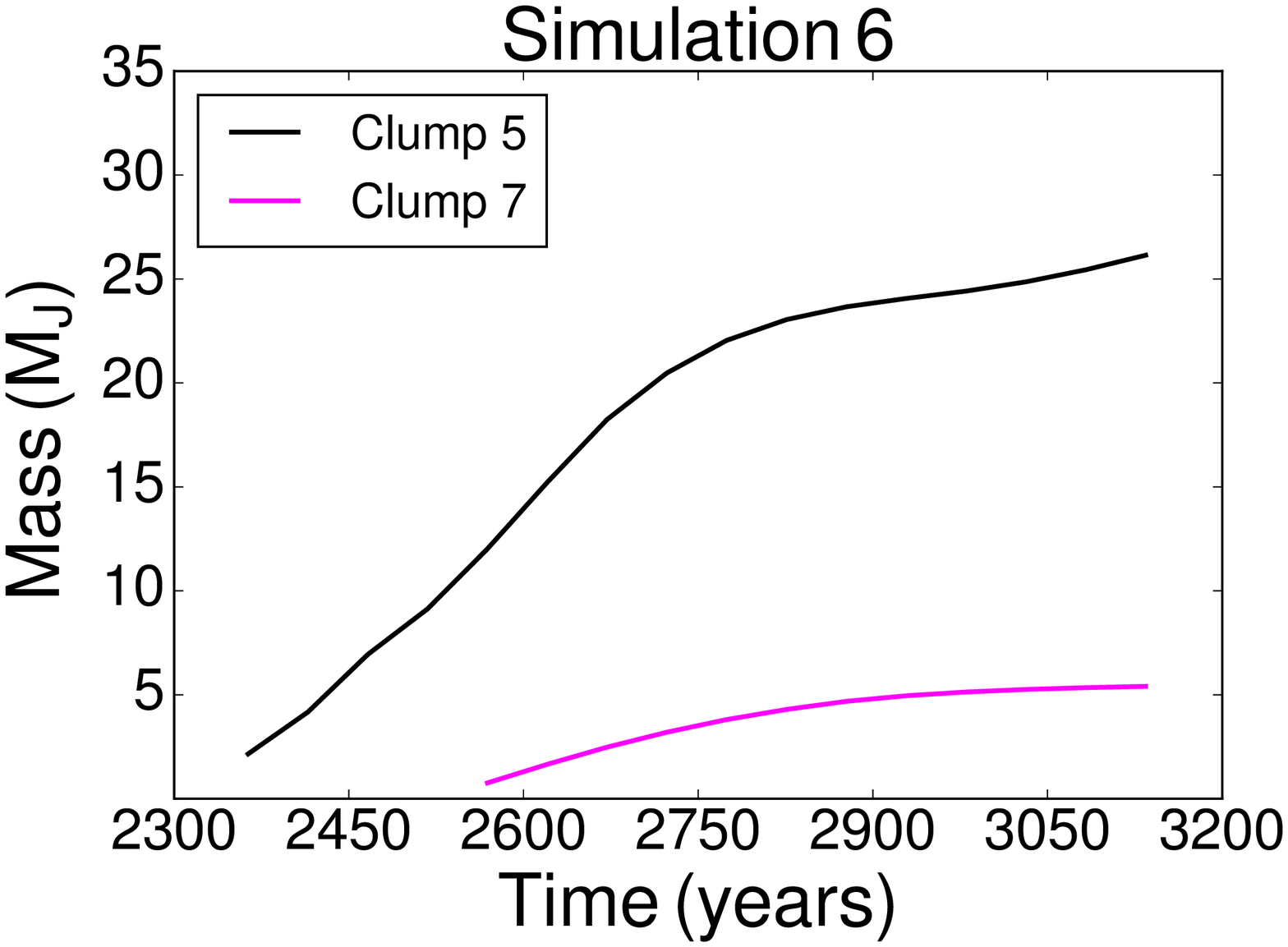}
        %\caption{$\dot{M}= 3.6\timeUonlys 10^{-7}\mdotunits$,T=10K,q=0.35}
       % \caption{}
        \label{fig:sim6massVtimeUonly}
    \end{subfigure}
    \begin{subfigure}[b]{0.3\textwidth}
        \includegraphics[width=\textwidth]{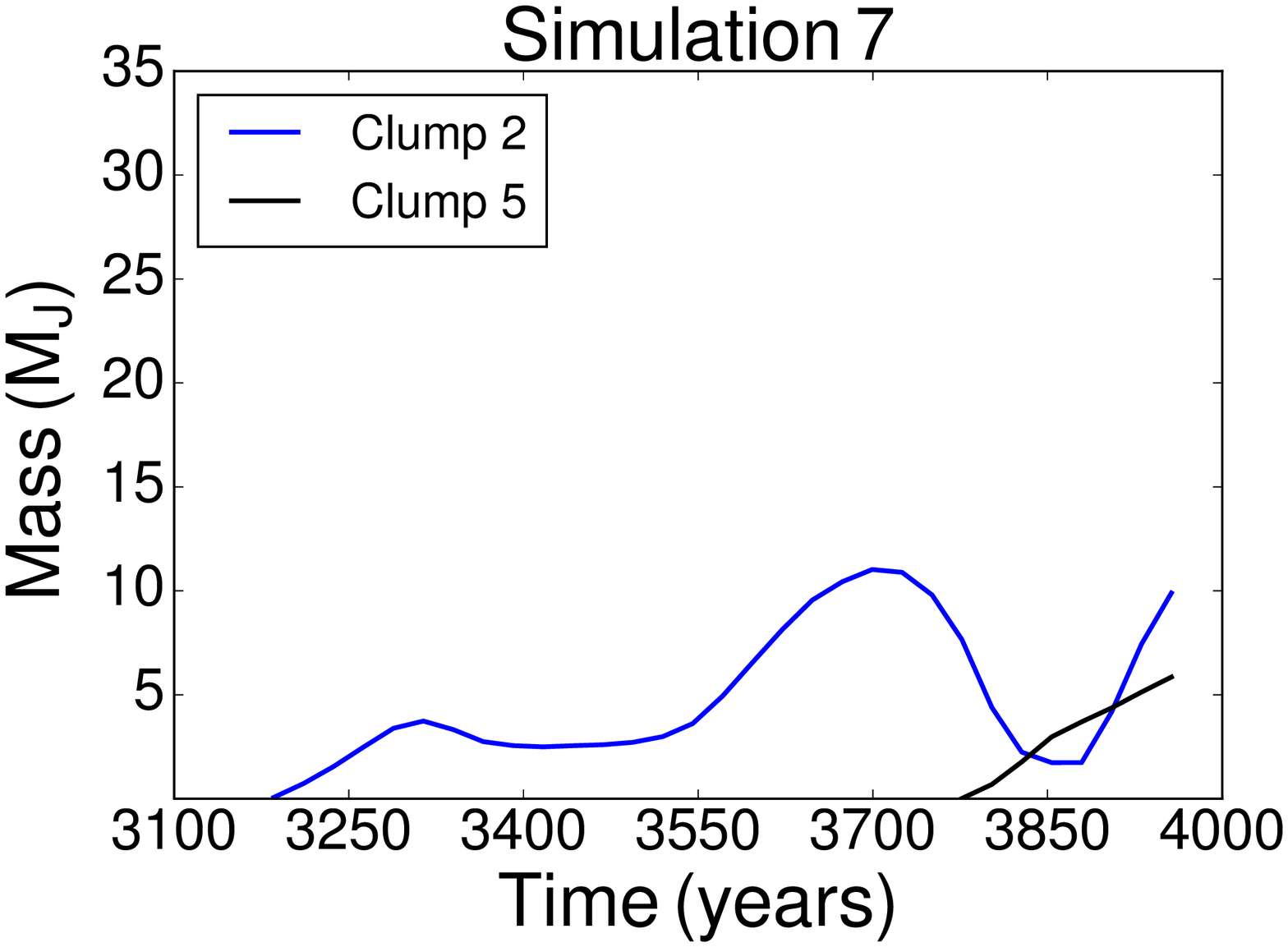}
        %\caption{$\dot{M}= 2.8\timeUonlys 10^{-7}\mdotunits$, T=30K,q=0.44}
       % \caption{}
        \label{fig:sim7massVtimeUonly}
    \end{subfigure}
    \begin{subfigure}[b]{0.3\textwidth}
        \includegraphics[width=\textwidth]{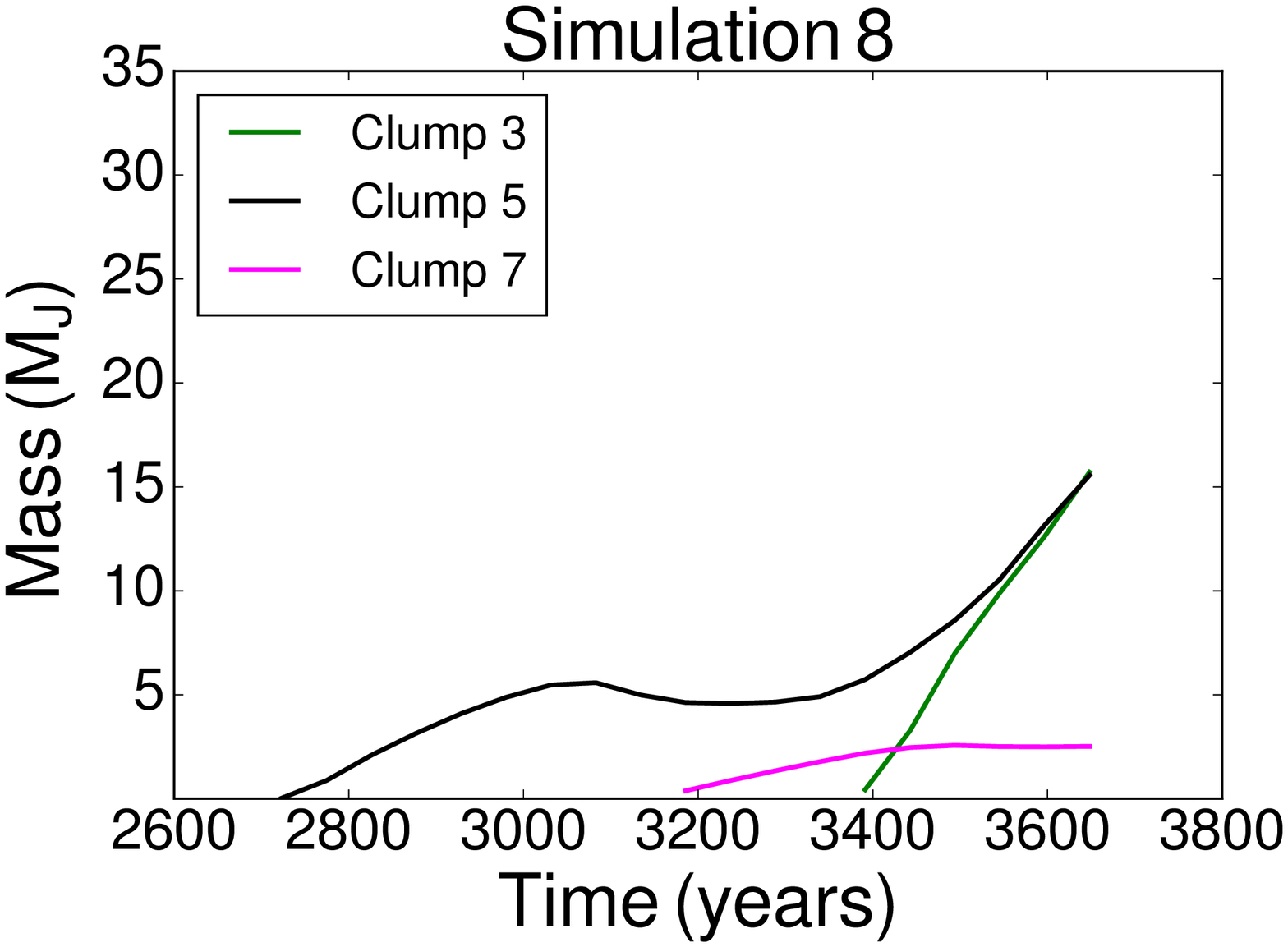}
       % \caption{$\dot{M}= 3.2\timeUonlys 10^{-7}\mdotunits$, T=30K,q=0.46}
       % \caption{}
        \label{fig:sim8massVtimeUonly}
    \end{subfigure}
    \begin{subfigure}[b]{0.3\textwidth}
        \includegraphics[width=1.\textwidth]{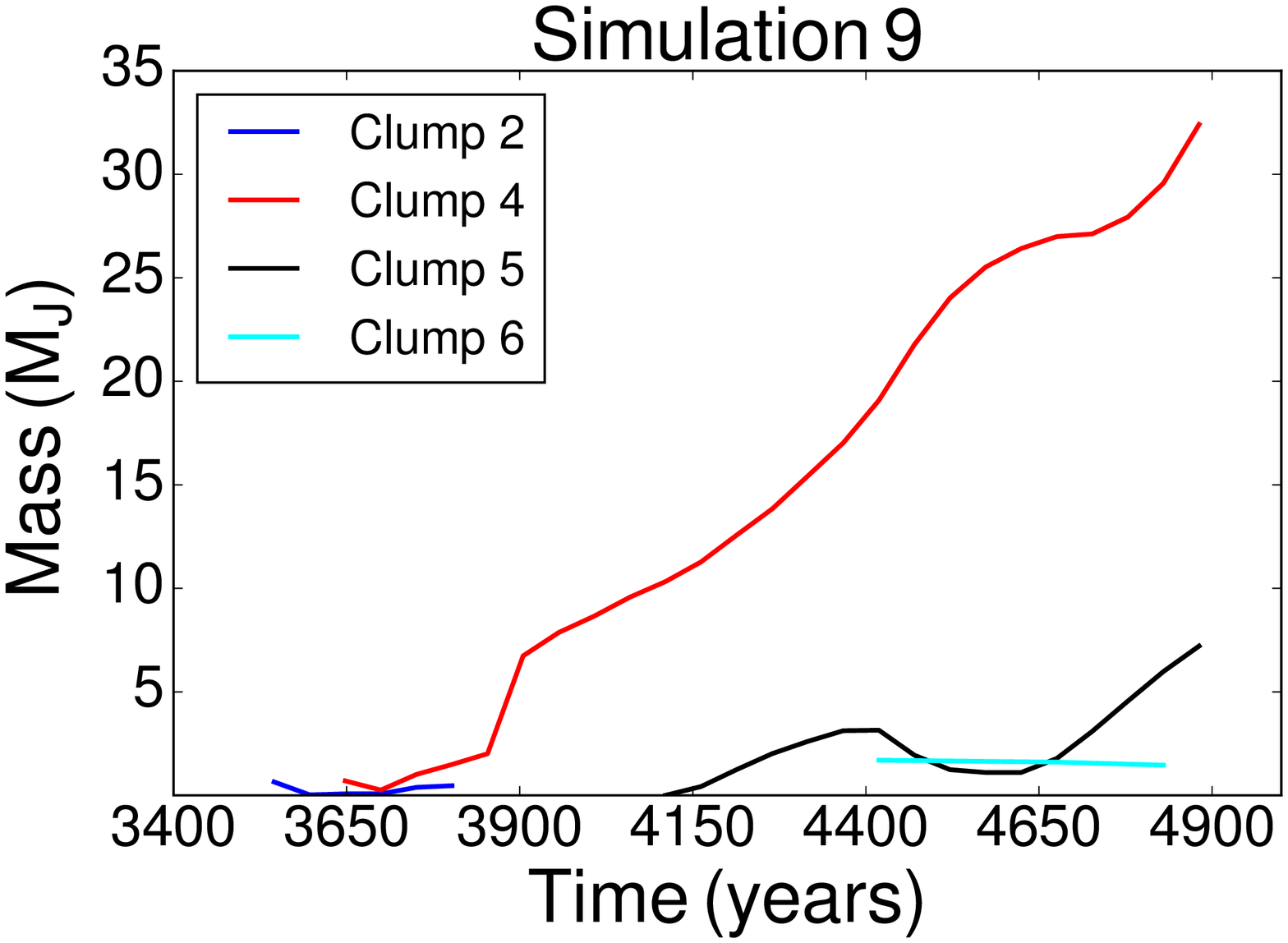}
        %\caption{$\dot{M}= 3.6\timeUonlys 10^{-7}\mdotunits$,T=30K,q=0.49}
        %\caption{}
        \label{fig:sim9massVtimeUonly}
    \end{subfigure}
    \caption{Mass accretion history for all clumps in all simulations, using only the ordered gravitational potential energy search. Fewer clumps are detected by this method than by using the density derivative, but those that are detected are likely to survive for a long time. Clumps are generally detected later in their evolution using this method, when their gravitational potential energy is negative enough to have neighbour particles assigned to them before they are assigned to the main body of the disc.}\label{fig:massVtimeUonly}
%\end{minipage}
\end{figure*}
%\end{comment}
%=================================================
% Mass plots of all simulations in density
%================================================
\begin{figure*}
  \centering
  \textbf{\large{Density derivative search}}\par\smallskip
    \begin{subfigure}[b]{0.3\textwidth}
        \includegraphics[width=\textwidth]{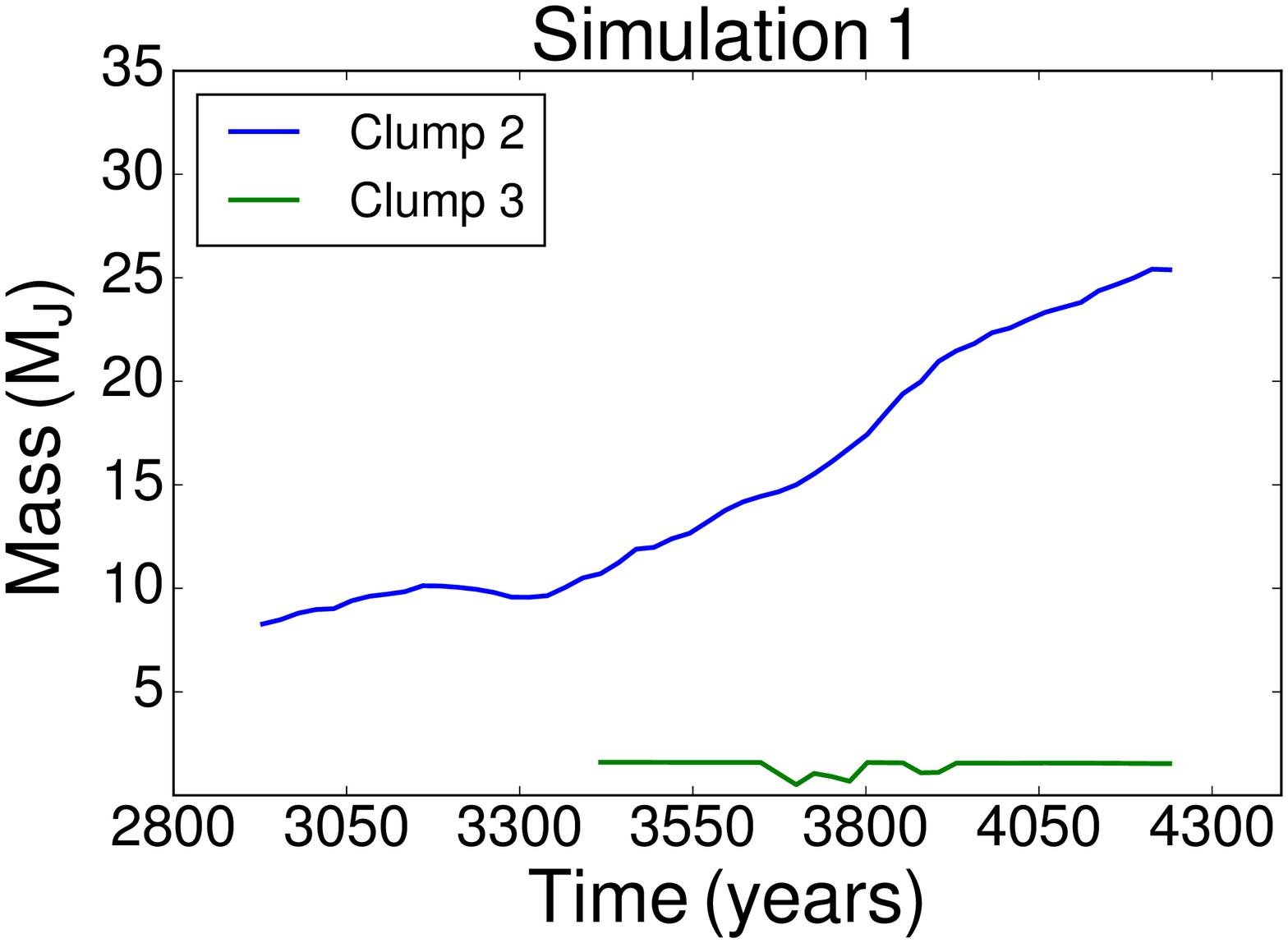}%{output}
       % \caption{$\dot{M}= 2.8\times 10^{-7}\mdotunits$}
       % \caption{}
       \vspace{0cm}
        \label{fig:sim1massVtime}
    \end{subfigure}
    \begin{subfigure}[b]{0.3\textwidth}
        \includegraphics[width=1.0\textwidth]{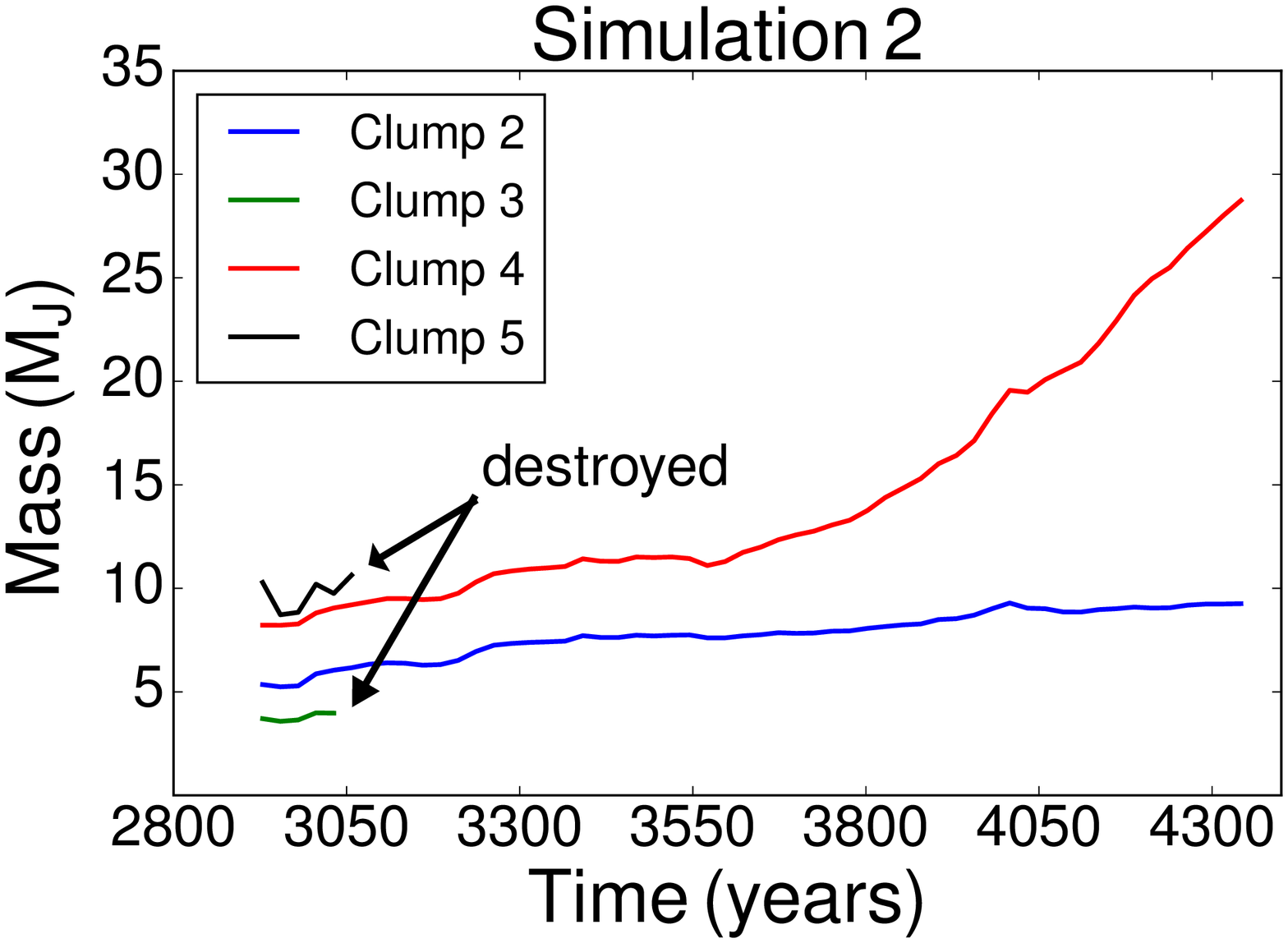}
        %\caption{$\dot{M}= 3.2\times 10^{-7}\mdotunits$}
        %\caption{}
        \label{fig:sim2massVtime}
    \end{subfigure}
    \begin{subfigure}[b]{0.3\textwidth}
        \includegraphics[width=1.\textwidth]{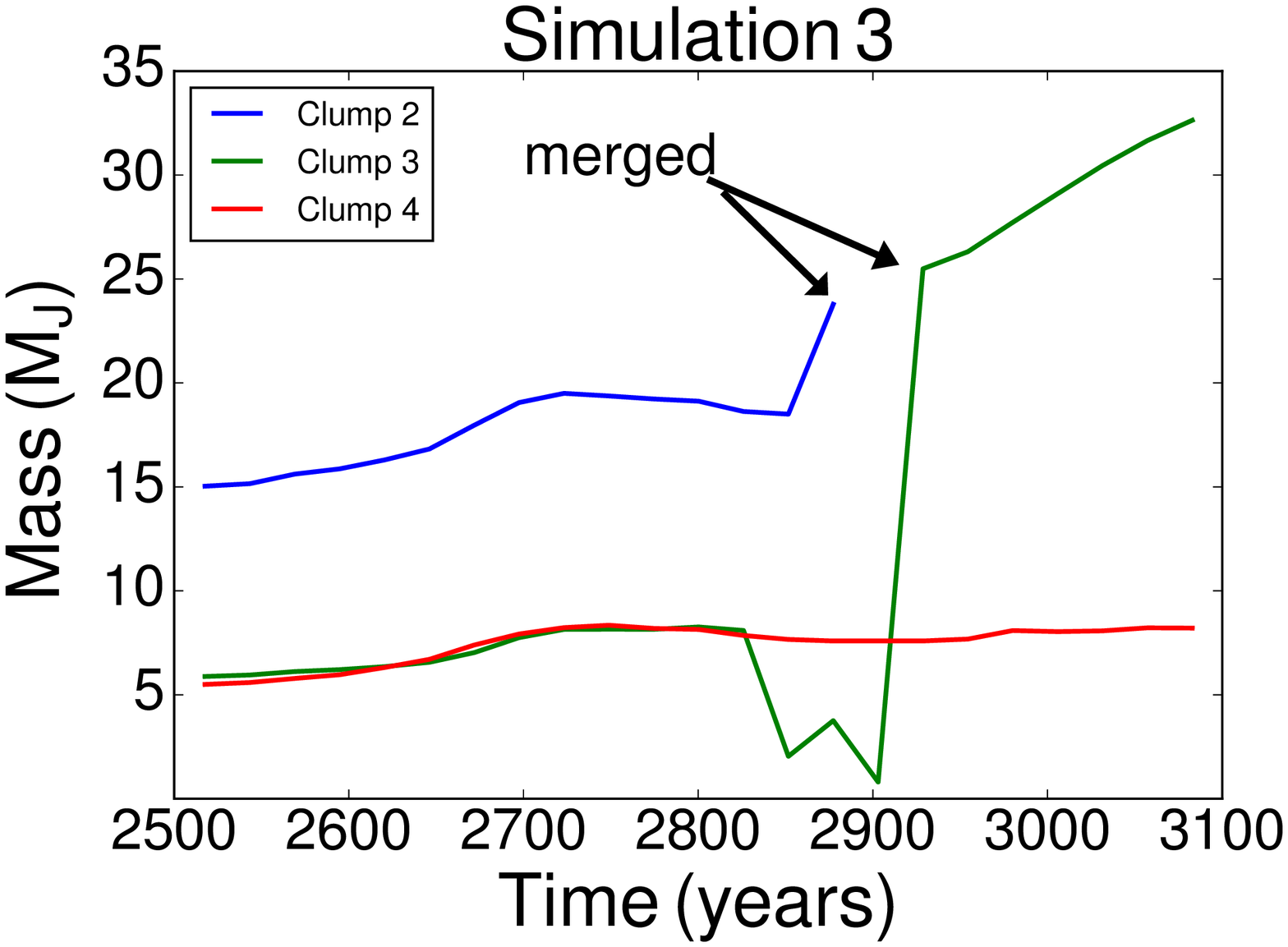}
        %\caption{$\dot{M}= 3.6\times 10^{-7}\mdotunits$}
       % \caption{}
        \label{fig:sim3massVtime}
    \end{subfigure}
    \begin{subfigure}[b]{0.3\textwidth}
        \includegraphics[width=\textwidth]{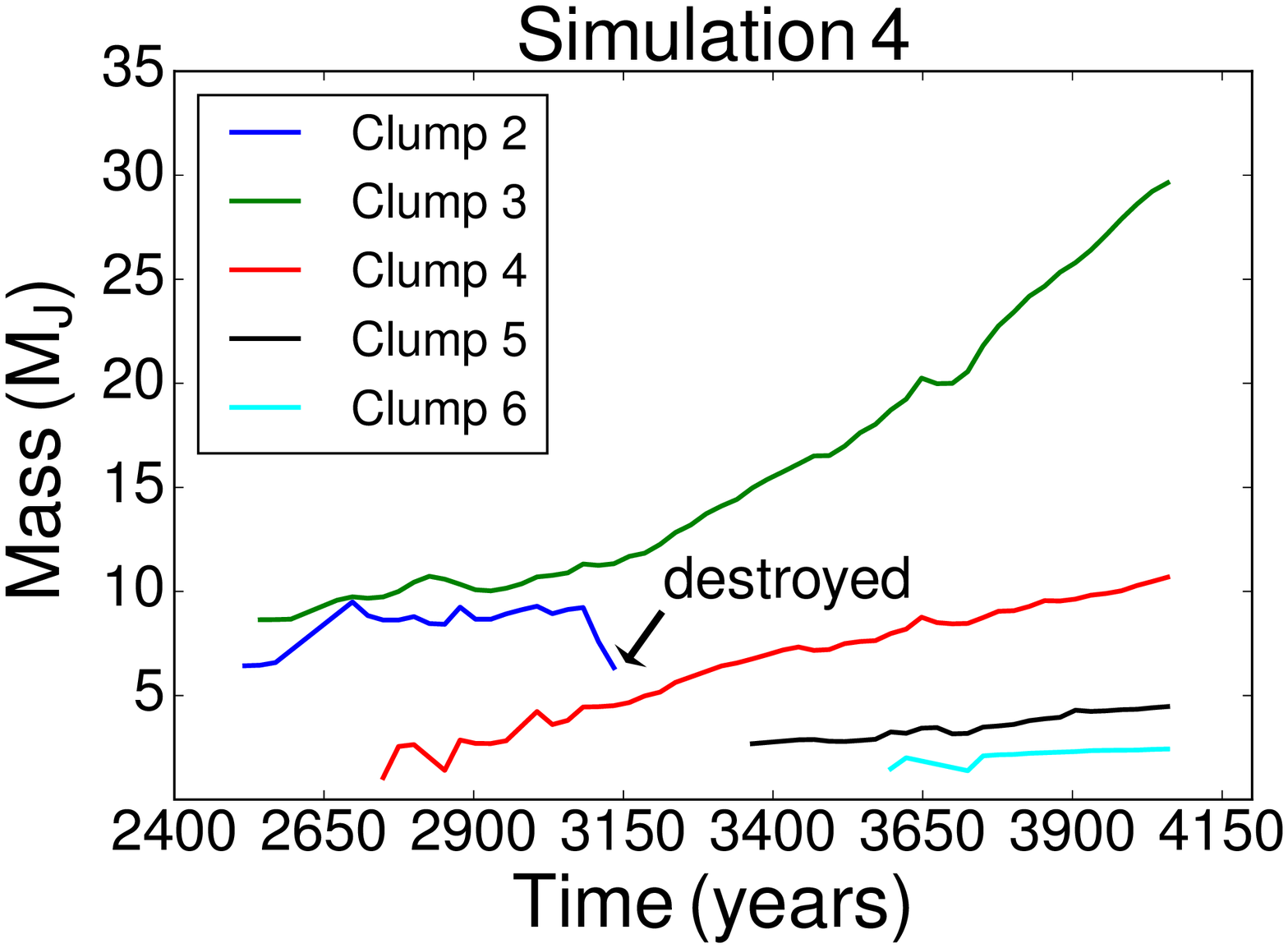}
      %  \caption{$\dot{M}= 2.8\times 10^{-7}\mdotunits$, T=10K, q=0.35}
        %\caption{}
        \label{fig:sim4massVtime}
    \end{subfigure}
    \begin{subfigure}[b]{0.3\textwidth}
        \includegraphics[width=\textwidth]{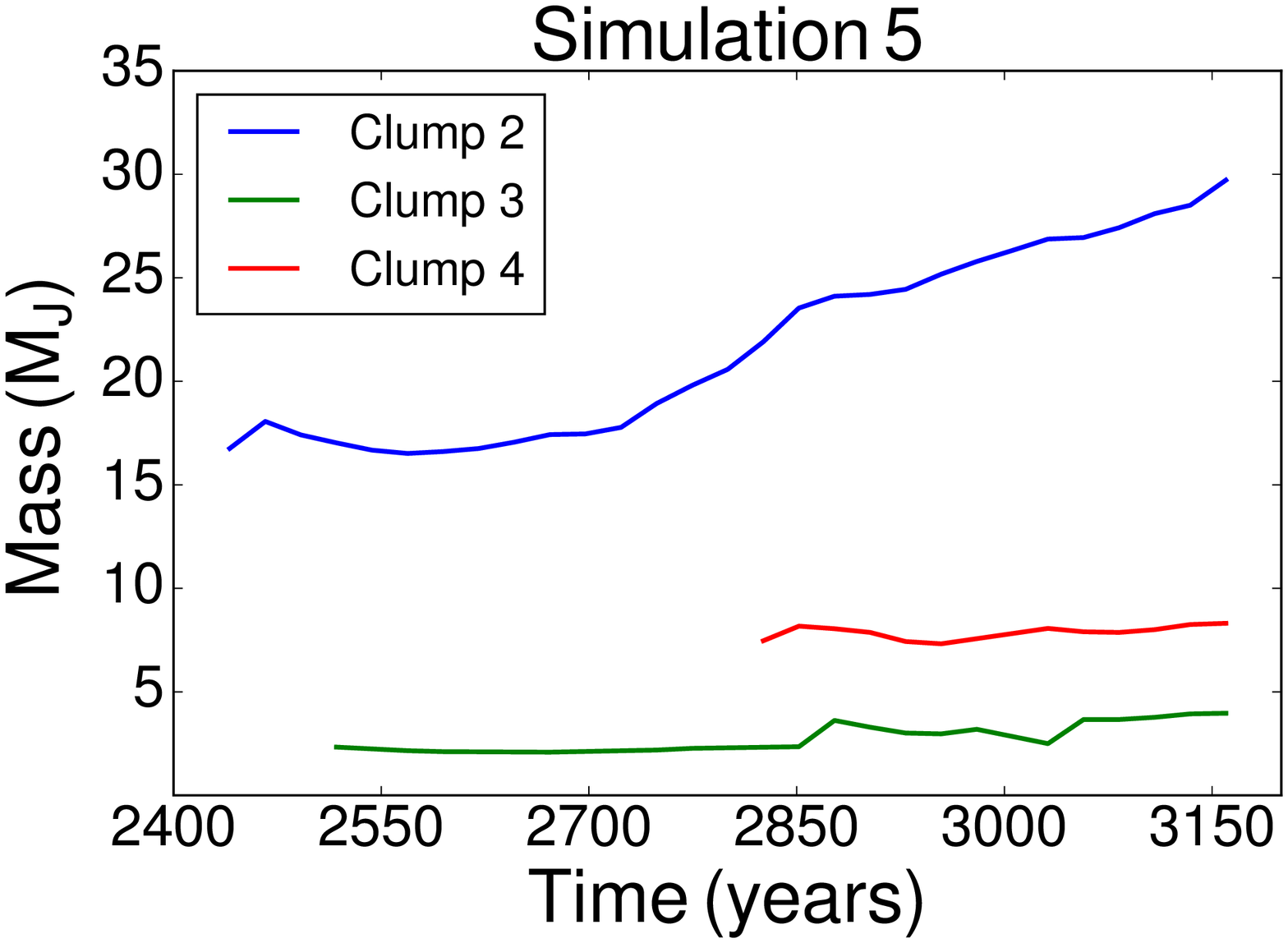}
       % \caption{$\dot{M}= 3.2\times 10^{-7}\mdotunits$, T=10K, q=0.37}
      %  \caption{}
        \label{fig:sim5massVtime}
    \end{subfigure}
    \begin{subfigure}[b]{0.3\textwidth}
        \includegraphics[width=1.\textwidth]{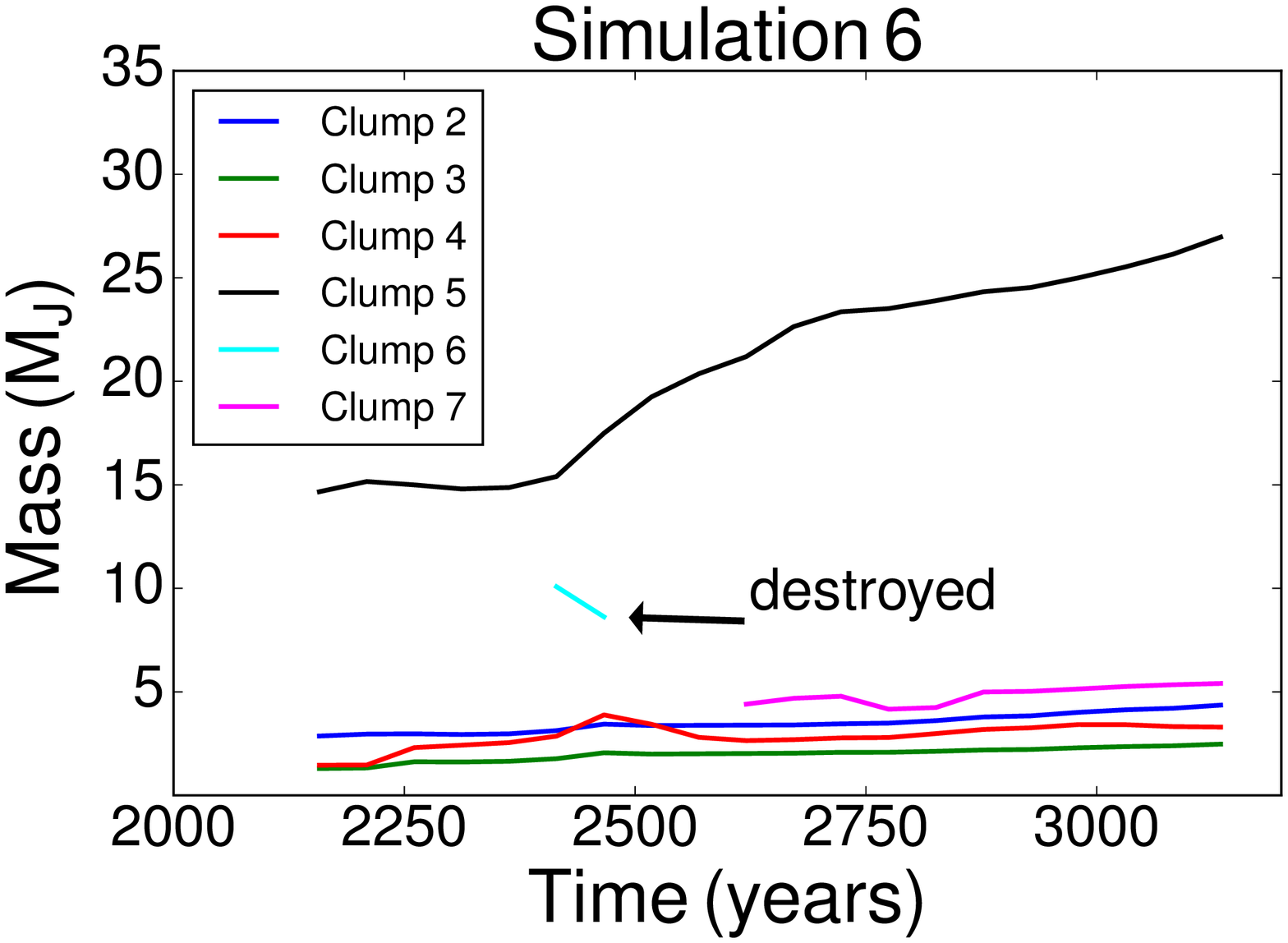}
        %\caption{$\dot{M}= 3.6\times 10^{-7}\mdotunits$,T=10K,q=0.35}
       % \caption{}
        \label{fig:sim6massVtime}
    \end{subfigure}
    \begin{subfigure}[b]{0.3\textwidth}
        \includegraphics[width=\textwidth]{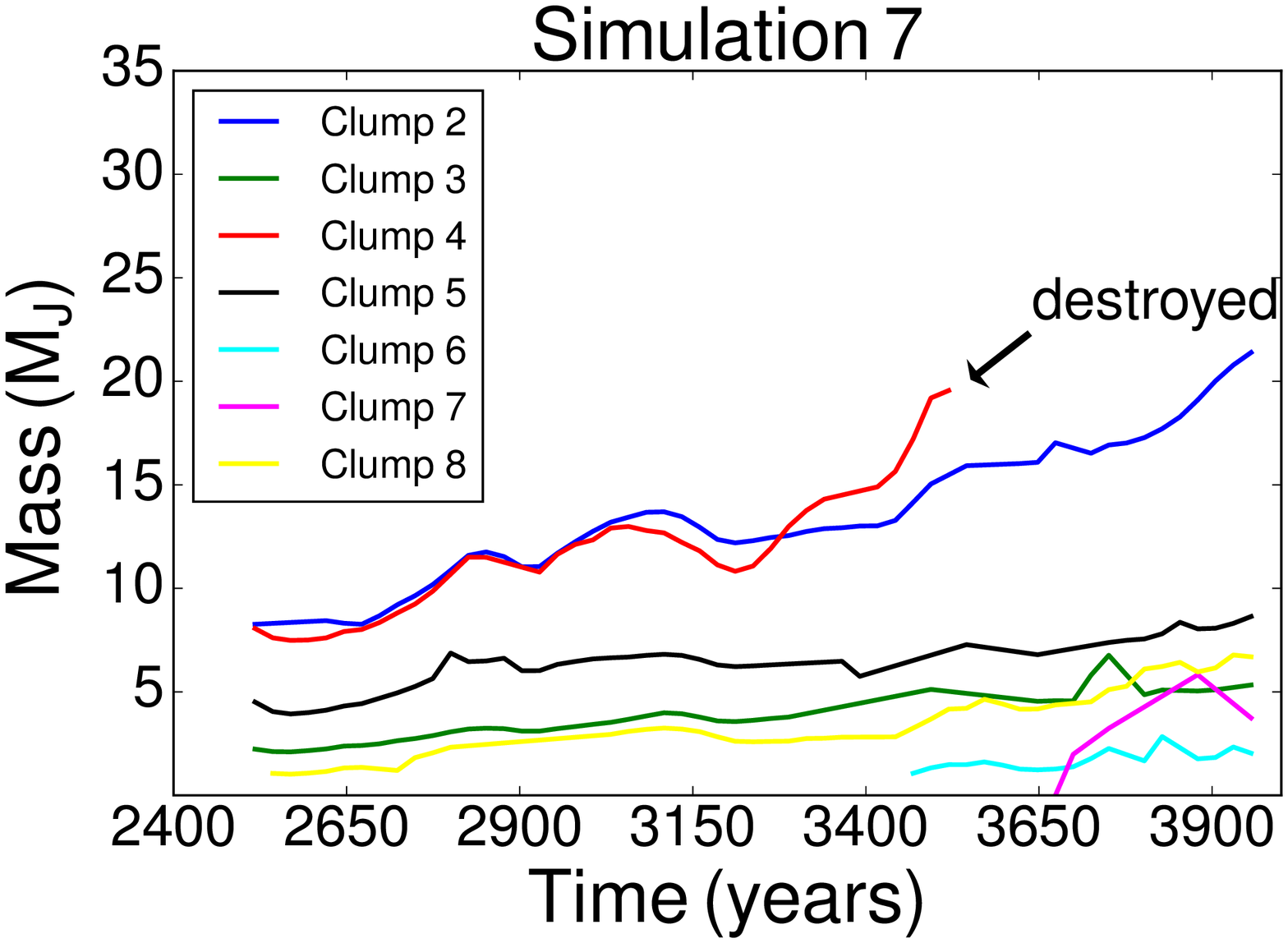}
        %\caption{$\dot{M}= 2.8\times 10^{-7}\mdotunits$, T=30K,q=0.44}
       % \caption{}
        \label{fig:sim7massVtime}
    \end{subfigure}
    \begin{subfigure}[b]{0.3\textwidth}
        \includegraphics[width=\textwidth]{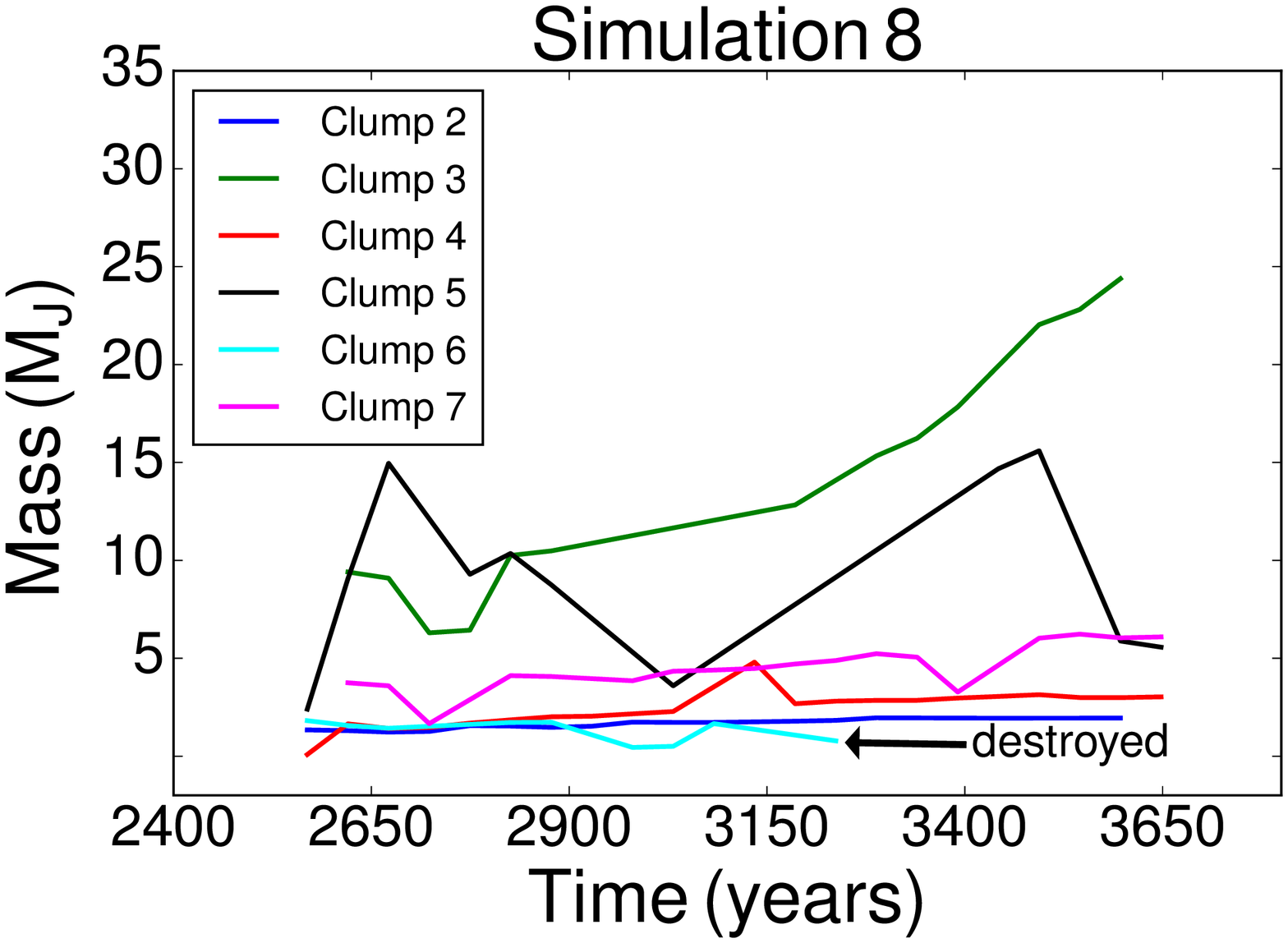}
       % \caption{$\dot{M}= 3.2\times 10^{-7}\mdotunits$, T=30K,q=0.46}
       % \caption{}
        \label{fig:sim8massVtime}
    \end{subfigure}
    \begin{subfigure}[b]{0.3\textwidth}
        \includegraphics[width=1.\textwidth]{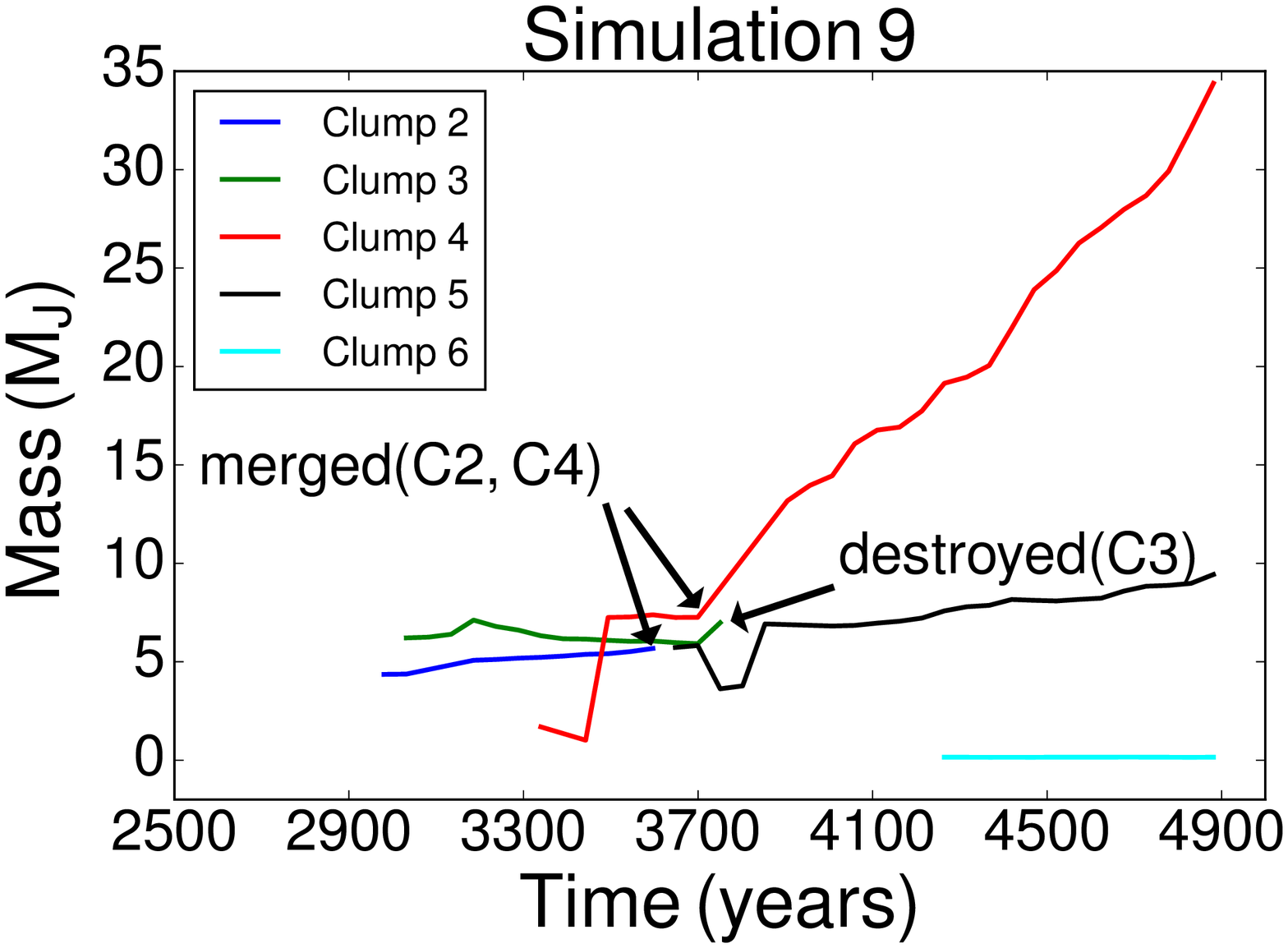}
        %\caption{$\dot{M}= 3.6\times 10^{-7}\mdotunits$,T=30K,q=0.49}
        %\caption{}
        \label{fig:sim9massVtime}
    \end{subfigure}
    \caption{Mass accretion history for all clumps in all simulations, found using the density derivative search method. More clumps are detected this way than by using the ordered potential search, as clumps buried in the potential well of the disc are identified early by their density peak. As can be seen by comparison with Figure \ref{fig:massVtimeUonly}, this search method is sensitive to low mass clumps, is sensitive to all clumps earlier in their evolution, and by comparison with Figure \ref{fig:migration} we can see this method is also able to detect clumps which are ultimately tidally destroyed.\label{fig:massVtime}}
%\end{minipage}
\end{figure*}
At this point, we have a set of clumps in each timestep, and we need to track them over the duration of the simulation. To do this, we use a standard algorithm from halo tracking in cosmological simulations (see, e.g., \citealt{springer2001}). Each clump, in each timestep, is given an integer ID by our algorithm. So that we can trace the evolution of this clump throughout the simulation, these IDs must be linked. Since we are modelling fluid through the use of pseudo-particles, the particles, that make any given clump, change between dumps, sometimes substantially. To link clumps, the crucial factors are the most-bound particle MBP, and shared member fraction (SMF). In our density-derivative search, we actually trace the most-dense particle, rather than the most-bound particle, but we use the term interchangeably to avoid the introduction of unnecessary acronyms. To be identified as the same clump between timesteps, they must share the MBP and have an SMF of > 50\%. In some particularly volatile simulations, when using the density derivative method, the MBP may change, and the SMF may be < 50\%. In this case, some of the clumps need to be manually linked during post processing by tracing a group of particles in each clump in each timestep. The basic algorithm is as follows:
\begin{enumerate}[label*=\arabic*.,leftmargin=.25in]
\item Loop over clumps $i$ in previous timestep \emph{lastdump}.
\item Find clump $j$, in this timestep \emph{thisdump}, which contains the MBP from clump $i$ in \emph{lastdump}.
\begin{enumerate}[label*=\roman*.,leftmargin=.35in]
\item If MBP$_{i}$ does not belong to any clump in \emph{thisdump}, clump $i$ is not present in \emph{thisdump}.
%\item If MBP does not have an ID in this timestep
\item If MBP$_{i}$ belongs to clump $j$ in \emph{thisdump}, and clump $i$ and $j$ share at least 50\% of particles, then ID$_{j}$ = ID$_{i}$, and the clumps are linked between the two timesteps.
\item If MBP$_i$ belongs to clump $j$ in \emph{thisdump}, but clumps $i$ and $j$ do not have SMF > 50\% of particles, then clump $j$ keeps its present ID.
\end{enumerate}
\item End loop over previous timestep.
\item Loop over clumps in \emph{thisdump}, checking for clumps with no progenitors in \emph{lastdump}. Increment the maximum number of distinct IDs by the number of new clumps, and assign each clump the correct ID.

\end{enumerate}
%
%=====================================================
%                   RESULTS
%=====================================================
\section{Results}
\label{sec:results}
\begin{figure}
\includegraphics[width=\linewidth]{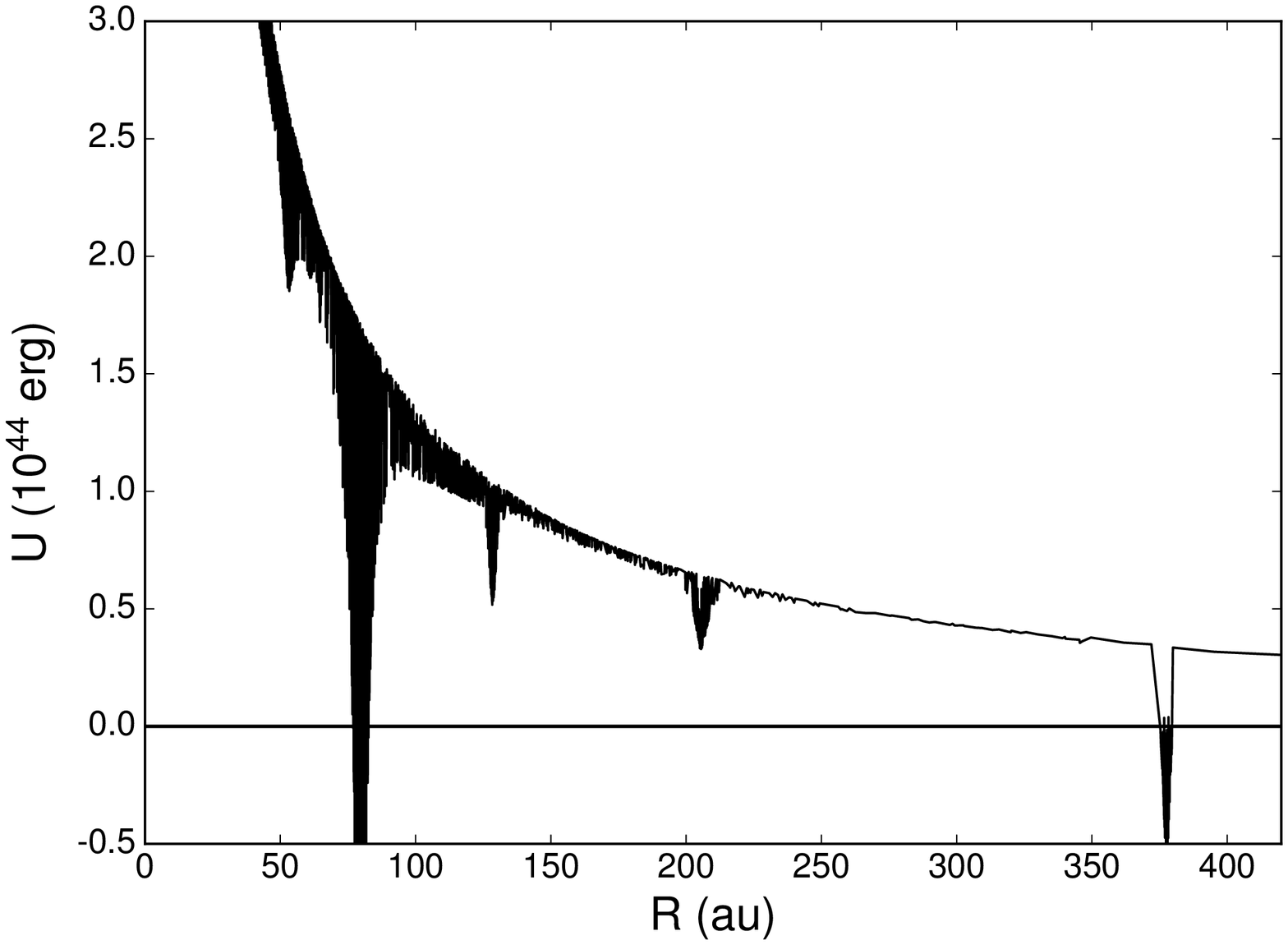}
\caption{Radial profile of gravitational potential energy for the disc shown in Figure \ref{fig:columndensity}, simulation 6. Only two clumps, at $r\sim 80$ au and $r\sim 375$ au have a sufficiently deep potential well to be identified by our ordered potential search. The other three clumps, at $r\sim 50$ au, $r\sim 125$ au and $r\sim 210$ au are identified as belonging to the main body of the disc. Their detection in the density derivative search, but not in the ordered potential search, indicates that they would likely be tidally destroyed.\label{fig:potential}}
\end{figure}
%=======================================
% Potential V r for disc 062
%=======================================
%=================================================
% Mass plots of all simulations in potential
%===============================================
%\begin{comment}
%===========================================================
% Histogram figure
%===========================================================
\begin{figure*}
\begin{center}
\begin{tabular}{cc}
\textbf{Semi-major axis distribution} & \textbf{Mass distribution}\\
\includegraphics[width=0.5\linewidth]{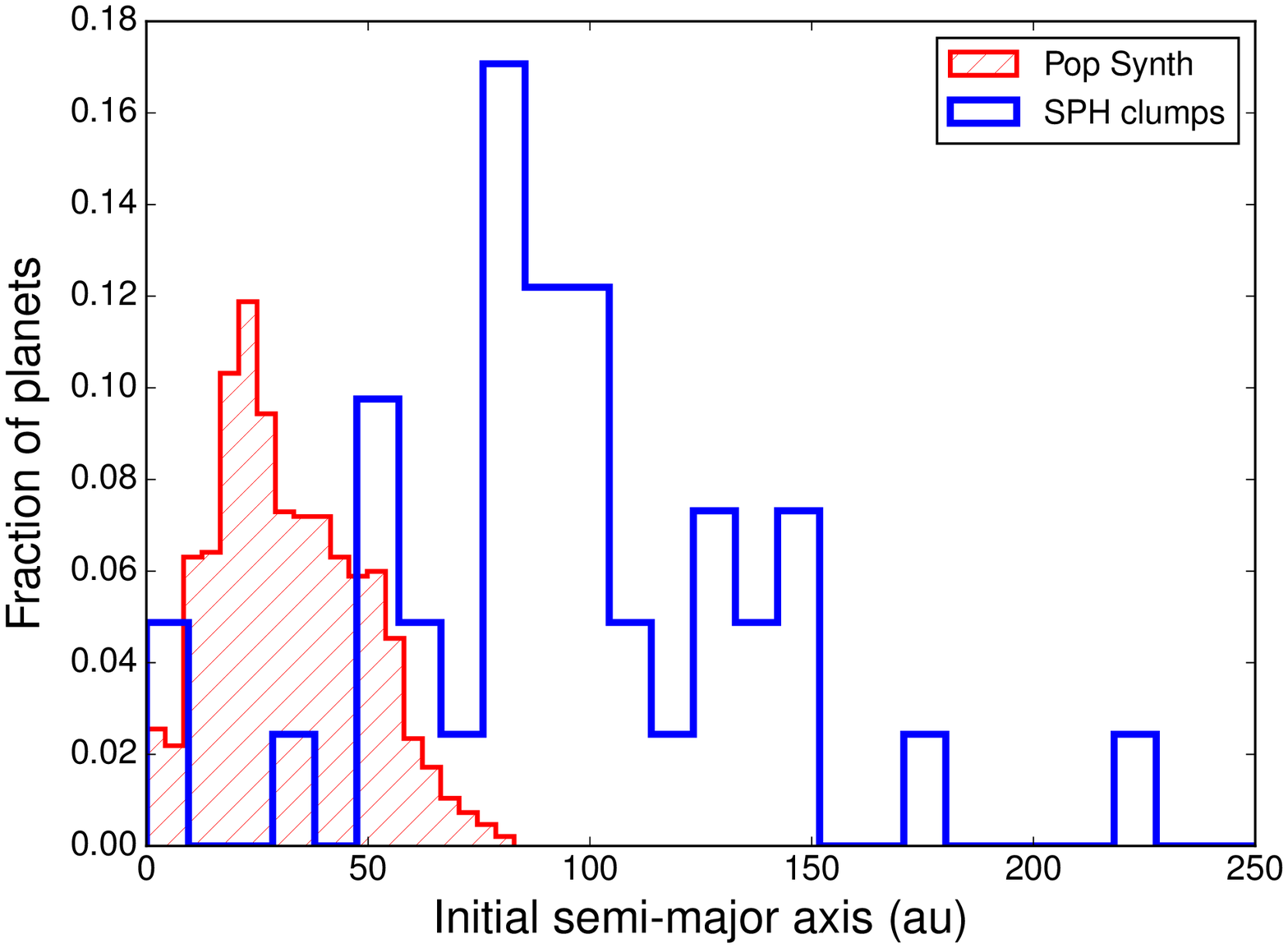} & \includegraphics[width=0.5\linewidth]{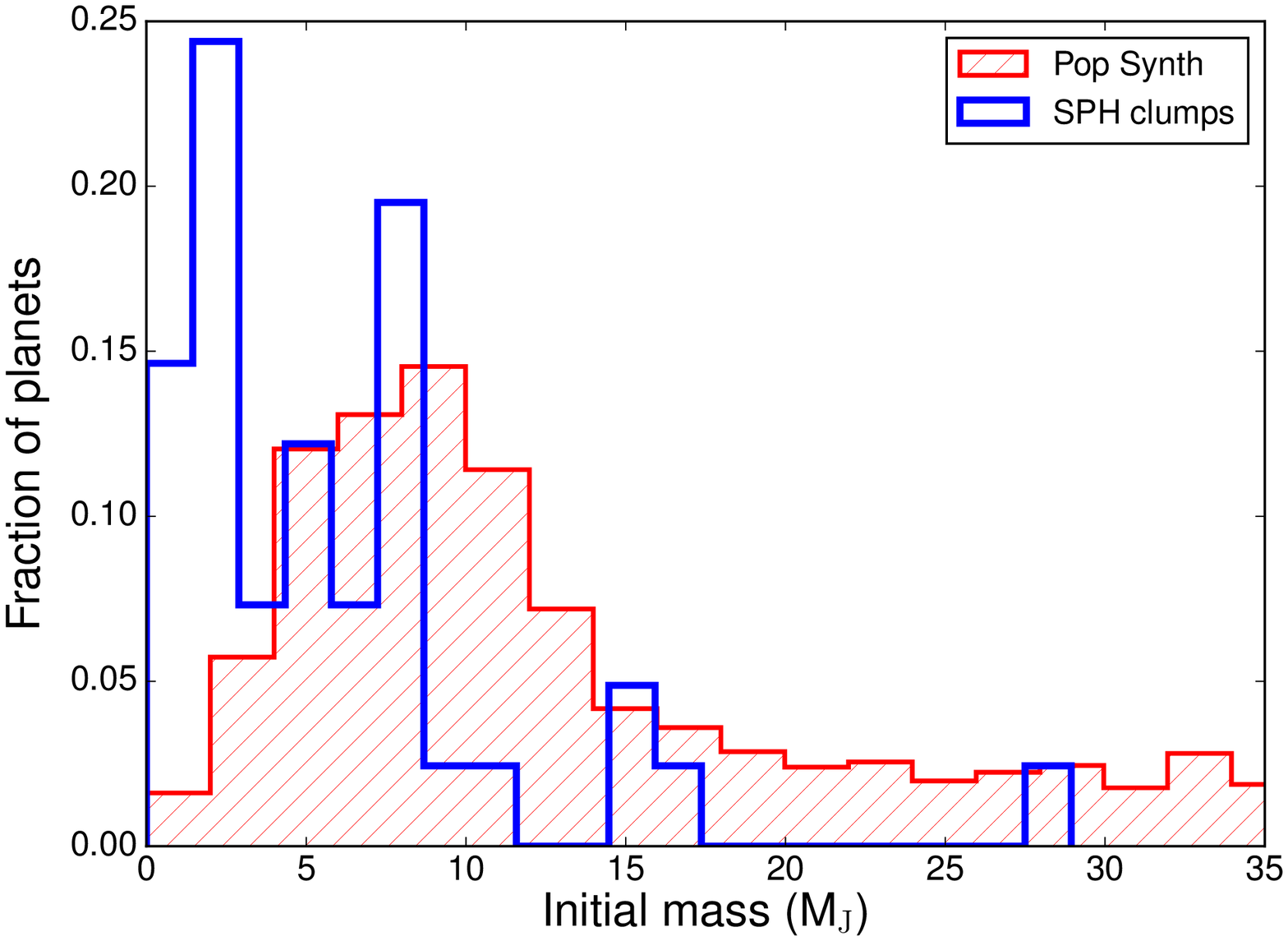} \\
\includegraphics[width=0.5\linewidth]{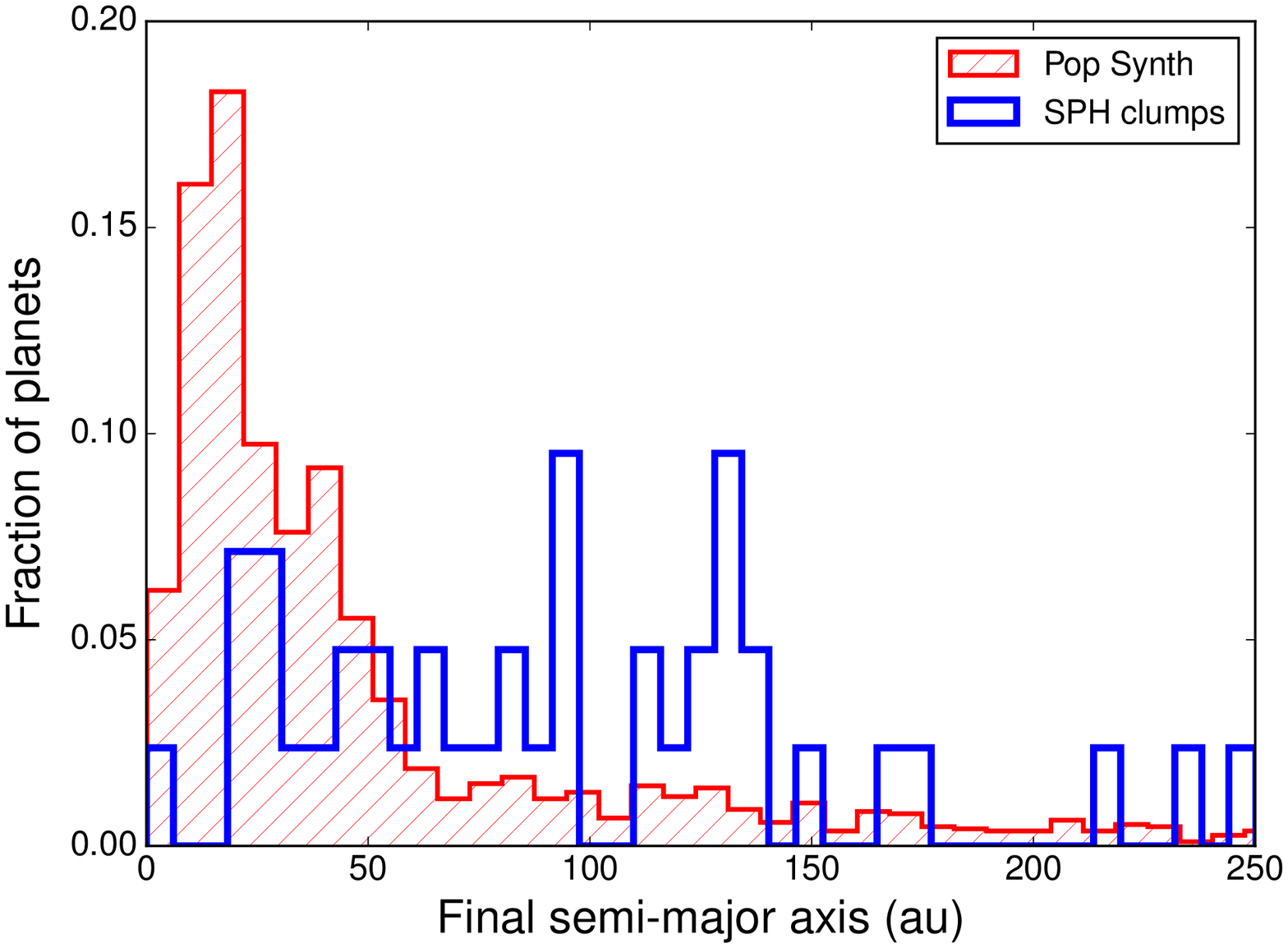} & \includegraphics[width=0.5\linewidth]{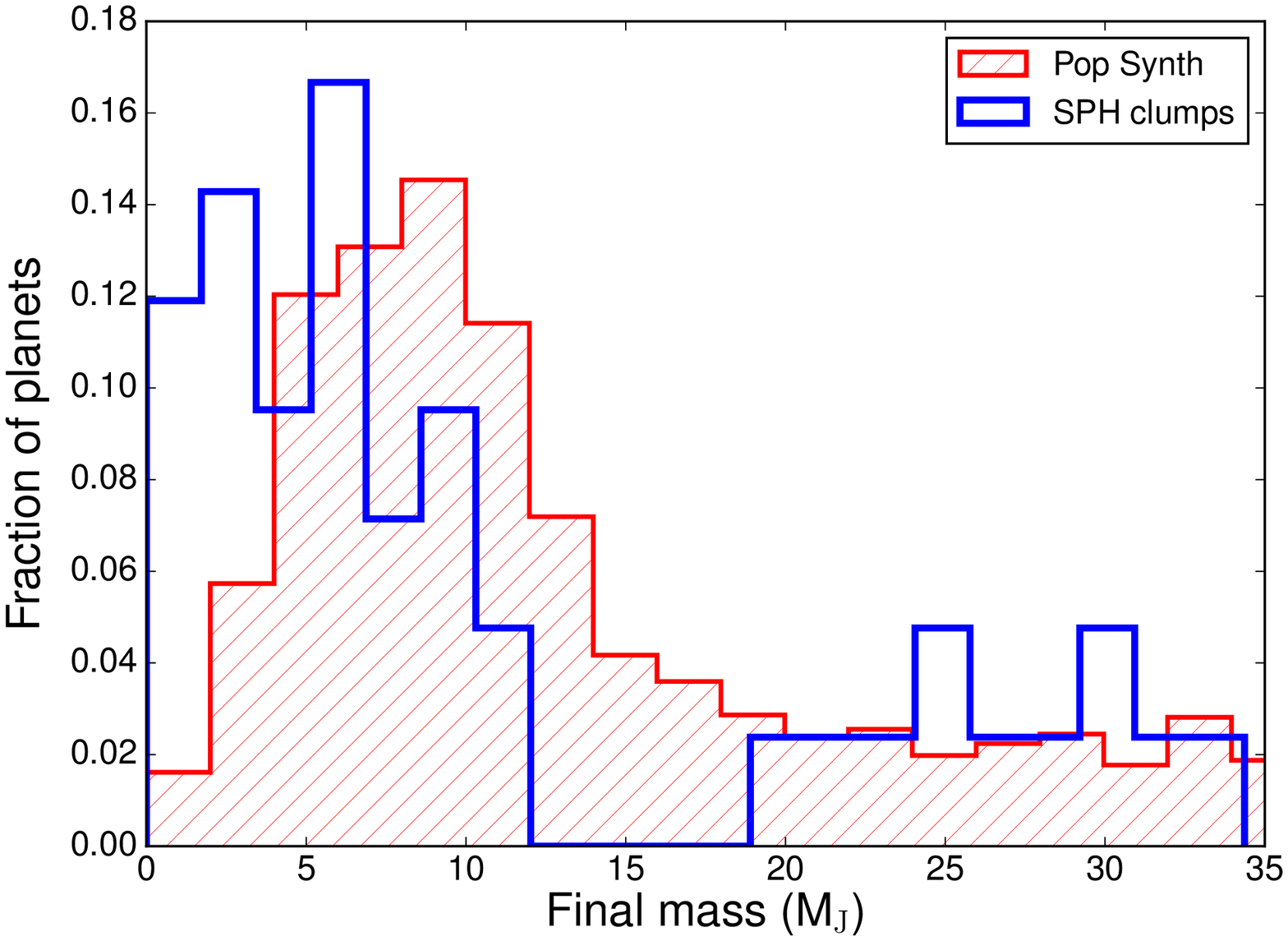} 
\end{tabular}
\caption{Left column shows initial (top) and final (bottom) semi-major axis distribution for our SPH clumps and the population synthesis model of \citet{forganricepopsynth2013}. Right column shows the initial (top) and final (bottom) mass distribution for the same. Population synthesis data is shown in hatched red, while blue outline shows SPH clumps. Our initial mass and semi-major axis distributions are not strictly accurate, due to limitations of the algorithm requiring a threshold to be met before identification. However, our algorithm is quite robust at later times, and our final semi-major axis distribution shows the importance of clump-clump interactions in the final configuration of a system, with many clumps at large separations due to interactions with each other. Our final masses below 5 M$_{\mathrm{J}}$ are typically underestimated by a factor of 2 what would be identified as mass belonging to the clump "by-eye", and accounting for this shows a final mass distribution not unreasonably dissimilar to what we should expect from GI population synthesis models, given the small $N$ statistics we are considering.\label{fig:histogram}}
\end{center}
\end{figure*}
%===============================================================
% Final mass-semi major axis distribution for SPH and pop synth
%===============================================================
\begin{figure*}
    \begin{minipage}[t]{.48\textwidth}
      \centering
      \includegraphics[width=\linewidth]{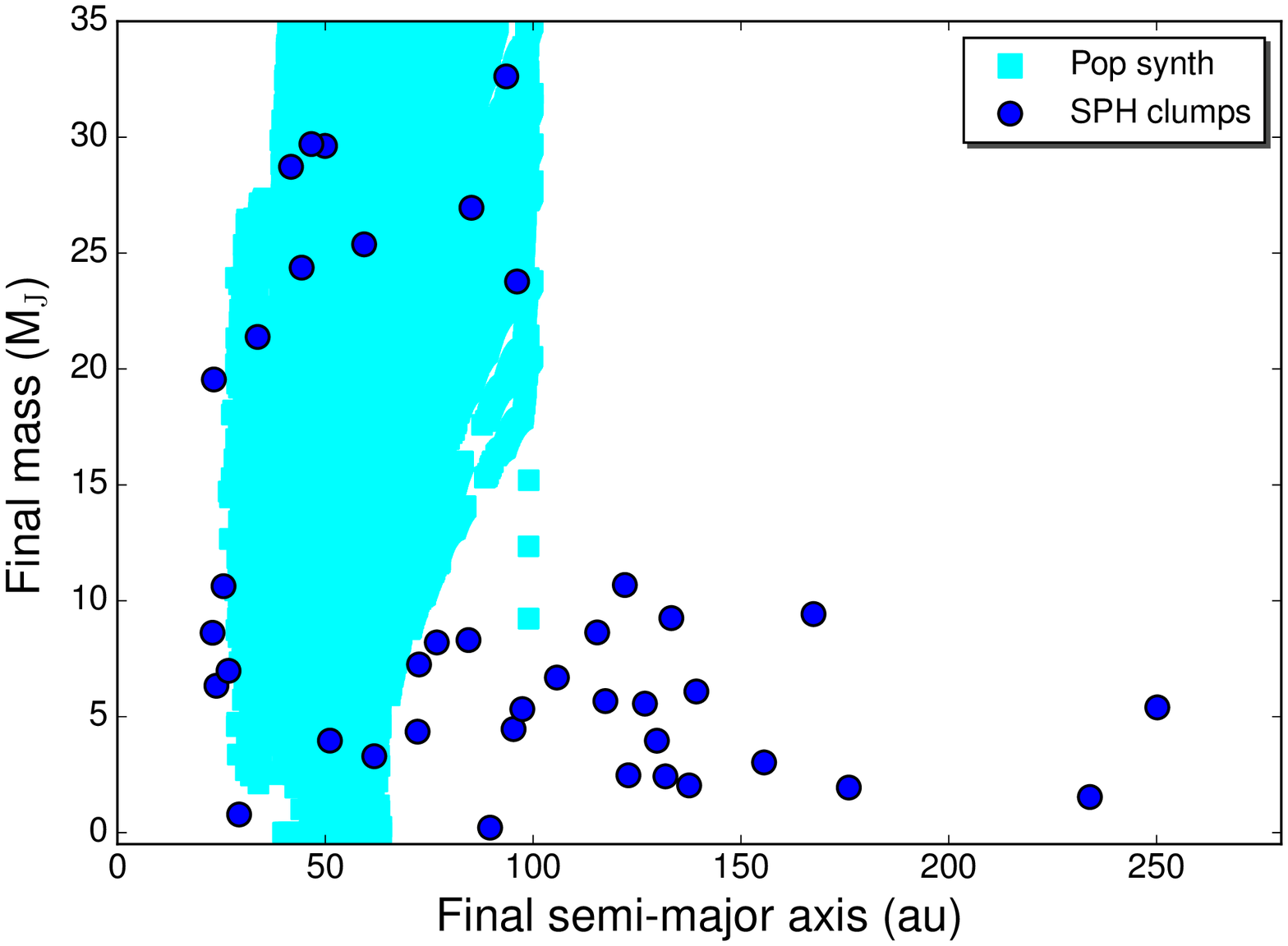}
\caption{ Final mass semi-major axis relation for our SPH clumps (dark blue circles) and \citet{forganricepopsynth2013} fragments (light blue squares). We can see a dearth of intermediate mass SPH clumps at $15-25$ M$_\mathrm{J}$. However, since we are dealing with a small sample size we cannot say for sure if this is statistically significant. Despite the small sample size, we can see that large separations for low-mass objects are much more common than suggested by population synthesis models.
%However, despite our small sample size, we can see that in our SPH simulations, we often end up with large separations for low mass objects, this is due to clump-clump interaction. These happened in all simulations, and indicate that they play an important role in the ultimate fate of these fragments, since at large separations they are unlikely to be tidally destroyed. As such, clump-clump interactions should probably be modelled in GI population synthesis models.
\label{fig:aVmassSPHGI}}
      \end{minipage}
      \hfill
      \begin{minipage}[t]{.48\textwidth}
      \includegraphics[width=\linewidth]{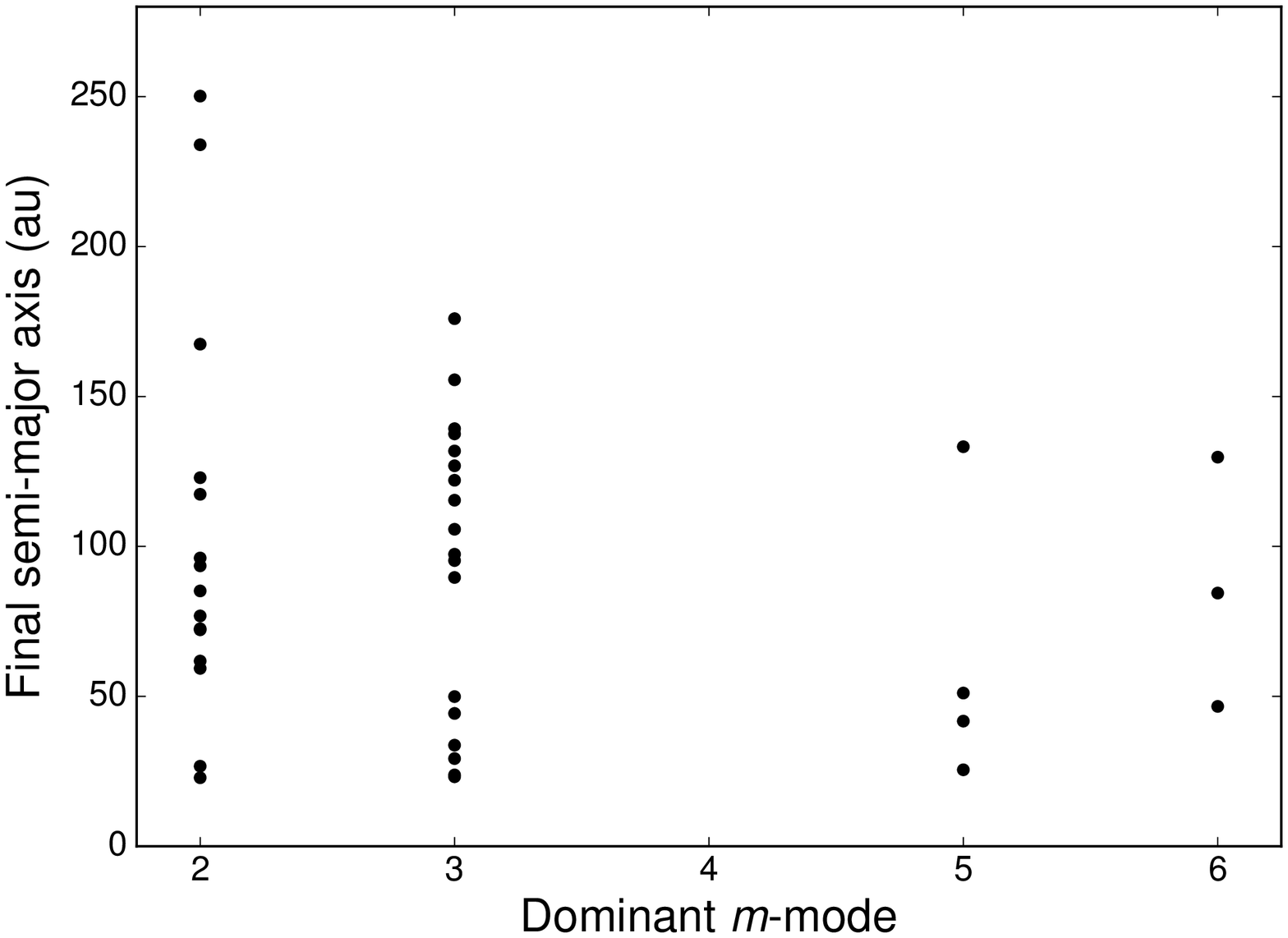}
      \caption{Final semi-major axis of all clumps in all simulations as detected by the density derivative search, as a function of the dominant $m$-mode in the disc. The largest semi-major axis require a 2 armed spiral, which is capable of exerting global torques. There appears to be a rough empirical relationship such that the maximum semi-major axis $a_{\mathrm{max}}\propto\nicefrac{1}{m}$, although this result is preliminary due to a small number of data points.\label{fig:mspiralva}}      
	\end{minipage}
\end{figure*}

\begin{figure*}
    \centering
    \begin{minipage}[t]{.48\textwidth}
      \centering
 \includegraphics[width=\linewidth]{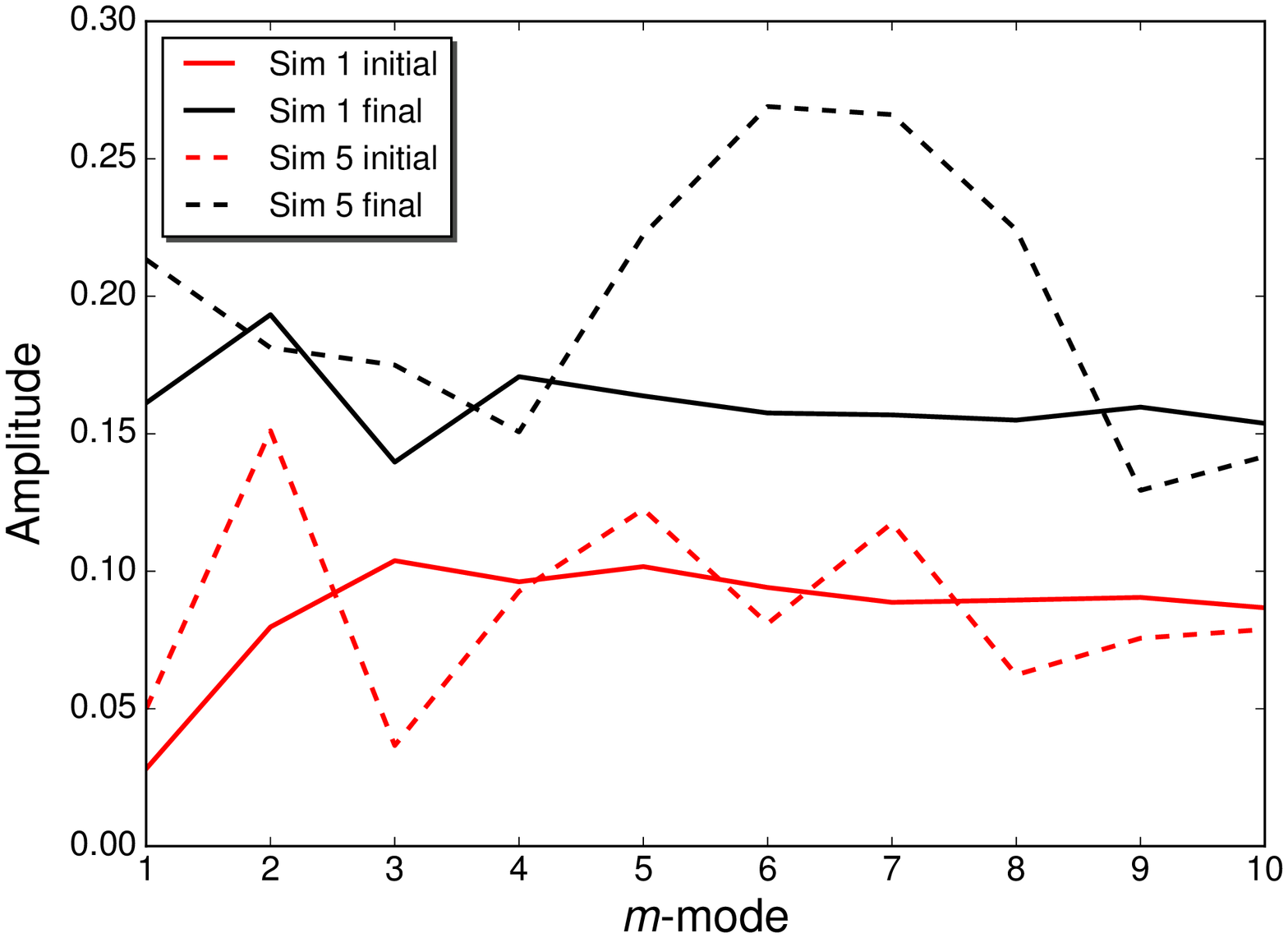}
\caption{Amplitude of the first 10 Fourier components of the density structure of the discs in simulation 1 and simulation 5, calculated between $R=20$ au and $R=100$ au, at their initial state (when they have begun to fragment), marked in solid lines, and their final state, marked in dashed lines, at the last timestep.
  \label{fig:fourieramp}}
      \end{minipage}
    \hfill
    \begin{minipage}[t]{.48\textwidth}
      \centering
  \includegraphics[width=\linewidth]{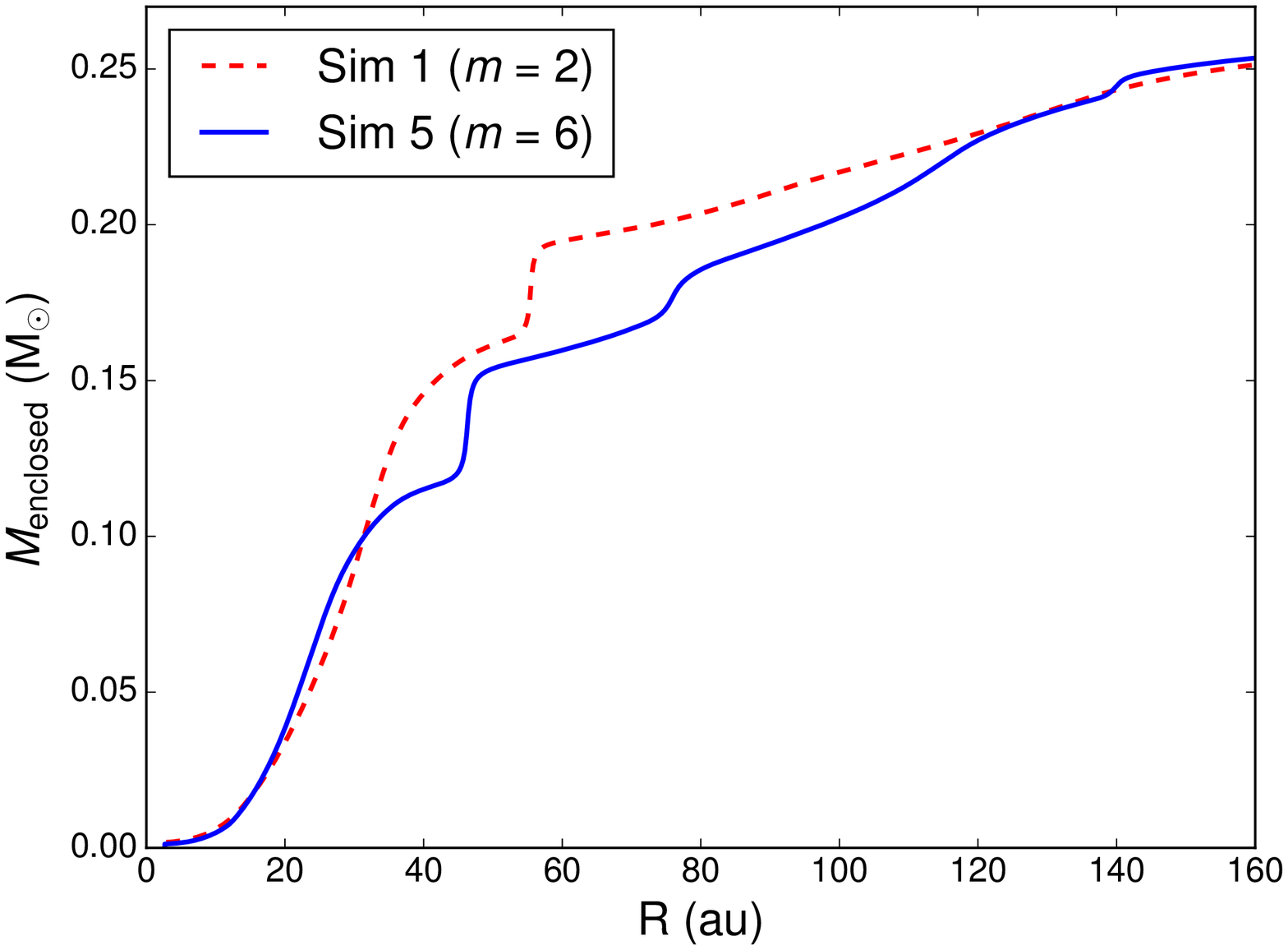}
  \caption{Disc mass enclosed as a function of radius for the final timesteps of simulation 1 (red dashed line) and simulation 5 (blue solid line). Since more mass is enclosed at shorter radii for simulation 1, more mass has been transported inwards in the disc, giving it a slightly steeper density profile.\label{fig:enclosedmassvr}}
    \end{minipage}
\end{figure*}
We ran a total of 9 SPH simulations with almost identical initial conditions, differing only in the random number seed used to initialise the disc.
%Each simulation was run for as long as possible, stopping only when the the largest clump reached a density too high to continue computationally without switching to sink particles.
Column density plots of the 9 simulations are shown in Figure \ref{fig:columndensity}. Despite the almost identical initial conditions, there is a large variation in final configurations and number of clumps in the system.

Our results sections fall into three broad categories- first, the relative merits of the two methods and the difference in clump detection between them. We show that, interestingly, our density derivative search detects all clumps that are detectable by eye, whereas our ordered potential search does not. In fact, generally speaking, the potential search does not detect any clump that is eventually destroyed, giving a good indicator of the likelihood of a clump's survival.

We next discuss the implications of our results for current population synthesis models, comparing our clump mass and semi-major axis functions to existing population synthesis models. We show the clump interaction needs to be included in GI population synthesis models as early as during the gas phase, as scattering plays an important role.

Finally, we discuss interesting events in the simulations themselves. We introduce a piece of nomenclature now, to avoid confusion, that S$a$C$b$ means simulation $a$, clump $b$. This abbreviation is given in the title of any plot of a specific clump. Note that our clump numbering begins at 2, since clump 1 is the star+disc system. We also state now, for clarity, that any mass stated for our clumps \textit{includes} unbound material. This is deliberate, in order to track more of the mass of the clump. Furthermore, what is currently unbound material around the clump, at these very early times, may eventually lose spin angular momentum through interactions with material in the disc, ultimately becoming part of the clump. By including the unbound material, we trace more of this process from earlier times. For high mass clumps ($>20$ M$_{\mathrm{J}}$), the amount of unbound material is small, typically around $10\%$ or so. This is larger for lower mass clumps, up to $\sim 25\%$ of material identified may be unbound, rising to 40$\%$ in clumps ocurring in particularly volatile simulations that have many clumps, as their formation is often disrupted by interactions with other.% clumps.

We look at the orbital properties of the clumps, and discuss clump mergers and tidal destructions by the central star. We show that: (1) destruction and merging are fairly common, (2) interactions between clumps can result in a clump changing its direction of spin from prograde rotation to retrograde rotation, and (3) retro-rotating clumps typically have more dramatic changes in their radial temperature profiles than prograde clumps.
\subsection{Relative merits of gravitational potential energy search and density derivative search}
\label{subsec:relative}
%===================================================================================
We identified clumps in our simulation using two different search methods, an ordered potential energy search (OPS) based on \citet{smith2008}, and a novel approach based on a 2D density derivative search (DDS). Our first conclusion is that searching using the ordered potential of the particles only detects clumps that survive the duration of the simulation. However, using the DDS method, destroyed clumps are also detected. This is shown in Figure \ref{fig:migration}, which shows the total change in semi major axis, from when the clump is first detected to when the clump is last detected (or destroyed), against the total time for which the clump exists, i.e. from initial identification to the end of the simulation, or the last timestep in which it is identified, if it is destroyed. Larger markers indicate more massive clumps. The left hand panel shows the DDS results, the right hand panel shows the OPS results. Circular markers indicate clumps that have survived until the end of the simulation, square markers indicate clumps that are destroyed, and triangle markers indicate a clump that merged into another clump. There are no identical markers in both plots because the different algorithms detect the clumps at different times, therefore they migrate different distances.

In addition to containing no destroyed clumps, the OPS sample also has a relative insensitivity to clumps which will have a final mass of less than $\sim$ 5 M$_\mathrm{J}$, and detects most of the clumps later in the simulation. This is shown in Figure \ref{fig:massVtimeUonly}, which shows mass accretion histories for clumps in each of the 9 simulations, as is detected the OPS. By comparison, Figure \ref{fig:massVtime} shows the same 9 simulations, but with the mass accretion histories of the clumps determined by using the DDS method. As can be seen, many low mass clumps evade detection entirely under the OPS method; for example, simulation 6 has an additional 3 clumps that are not detected by the OPS, and those that are detected are generally detected later, such as clump 2 in simulation 2, which is detected $\sim 400$ years later in the OPS than in the DDS.

This is due to the nature of the potential search. Figure \ref{fig:potential} shows the radial gravitational potential energy profile of the disc in Figure \ref{fig:columndensity}, simulation 6. Since the OPS proceeds from the particle with the most negative gravitational energy to the most positive, the OPS detects the clump at $\sim$ 80 au in Figure \ref{fig:potential} first, and then detects the clump at $\sim$ 375 au second. However, it fails to detect the clumps at $\sim$ 50 au, $\sim$ 125 au and $\sim$ 200 au. This is because particles at the potential energy of the main body of the disc are identified as belonging to the clumps with the most negative potential energy during the neighbour check described in section \ref{subsubsec:clumpfind}. Then, when the particles belonging to clumps at 50, 125 and 200 au are checked, they are found to already belong to either the main body of the disc (i.e. the central sink) or one of the clumps with the deepest potential well.

Although this could be fixed by adopting a gridded approach to the potential search (thereby eliminating the dominating effect from the clumps with the largest potential well), OPS has a desirable feature, namely demarcating clumps that are likely to survive the simulation, and those that are not. 

The OPS method's insensitivity to small clumps, and reliance on deep potential wells for identifying the body of the clump, mean that fewer clumps are detected by the OPS method, and those that are identified are often initially identified at artificially small masses ($\sim 10^{-3}$ M$_{\mathrm{J}}$), as there is only a small amount of mass with a potential well deep enough to be identified. This can be seen in Figure \ref{fig:massVtimeUonly}, which shows mass accretion histories for all 9 simulations using the OPS method. In every simulation, what ultimately grows to be the largest clump is initially identified with a mass well below the Jeans mass. The DDS search method, however, does a better job of correctly identifying the mass associated with the young clumps, typically identifying between 5 M$_{\mathrm{J}}$ and 10 M$_{\mathrm{J}}$ of mass. This is shown in Figure \ref{fig:massVtime}, which shows the mass accretion histories for all 9 simulations, as identified using the DDS method. 

In addition to the difference in measured initial clump mass, both methods differ in the final mass attributed to clumps. Typically, the growth is smoother for the OPS method, since once mass is deep in the potential well it is unlikely to change. However, if the gravitational potential energy profile of the disc changes then so too will the attributed mass of the clumps.

Another feature of the DDS method is its ability to identify low mass, low density clumps that do not have a strong signal in their potential. This can be seen in Figure \ref{fig:massVtime}, simulation 6, which identifies an additional three clumps compared to Figure \ref{fig:massVtimeUonly}, simulation 6. Comparing Figure \ref{fig:potential} and Figure \ref{fig:derivative}, we can see that these clumps have much stronger signals in their radial density profiles compared to their radial potential energy profiles. This is a useful predictive feature, since the left hand panel of Figure \ref{fig:migration} and all of Figure \ref{fig:massVtime} show that $\sim20\%$ of the fragments in our simulations are tidally destroyed (we discuss the implications of this for population synthesis in section \ref{subsec:compareGI}), none of which are detected in the OPS method.

Therefore, if a clump is detected in the DDS and not in the OPS, it is indicative that either the clump will stay relatively low mass and not accrete further, or that it will be tidally destroyed. 
%\subsection{Comparison to gravitational instability population synthesis models}
\subsection{Comparison to gravitational instability population synthesis models}
\label{subsec:compareGI}
We ran the \citet{forganricepopsynth2013} GI population synthesis (GIPS) models for 4000 years, which is comparable to the timescale for which our SPH simulations are able to run until no longer computationally feasible. We dictated that the opacity power law be $p_k=1$, and that the disc is not truncated after fragmentation. Figure \ref{fig:histogram} shows initial and final mass and semi-major axis distributions for the GI population synthesis model and for the clumps in our SPH simulation. The red, hatched histogram is the population synthesis, blue outline is the SPH clumps. The left hand panel of Figure \ref{fig:histogram} shows the initial and final semi-major axis distributions for both samples, and the right hand panel shows initial and final mass distribution for both samples. We cut off the tail of the initial and final masses beyond $>35$ M$_{\mathrm{J}}$, since we cannot feasibly simulate masses above this without switching to sink particles. 

Comparing the initial mass and semi-major axis distributions in the GIPS models to our SPH clumps, it would initially appear that clumps in SPH simulations form much further out, and at much lower masses, than in the population synthesis models. In reality, this is somewhat a limitation of our identification algorithms, as at very early times, the clumps can escape detection because of their low density/less negative gravitational potential energy, such that they have already undergone some radial migration before they are detected. If they have undergone sufficient radial migration, they may be far enough out in the disc to not accrete much material, hence remaining low mass.

Having established that the initial mass and semi-major axis distributions for our SPH clumps are subject to some limitations of our detection algorithm, we now compare the final mass and final semi-major axis distributions of our SPH clumps to those in the GI population synthesis model. First, the bottom left hand panel of Figure \ref{fig:histogram} shows a dearth of clumps at $R< 25$ au, when compared to the GIPS model - this is simply due to the last measured value of $a$ before the fragment is destroyed. This figure also shows that the distribution of semi-major axes is different to what is expected, given the GIPS model data, and therefore the mechanism that allows these separations to exist at early times (i.e. clump-clump interactions) plays an important role in the ultimate orbital distribution function of the sample.

Second, the bottom right hand panel of Figure \ref{fig:histogram} shows that our final mass distribution is bimodal, with peaks at $\sim 5$ M$_{\mathrm{J}}$ and $\sim 30$ M$_{\mathrm{J}}$. This is somewhat consistent with previous measurements of mass distributions of fragmenting discs by \citet{vorobyov2013}, who find that there are two maxima in their mass distribution, at $\sim 5$ M$_{\mathrm{J}}$ and $\sim$ 60 M$_{\mathrm{J}}$. Unlike \citet{vorobyov2013}, we find a gap at $\sim$ 15 M$_{\mathrm{J}}$, whereas they find a minima at $\sim$ 25 M$_{\mathrm{J}}$, and our second maxima is at $\sim$ 30 M$_{\mathrm{J}}$ rather than $\sim 60$ M$_{\mathrm{J}}$. Given our small $N$ statistics, we would probably expect our distribution to converge on a minima at $\sim$ 15 M$_{\mathrm{J}}$, rather than the gap that is currently present. Additionally, our second peak is capped at $\sim$ 30 M$_{\mathrm{J}}$ in our simulations since this is typically when the density in a fragment becomes so high that it is computationally unfeasible to continue the simulation. With increased computation time, the mass of our largest clumps would probably increase.

Since our algorithm is quite robust at later times, detecting to within a factor of 2 the "by eye" clump mass in low-mass clumps (considered only in the bound region of the clump), we can see, in the final mass distribution shown in the bottom right panel of Figure \ref{fig:histogram}, that the GI population synthesis model is significantly underestimating the fraction of planets at $<5$ M$_\mathrm{J}$, even accounting for under-estimating low mass clumps by a factor of 2. 

This can be explained by Figure \ref{fig:aVmassSPHGI}, which shows the mass semi-major axis distribution for the final values of the SPH clumps and the population synthesis fragments. The SPH clumps are the dark circles, and the population synthesis fragments are the light squares. As can be seen, low mass clumps in our simulations are scattered out to large semi-major axis at these early times, and fragment-fragment interactions are likely to play an important role in the ultimate fate of a fragment. If it is scattered out to large $a$, it is much less likely to be tidally destroyed and far more likely to survive the duration of the simulation. This would suggest that GI population synthesis models need to include fragment-fragment interactions in this early gas phase, since current models suggest that $\sim 40\%$ of initial fragments are tidally destroyed. If a significant fraction of these are scattered out to large radii, their survival rate could potentially be much higher.

\subsection{Orbital properties}
\label{subsec:orbital}
%==========================================
% Eccentricity semi-major axis
%==========================================
\begin{figure}
      \centering
\includegraphics[width=\linewidth]{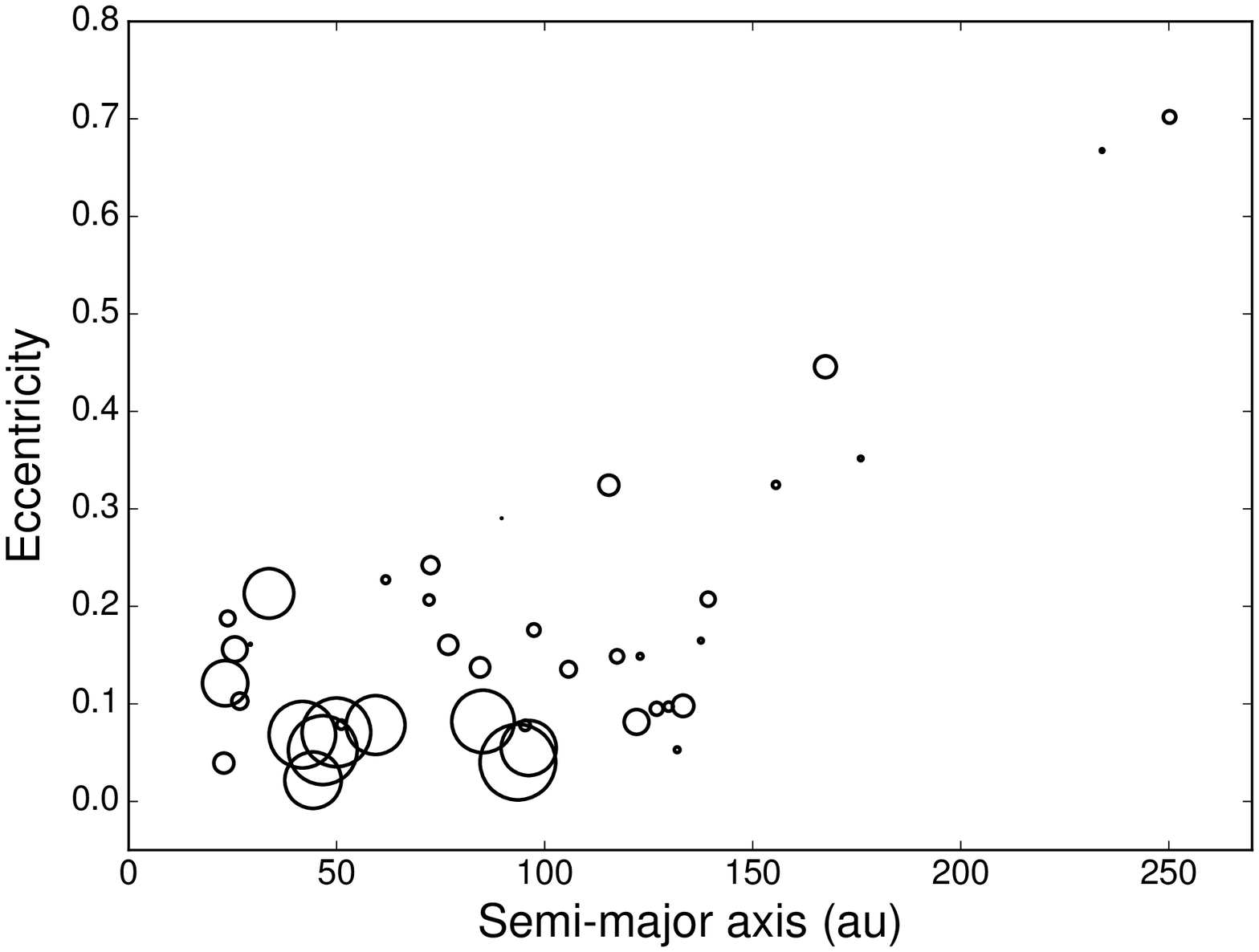}
\caption{Final semi-major axis and eccentricity relation for our SPH clumps. Larger markers represent more massive clumps. The population synthesis model does not contain eccentricity data. We can see, for the most part, our clumps have a roughly linear relationship between eccentricity and semi-major axis, although there is a large amount of scatter.\label{fig:eccentricity}}
\end{figure}
%=========================================
% Eccentricity histogram figure
%=========================================
\begin{figure*}
\begin{center}
\begin{tabular}{cc}
\textbf{Eccentricity} & \textbf{Inclination}\\
\includegraphics[width=0.5\linewidth]{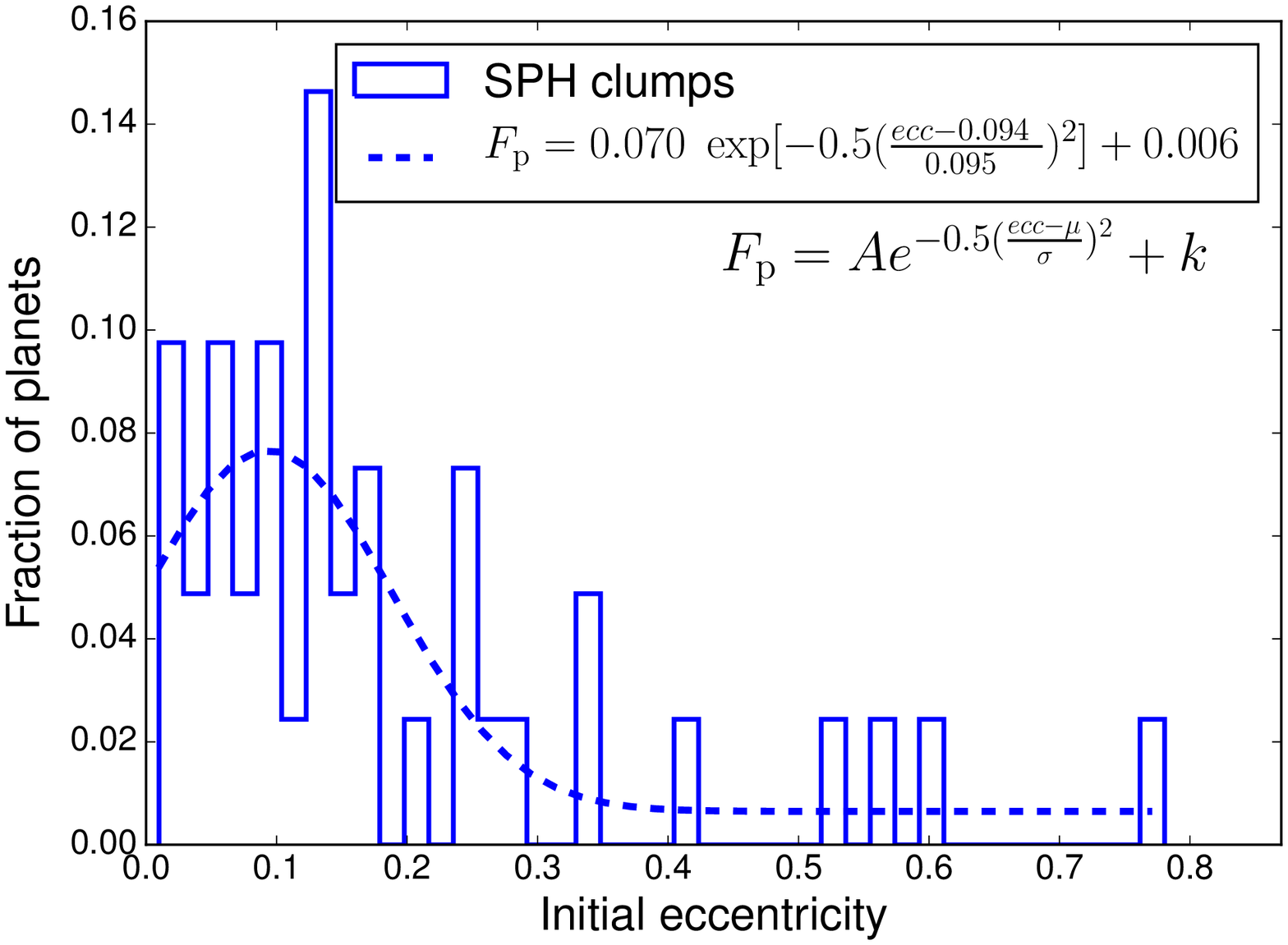} & \includegraphics[width=0.5\linewidth]{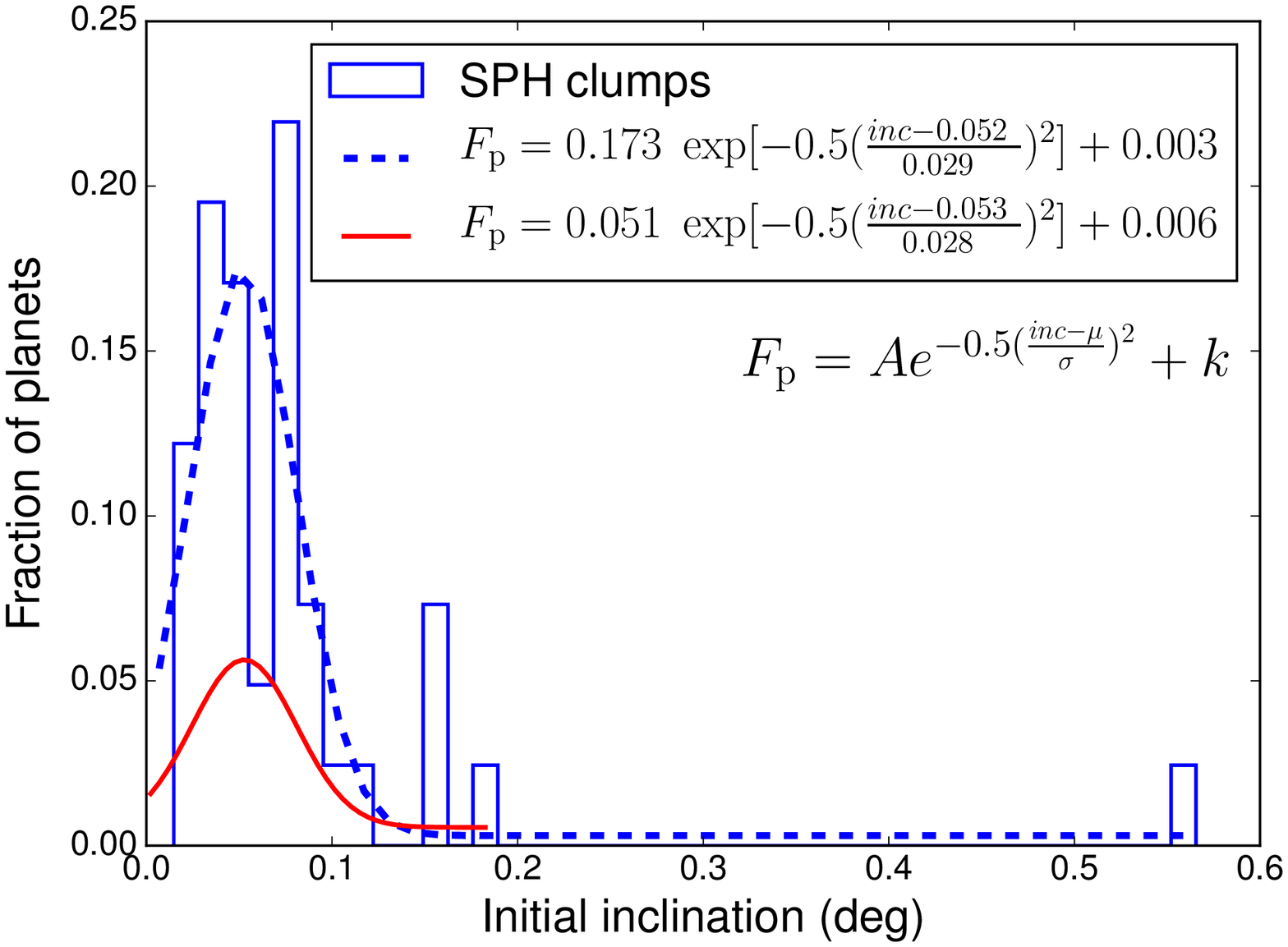} \\
\includegraphics[width=0.5\linewidth]{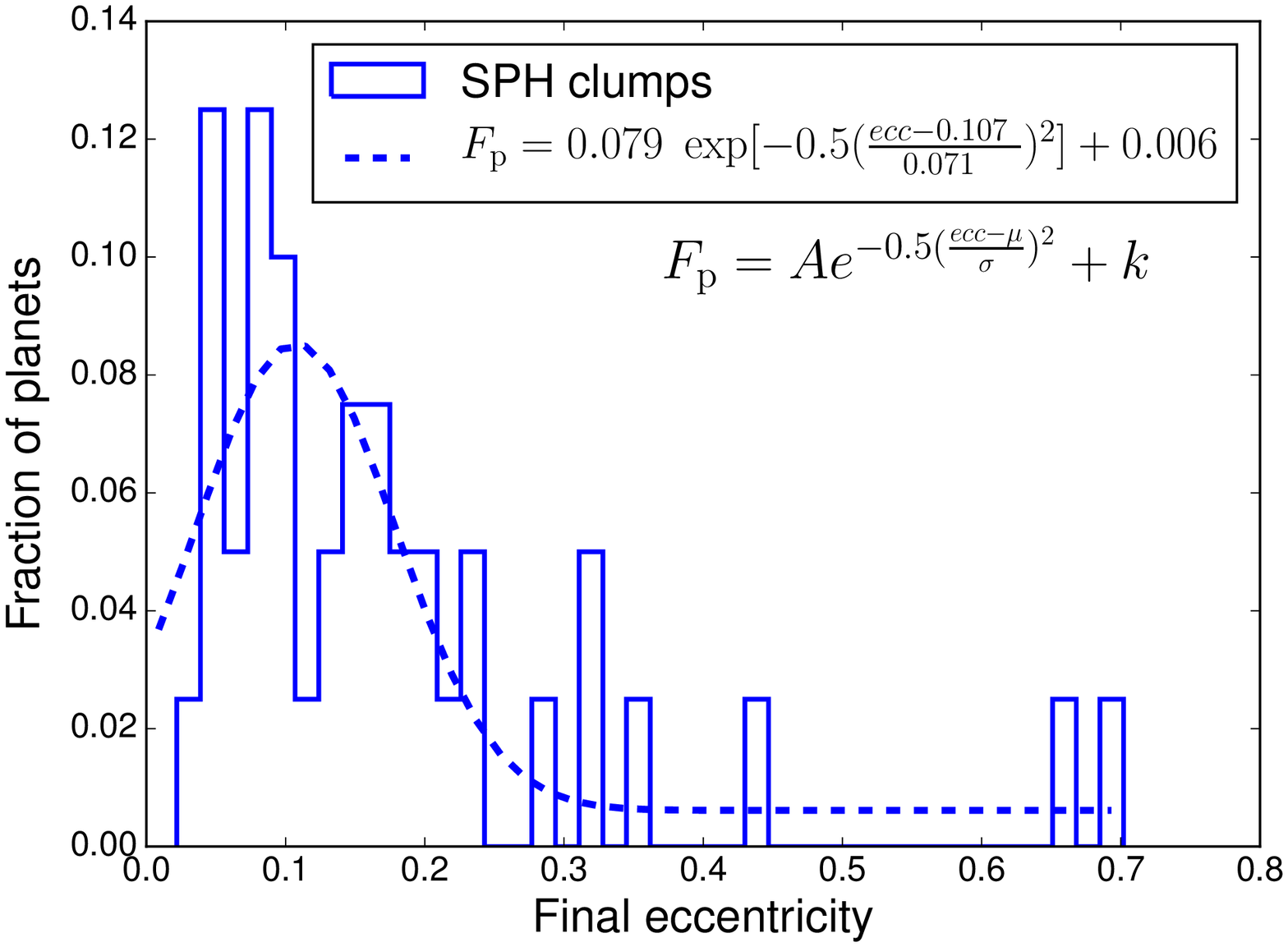} & \includegraphics[width=0.5\linewidth]{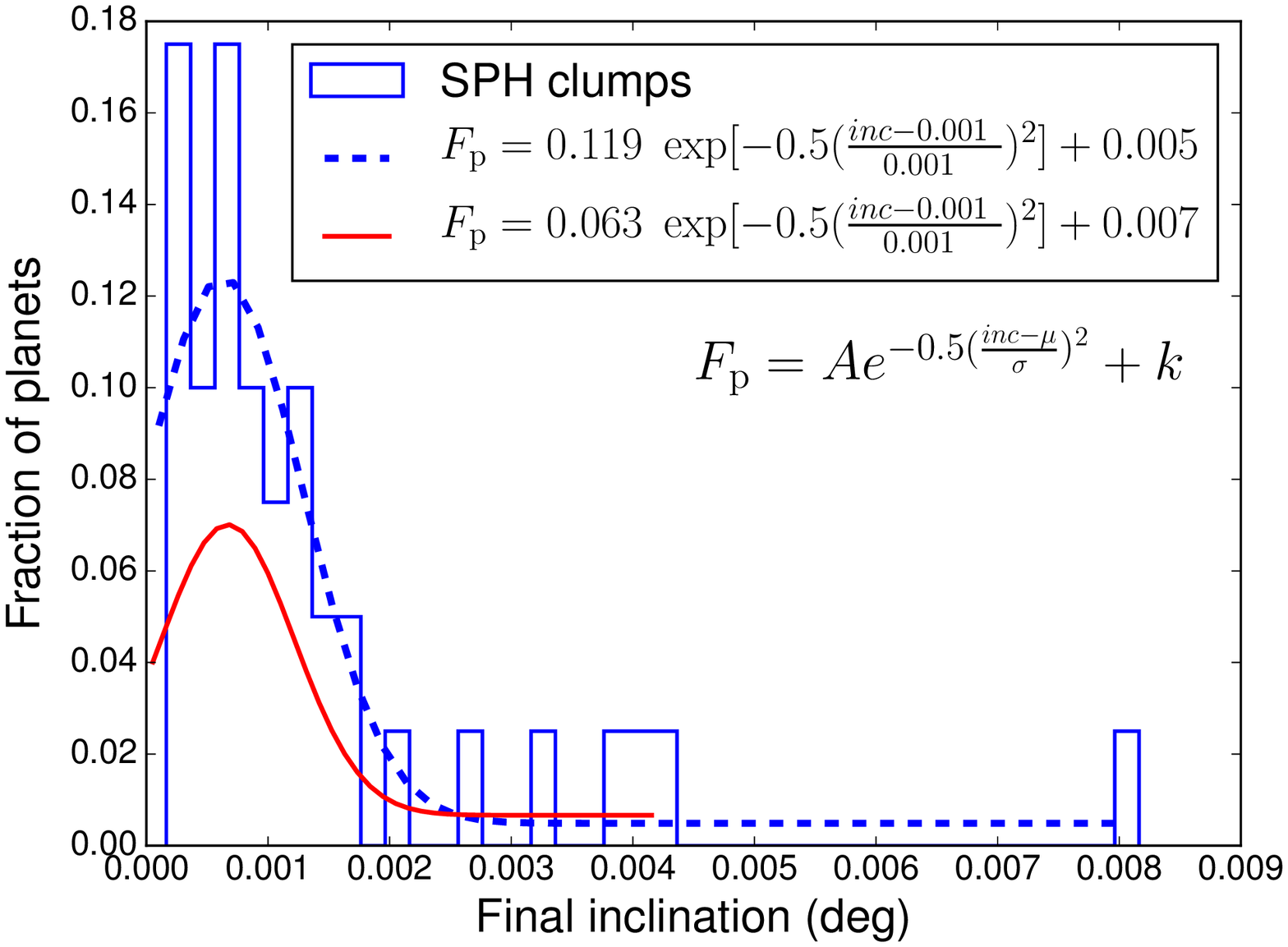} 
\end{tabular}
\caption{Left column shows initial (top) and final (bottom) eccentricity distribution for our SPH clumps. Right column shows the initial (top) and final (bottom) inclination distribution for the same. Inclination is calculated relative to the orbital plane of the central star. Since the population synthesis models of \citet{forganricepopsynth2013} do not contain eccentricity or inclination information, we do not plot them here. Using least-squares regression, we have fitted our distributions with a Gaussian of the form $F_{\mathrm{p}}=A\exp[-0.5(\frac{x-\mu}{\sigma})^2]+k$, where $F_{\mathrm{p}}$ is the fraction of planets, $A$ is the amplitude of the curve (without offset), $x$ is eccentricity or inclination, $\mu$ is the mean of the distribution, $\sigma$ is the standard deviation and $k$ is the offset constant. The fitted values are given in Table \ref{table:fitvalues} and in each plot legend. We have included these fits since we consider they may be useful in developing future GI population synthesis models, but caution that our sample size is small. Our inclination histograms have been fitted with two distributions. The dotted blue line includes all data points, and the solid red line does not include the most inclined point in each distribution, since with a small sample size, and an apparent gap between the rest of the clumps, it is unclear whether or not this is an outlier. There is little change between our initial and final eccentricity distribution, both peaking at $e\sim 0.1$, and a slightly smaller standard deviation for the final configuration. The inclinations of our clumps are decreased by a factor of $\sim$ 100 between their initial and final states, showing that clump orbital inclination is rapidly adjusted after formation, until it orbits almost entirely in the plane of motion of the central star.
\label{fig:histogramecc}}
\end{center}
\end{figure*}
%========================================
% Disalignment figure
%========================================
\begin{figure}
%\begin{minipage}{0.5\textwidth}
\centering
%\begin{tabular}{ll}
\begin{tabular}{c}
%trim={<left> <lower> <right> <upper>}
%\includegraphics[trim={5cm 0 0 0},clip]{example-image-a}
\includegraphics[width=1.0\linewidth,trim={1.cm 1cm 0 0},clip]{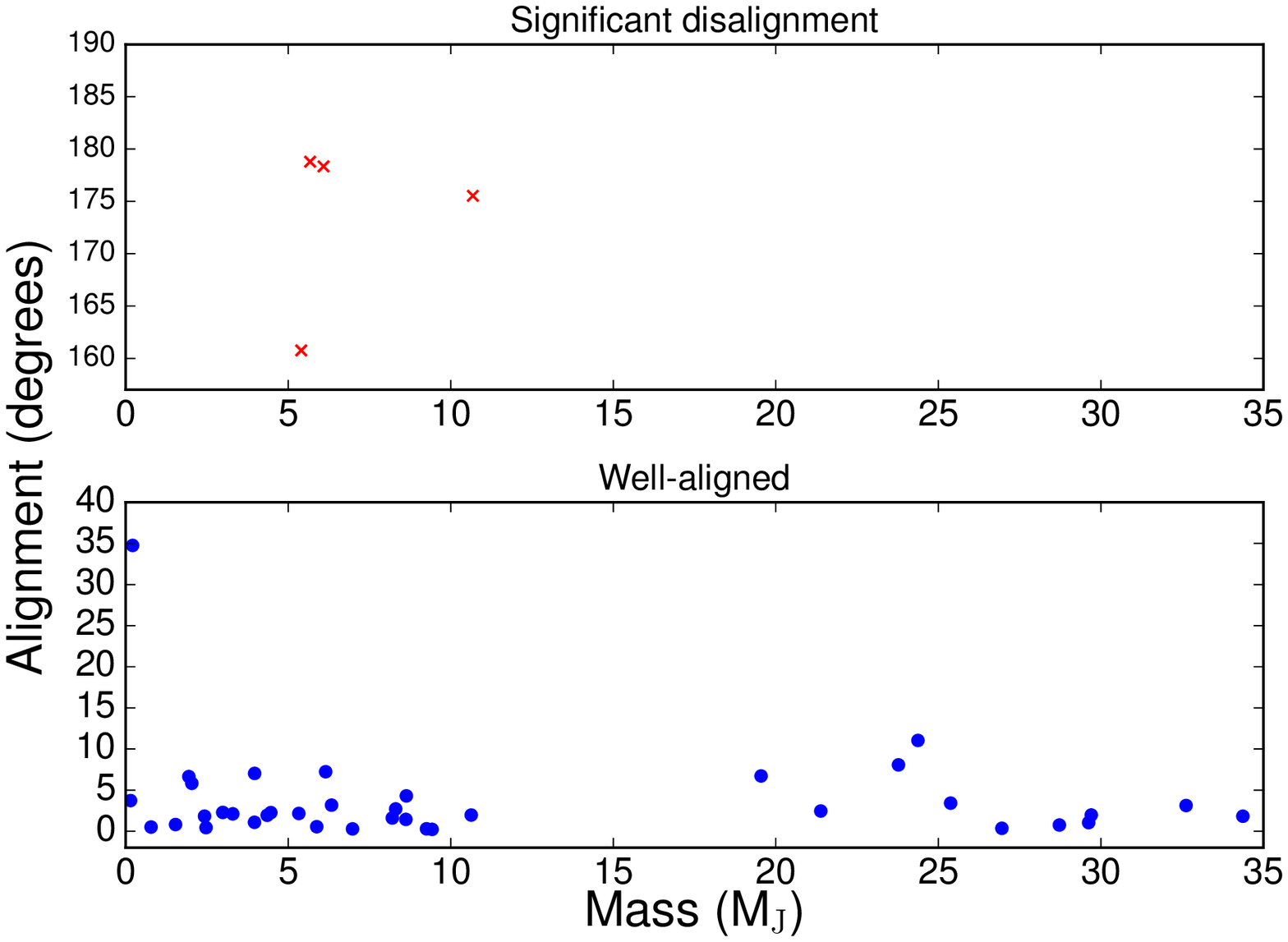}\\
 \includegraphics[width=1.0\linewidth,trim={1.cm 1cm 0 0},clip]{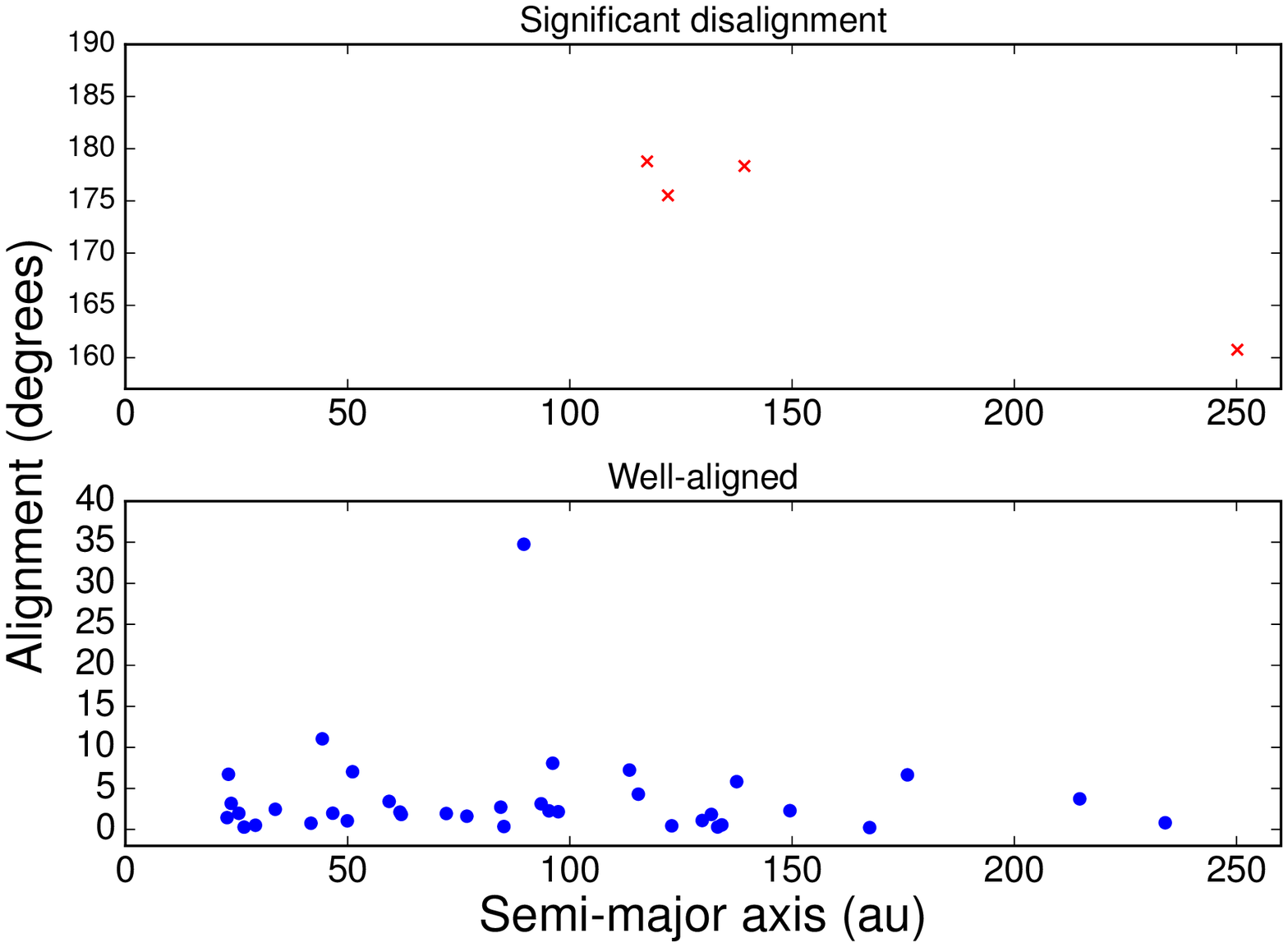}
\end{tabular}
\caption{All plots show alignment, on the y-axis, defined as the angle between the orbital angular momentum vector and the rotational angular momentum vector. Top two panels show alignment as a function of mass, bottom two panels show alignment as a function of semi-major axis. The plots are split, for clarity, into two different regions, between $0^o $ and $40^o$ and between $155^o$ and $190^o$. Well aligned clumps are marked by blue circles, clumps with a significant degree of disalignment are marked with red crosses. Both plots show that four low-mass, high-separation clumps are retro-rotators, given that all of them or orbiting in prograde motion. \label{fig:spinorbit}}
%\end{minipage}
\end{figure}
\begin{figure*}
%used this tool https://ff.cx/latex-overlay-generator/#/
\begin{tabular}{ccc}
\begin{tikzpicture}
%\node [anchor=west] (note) at (-1,3) {\Large Note};
%\node [anchor=west] (water) at (-1,1) {\Large Water};
\begin{scope}[xshift=1.5cm]
    \node[anchor=south west,inner sep=0] (image) at (0,0) {\includegraphics[width=0.33\linewidth]{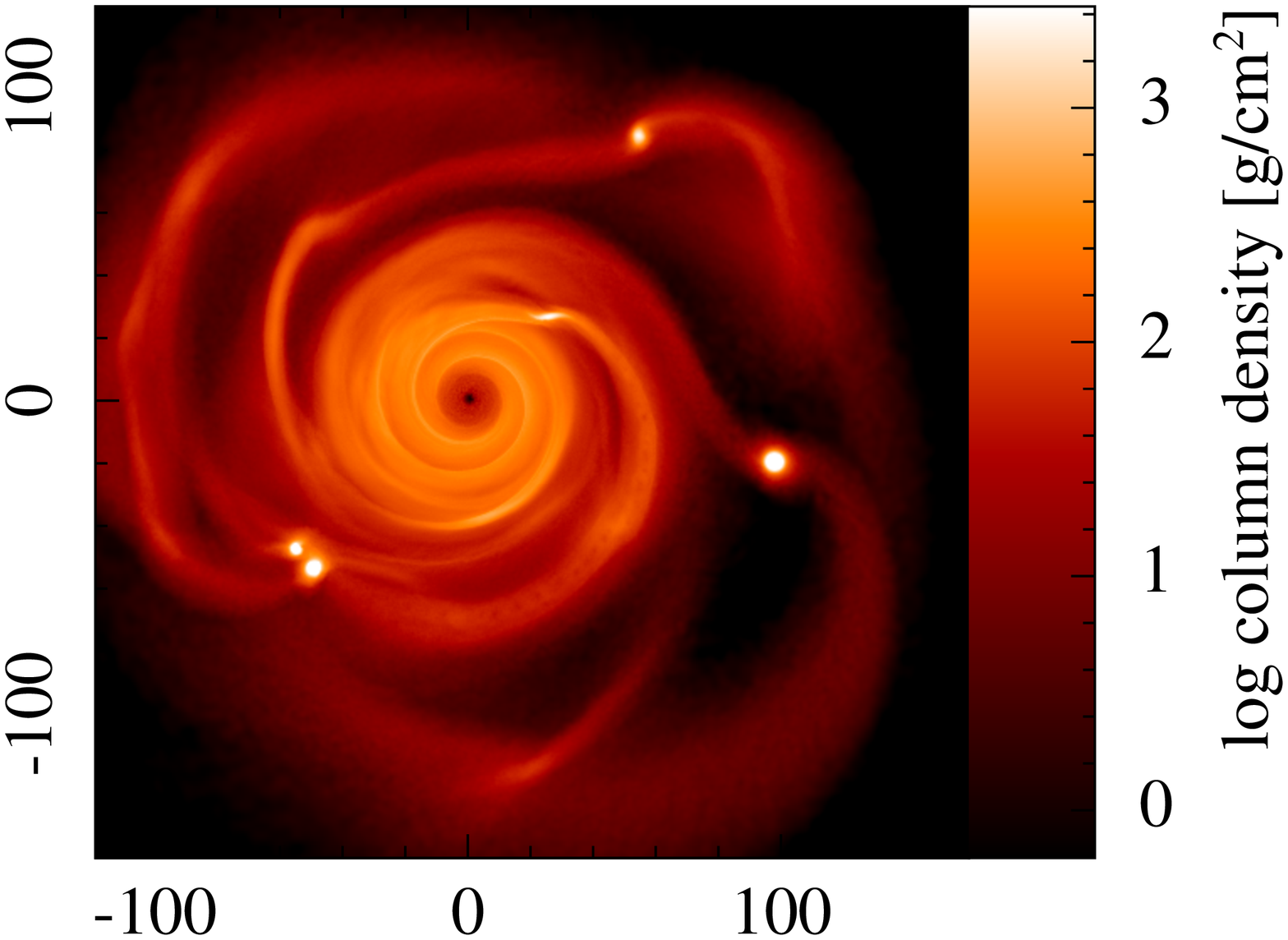}};
    \begin{scope}[x={(image.south east)},y={(image.north west)}]
    %{0.241,0.3057}{0.2766,0.3546}
      % {0.2489,0.3148}{0.2715,0.3437}
            \node[above left,text width=2cm,align=center,white] at (0.8,0.88){T=2456 yrs};
        %\draw[green,thick] (0.2489,0.3148) rectangle(0.2715,0.3437);
        	\draw[green,thick](0.209,0.3587) rectangle(0.2615,0.4516);%bl
        %\draw [-latex, ultra thick, red] (note) to[out=0, in=-120] (0.48,0.80);
        %\draw [-stealth, line width=5pt, cyan] (water) -- ++(0.4,0.0);
    \end{scope}
\end{scope}
\end{tikzpicture}%
&
\begin{tikzpicture}
%\node [anchor=west] (note) at (-1,3) {\Large Note};
%\node [anchor=west] (water) at (-1,1) {\Large Water};
\begin{scope}[xshift=1.5cm]

    \node[anchor=south west,inner sep=0] (image) at (0,0) {\includegraphics[width=0.33\linewidth]{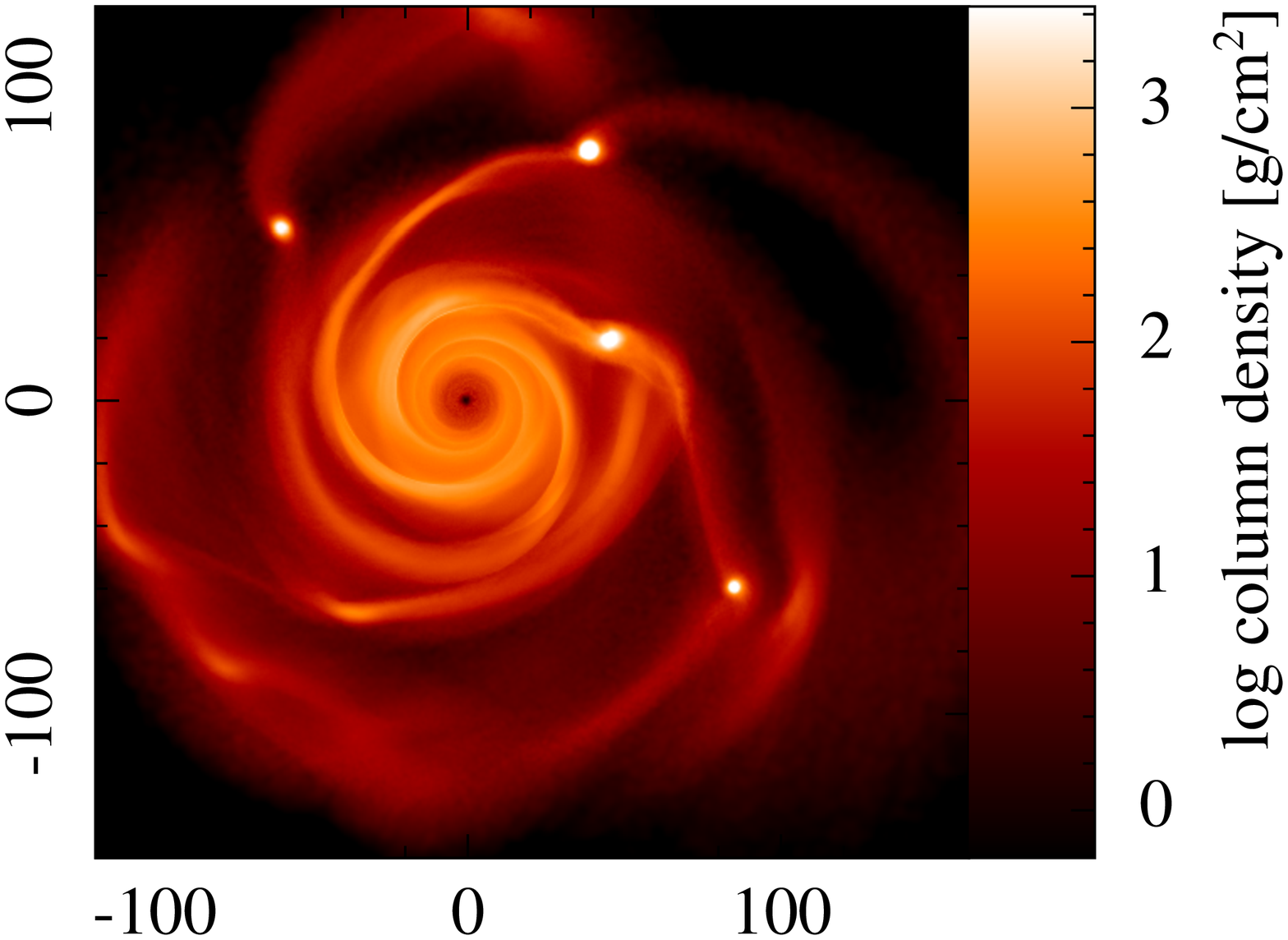}};
    \begin{scope}[x={(image.south east)},y={(image.north west)}]
    %{0.241,0.3057}{0.2766,0.3546}
      % {0.2489,0.3148}{0.2715,0.3437}
            \node[above left,text width=2cm,align=center,white] at (0.8,0.88){T=2646 yrs};
        %\draw[green,thick] (0.5765,0.2476) rectangle(0.6195,0.3142);
        \draw[green,thick] (0.545,0.3367)rectangle (0.599,0.4111);
        %\draw [-latex, ultra thick, red] (note) to[out=0, in=-120] (0.48,0.80);
        %\draw [-stealth, line width=5pt, cyan] (water) -- ++(0.4,0.0);
    \end{scope}
\end{scope}
\end{tikzpicture}%
&
\begin{tikzpicture}
  %\node[above left,text width=2cm,align=center,white] at (img.north east)
             %{\\~\\ \hspace{20mm}T=4238 yrs};
%\node [anchor=west] (note) at (-1,3) {\Large Note};
  %\node [anchor=west] (water) at (-1,1) {\Large Water};
  %\node[above left](water) at (0.5,0.5){ \hspace{20mm}T=4238 yrs};
\begin{scope}[xshift=1.5cm]
  \node[anchor=south west,inner sep=0] (image) at (0,0) {\includegraphics[width=0.33\linewidth]{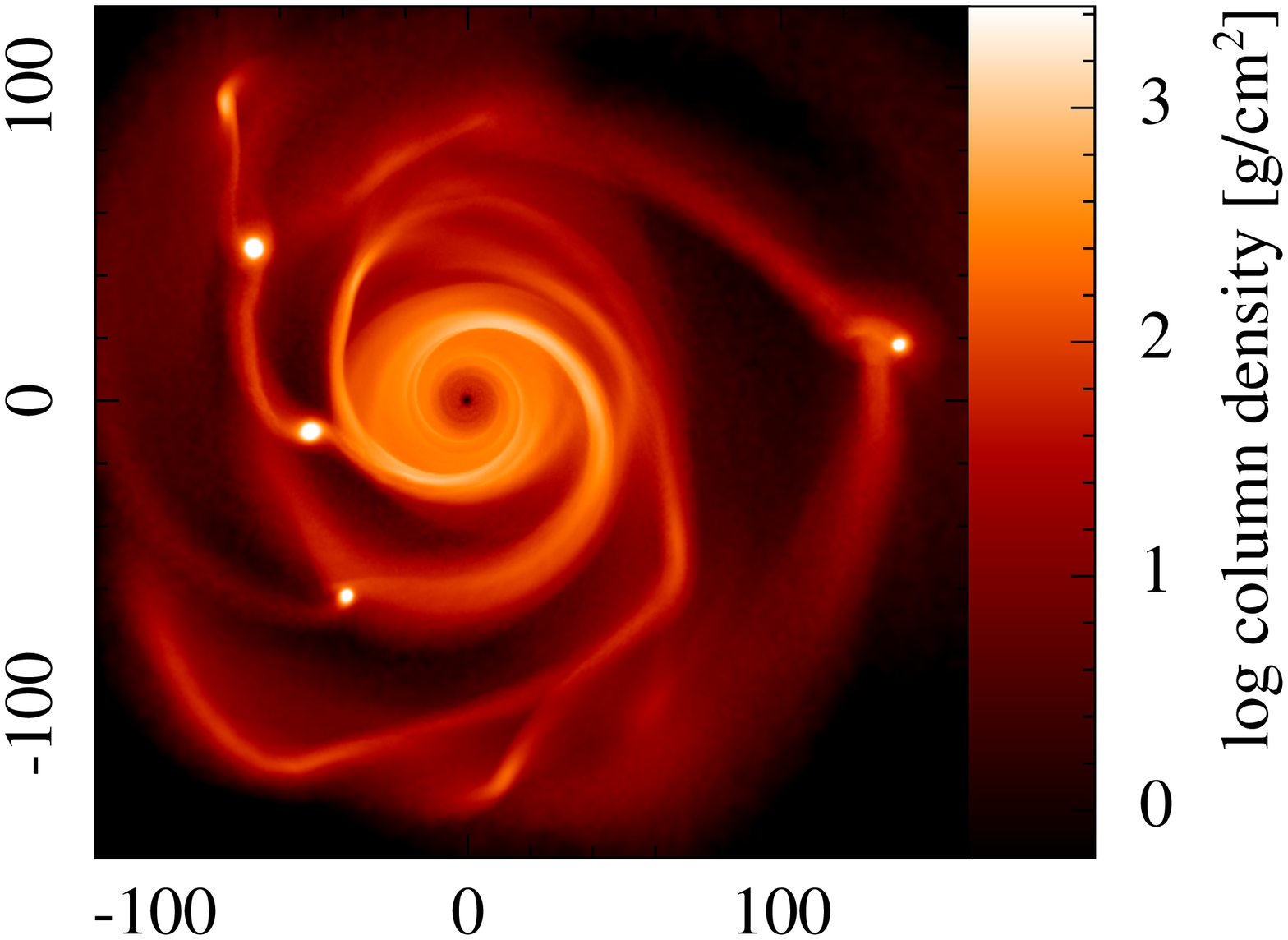}};
          %\node[above left,text width=2cm,align=center,white] at (img.north east)
             %{\\ \hspace{20mm}T=4238 yrs};
    \begin{scope}[x={(image.south east)},y={(image.north west)}]
    %{0.241,0.3057}{0.2766,0.3546}
      % {0.2489,0.3148}{0.2715,0.3437}
      \node[above left,text width=2cm,align=center,white] at (0.8,0.88){T=2800 yrs};
        %\draw[green,thick] (0.6545,0.5177) rectangle(0.7595,0.6789);
         \draw[green,thick](0.673,0.6016) rectangle (0.727,0.676);
        %\draw [-latex, ultra thick, red] (note) to[out=0, in=-120] (0.48,0.80);
        %\draw [-stealth, line width=5pt, cyan] (water) -- ++(0.4,0.0);
    \end{scope}
\end{scope}
\end{tikzpicture}%
\end{tabular}
\caption{Column density plots of simulation 4, increasing in time from left to right, showing clump 4 (marked by green square), an initially prograde-rotating clump, undergoing an encounter with another clump to become a retro-rotating clump.\label{fig:encounter}}
\end{figure*}
%=============================================
\begin{figure*}
    \centering
    \begin{minipage}[t]{.48\textwidth}
        \centering
        \begin{tabular}{c}
          \includegraphics[width=\linewidth]{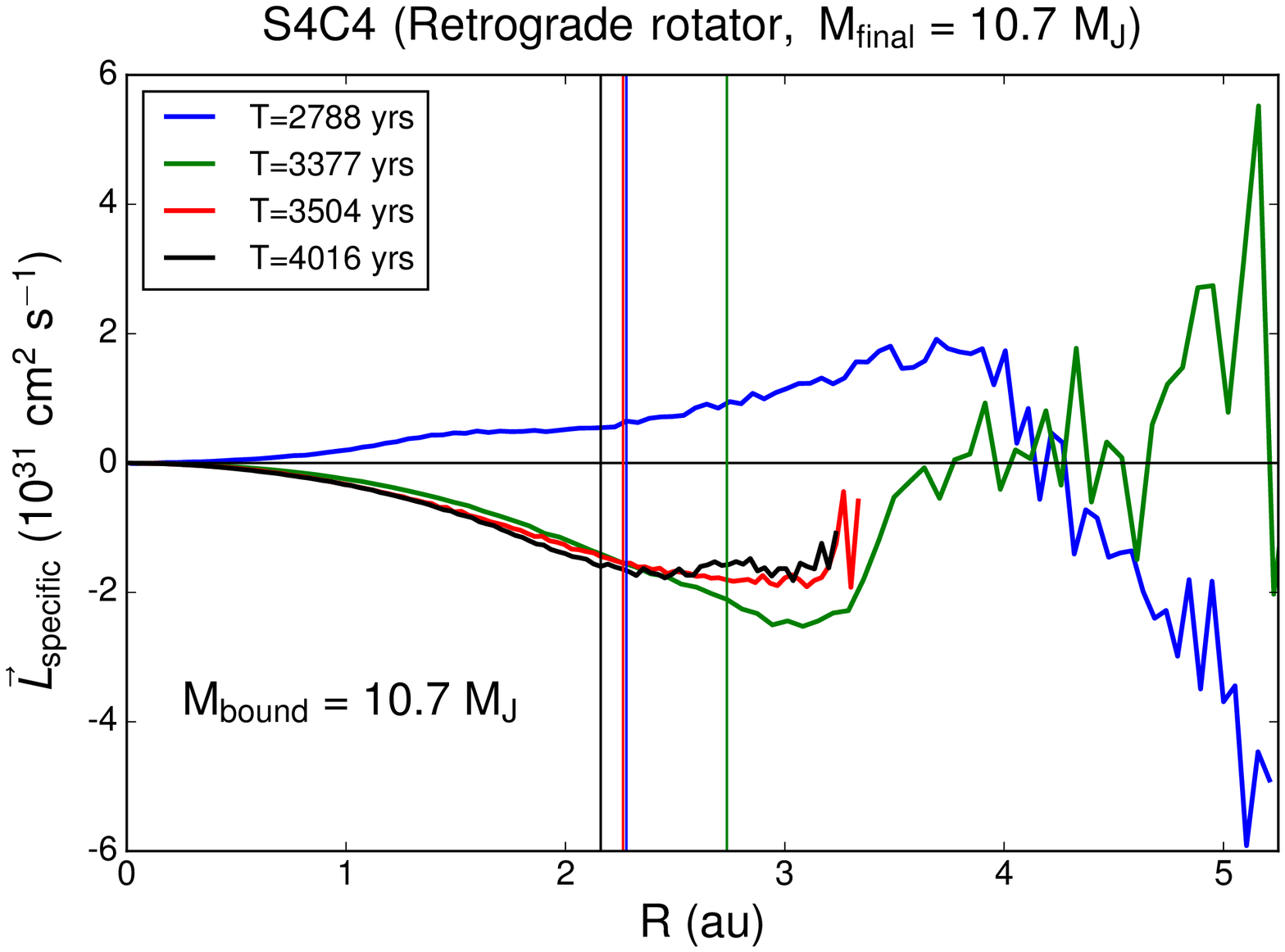}\\
          \includegraphics[width=\linewidth]{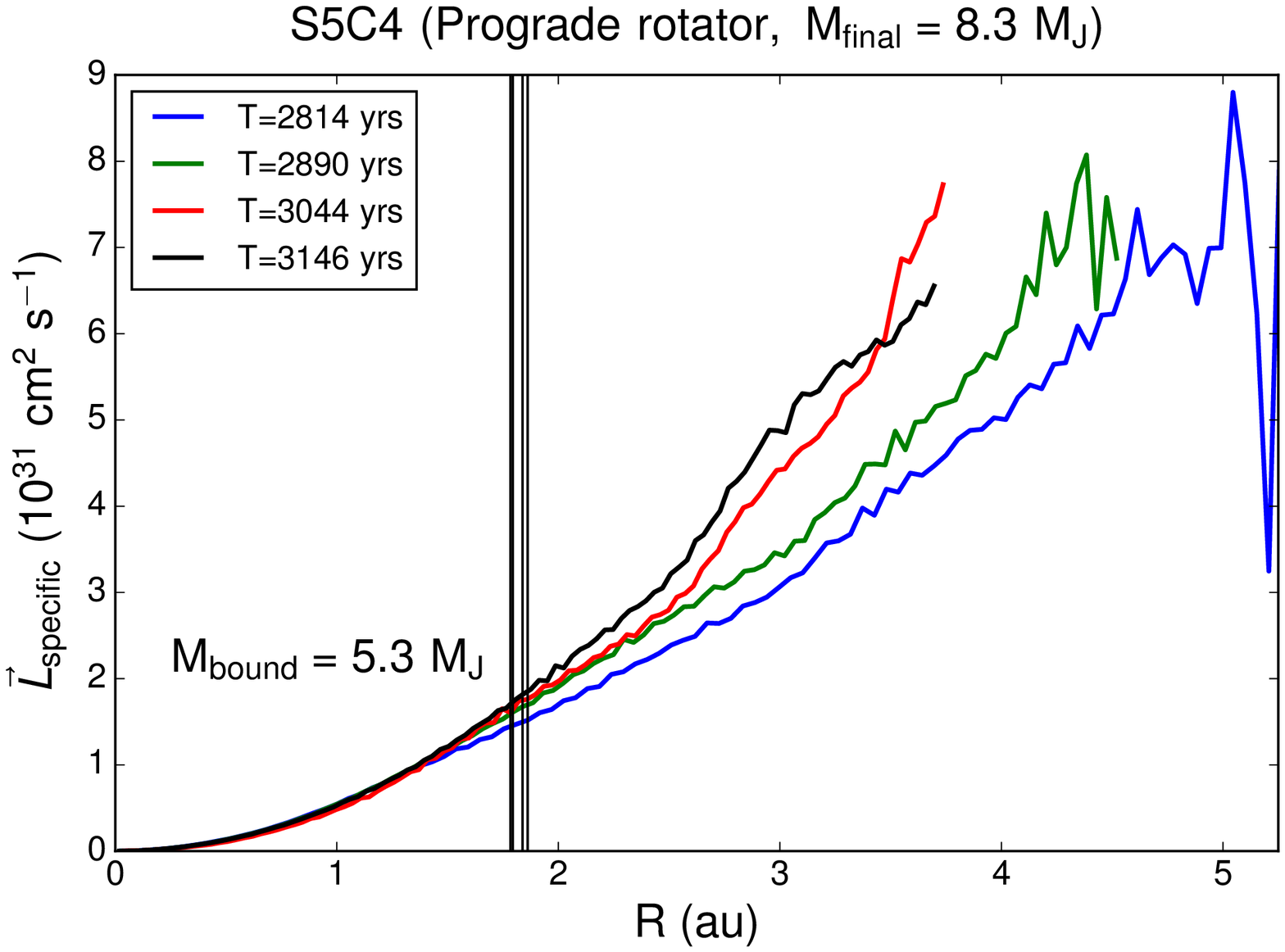}
        \end{tabular}
        \caption{Top: radial profile of specific angular momentum for an example retro-rotating clump (clump 4, simulation 4), at four different times. Horizontal lines indicate the last bound point of the clump at times indicate in the legend. The final bound mass is marked on the plot, and in this case, all of the mass that is identified as belonging to the clump is ultimately bound to the clump. Positive values for $\vec{L}_{\mathrm{specific}}$ indicate that the rotational angular momentum vector of the material and the orbital angular momentum vector of the whole clump are aligned (or at least inclined at less than $90^{\circ}$), and negative values indicate that the two vectors are anti-aligned. The blue line is when the clump is first detected, and we can see that the majority of the clump is in prograde rotation, and the outer $\sim$ 1 au is in retrograde rotation. The material came to be retro-rotating due to a close encounter with another clump, which is shown in Figure \ref{fig:encounter}. As the clump continues to accrete material, we can see that angular momentum is exchanged between the inner material and the outer material (green line, $T= 3377$ years). As the clump contracts, this positive angular momentum material is no longer considered part of the clump. Bottom: for comparison, the specific angular momentum profile of a clump of comparable mass undergoing prograde rotation (clump 4, simulation 5). The horizontal line indicates the last bound radius in the clump for the four clumps, only one line is plotted here as the four are so close together. The final bound mass of the clump is stated on the plot.\label{fig:lspecific}}
    \end{minipage}%
    \hfill
    \begin{minipage}[t]{0.48\textwidth}
        \centering
\begin{tabular}{c}
\includegraphics[width=\textwidth]{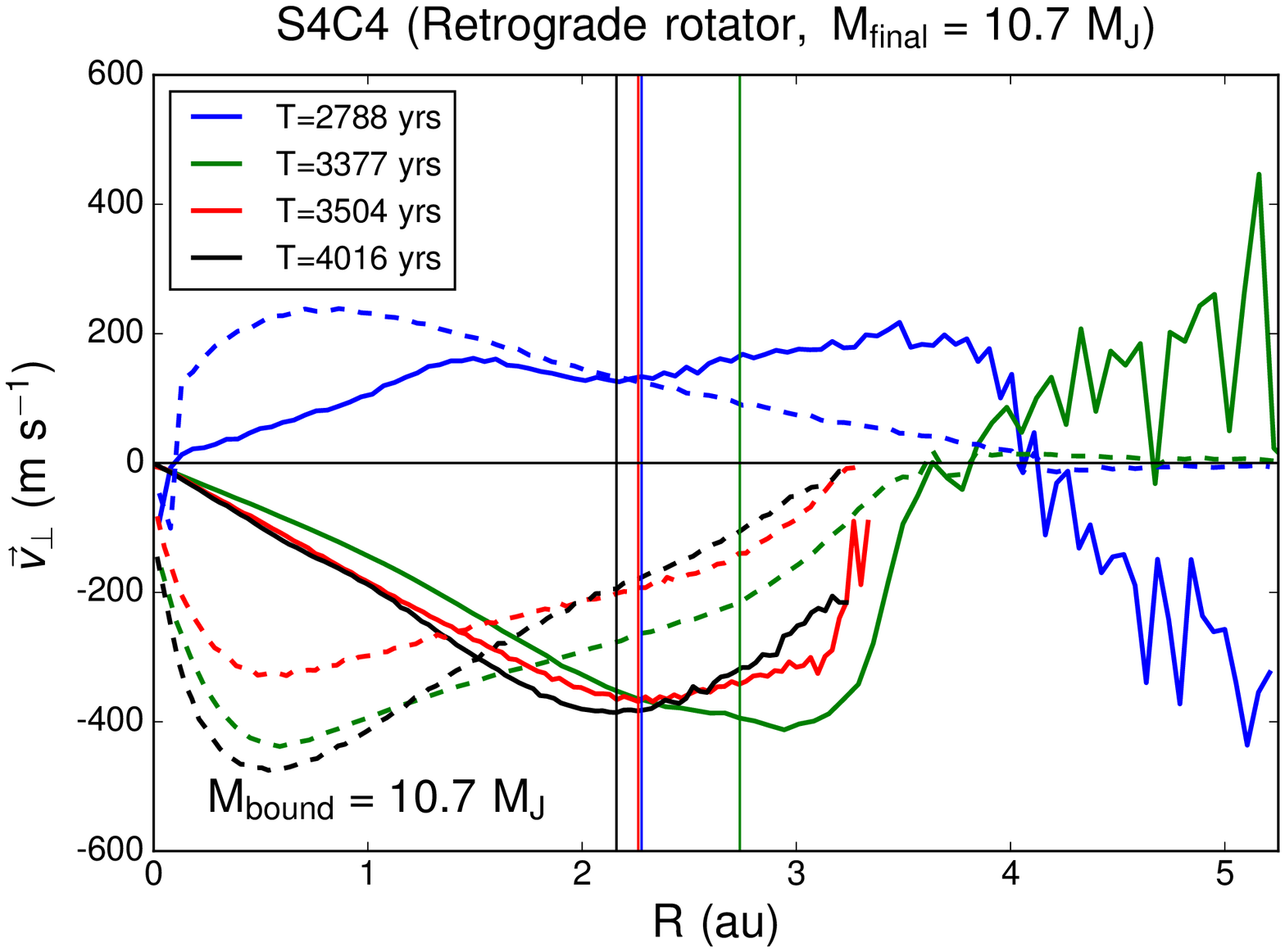}\\
\includegraphics[width=\textwidth]{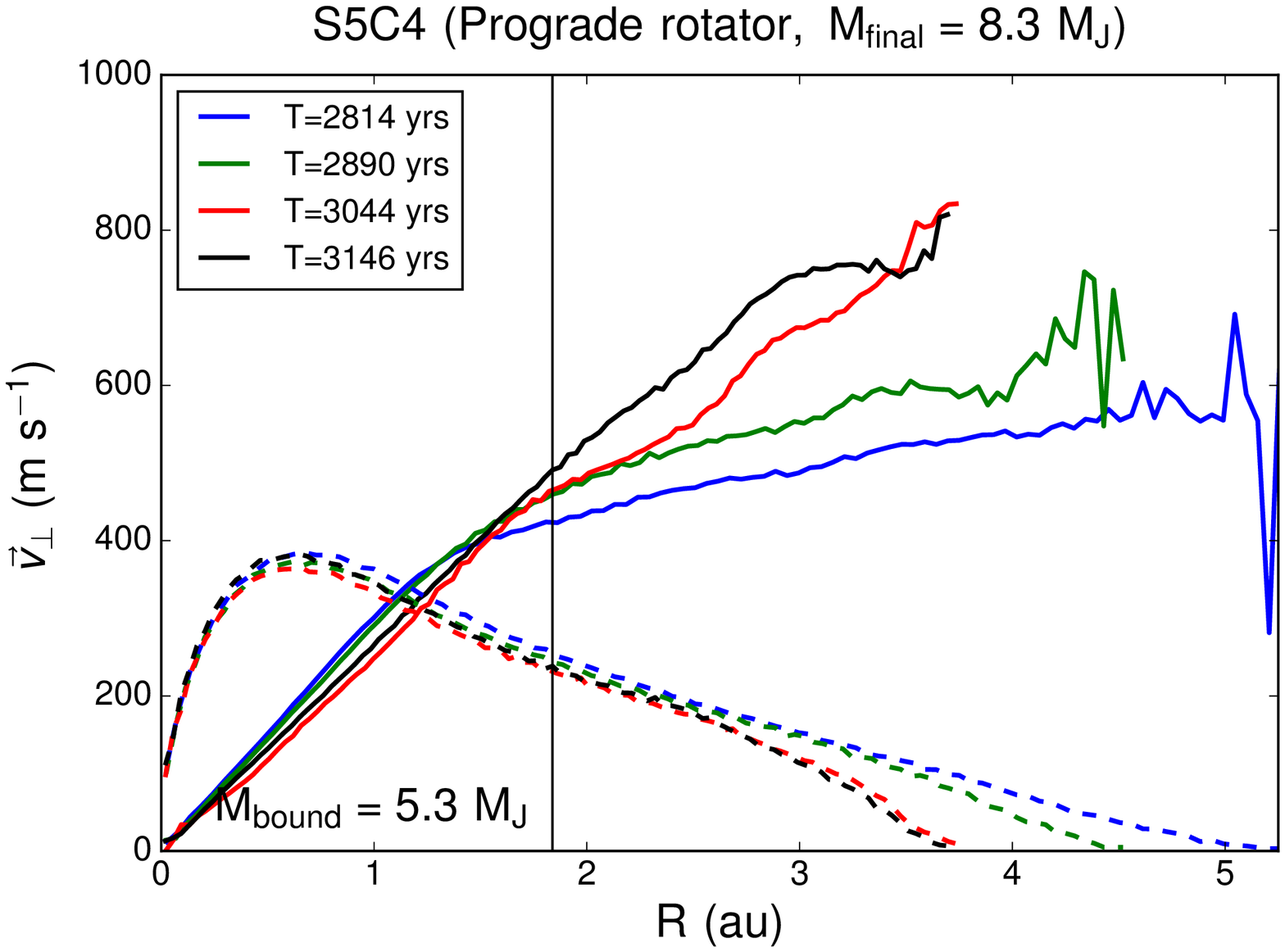}
\end{tabular}
\caption{Top: rotation velocity curve for an example retro-rotating clump (clump 4, simulation 4), at four different times. Horizontal lines mark the last bound radius in the clump at times given in the legend. In this case, the total bound final mass is equal to the total mass ultimately identified for the clump. Dashed lines correspond to breakup velocities at these times. Negative velocities indicate that the rotational angular momentum vector of the material and the orbital angular momentum vector of the whole clump are anti-aligned (or at least inclined at more than $90^{\circ}$) relative to each other. We can see that beyond $\sim$ 1.5 au, the material is rotating at faster than its breakup velocity, which would perhaps suggest that material is spreading outwards from the clump, in a disc-like, or toroidal, manner. Bottom: for comparison, a clump of comparable mass undergoing prograde rotation (clump 4, simulation 5). Again, beyond $\sim$ 1 au, rotation velocity exceeds breakup velocity, and so we may expect to see a considerable spread of material around such an object, morphologically similar to a toroid.\label{fig:rotvel}}
    \end{minipage}
\end{figure*}
%================================================

We carry out analysis of the orbital properties of our clumps only using the sample as detected by the density derivative search, as this method is sensitive to most clump masses and semi-major axes. The total semi-major axis evolution of all clumps is shown in the left hand panel of Figure \ref{fig:migration}, which we have already discussed, and refer the reader back to. Circles mark surviving clumps (including clumps that subsume another clump), squares mark destroyed clumps, and triangles mark merged clumps. Larger markers correspond to more massive clumps. For destroyed clumps, we take the last measured mass. Roughly half of our most massive clumps migrate radially inwards, which is consistent with migration in locally isothermal discs, as objects exchange angular momentum with the surrounding gas and move inwards. However, about half of our most massive clumps migrate radially outwards. This is known to be possible in radiative discs \citep{kleynelson2012}, but requires either large torques or steep surface density gradients \citep{angelolubow2010}. Large torques can have many sources, but in massive, self-gravitating discs they are likely to be in the form of global spiral arms. We carried out a Fourier analysis on the density structure of our discs, to determine the Fourier amplitude of each $m$-mode (where $m$ is the number of spiral arms). The amplitude, $A_{\mathrm{m}}$, of each mode, $m$, is calculated by
\begin{equation}
A_{\mathrm{m}} = \Bigg| \sum_{i=1}^{N_{\mathrm{region}}} \frac{e^{-im\phi_i}}{N_{\mathrm{region}}} \Bigg|,
\end{equation}  
where ${N_{\mathrm{region}}}$ is the number of particles in the region we are considering (for our case, $R=20$ to $R=100$ au), and $\phi_i$ is the azimuthal angle of the $i^{\mathrm{th}}$ particle. Some example amplitudes are shown in Figure \ref{fig:fourieramp}, which shows the first 10 Fourier components of the density structure of 2 discs in their initial state (i.e. when they have just begun to fragment), marked in red, and the same 2 discs in their final state, marked in black. The discs are from simulation 1 and simulation 5, and their final state can be seen in their column density plots, shown in Figure \ref{fig:columndensity}. These discs were selected because they ran for the same length of time, and they have contrasting final $m$-modes states, simulation 1 ultimately peaks in the $m=2$ mode, and simulation 5 ultimately peaks in the $m=6$ mode.

In this fashion, we determine the dominant final $m$-mode in each disc, and plot the final semi-major axis of our clumps as a function of this $m$-mode, in Figure \ref{fig:mspiralva}. We note that our decision to name a dominant mode, based on a relatively low amplitude difference, may be questioned. However, these discs are not in a quasi-steady state, having undergone fragmentation, so persistent spiral modes may be unlikely to form due to tidal disruption from these clumps. Despite the transient nature of the spiral modes, global, low $m$-mode spiral arms can exert considerable torque, and this is clearly important for the final orbital configuration of our clumps, as shown in Figure \ref{fig:mspiralva}, which displays a rough empirical relationship between the maximum semi-major axis of a clump, $a_{\mathrm{max}}$, and $m$, such that
\begin{equation}
\label{eq:apropm}
a_{\mathrm{max}}\propto\frac{1}{m}.
\end{equation}
Of course, this relationship is preliminary, since we only consider 9 discs, all of the same mass, and it has been shown that more massive discs are dominated by low $m$ spirals \citep{lodatorice2004,lodatorice2005}. Indeed, since it has been shown that in discs without fragmentation we expect the number of spiral modes to be related to the disc-to-star mass ratio, $q$, such that $m\propto \nicefrac{1}{q}$ \citep{donghallrice2015}, to examine the full parameter space of spiral modes requires a range of $q$ values. Since we consider discs with identical $q$ values, \textit{why} these discs have different dominant $m$-modes is a valid question. Again, this may be explained by these discs having fragmented into bound clumps. Bound objects in a gaseous disc produce stronger, more persistent torques than spiral density fluctuations alone. If one of our clumps is scattered out of the main body of the disc, say, by interaction with another clump, it may exert a tidal torque on the material in the disc (and by Newton's third law, the material in the disc will also exert a force on the clump). Larger torques are associated with lower $m$-modes, and if these tidal responses from the discs to the clumps is responsible for the low $m$-mode domination in the disc, then we expect to see steeper surface density profiles in discs with low $m$-modes, since more mass will have been redistributed inwards as a result of this torque. This is difficult to unequivocally demonstrate in our set of simulations, since each simulation was run for a different length of time, and the amount of mass redistributed increases with time. 

However, Figure \ref{fig:enclosedmassvr} shows the mass enclosed as a function of radius for the two discs plotted in Figure \ref{fig:fourieramp}, that ran for the same length of time. Although by no means conclusive, the slight increase in mass for a given radius between 40 au and 160 au for the $m=2$ disc in simulation 1 is consistent with tidal forces being responsible for the low $m$-mode becoming more dominant. To properly establish the nature of the preliminary relationship detailed in equation \ref{eq:apropm} therefore requires a range of disc masses and fragmentation scenarios, and we leave this to future work.
%
%\label{subsec:spin}
Figure \ref{fig:eccentricity} shows the relationship between eccentricity, $e$, and semi-major axis for our SPH clumps. Larger markers indicate more massive clumps. For the most part, the more massive clumps are located on close in ($a\sim50$ au), low eccentricity ($e\sim 0.1$) orbits, while lower mass clumps are at larger separations and with higher eccentricity. Since disc fragmentation forms objects on low eccentricity orbits ($e<0.1$), we can see the importance of clump-clump interactions in determining the final orbital properties of a clump. Very large eccentricities ($e\sim 0.7$) at large $a$ indicate that a clump is close to ejection, as excitations beyond unity ensure a clump is ejected from the disc. The top two panels of Figure \ref{fig:histogramecc} show the initial eccentricity distribution (left) and initial inclination distribution (right) of our SPH clumps. The bottom two panels of Figure \ref{fig:histogramecc} show the final eccentricity distribution (left) and final inclination distribution of the same. Inclination is calculated relative to the orbital plane of the central star, such that
\begin{equation}
i = \arccos\Bigg(\frac{L_{\mathrm{z}}}{|\vec{L}|} \Bigg),
\end{equation}
where $i$ is the orbital inclination of the clump, $\vec{L}$ is the orbital angular momentum vector of the clump (calculated relative to the centre of mass and centre of velocity of the central star), and $L_{\mathrm{z}}$ is the $z$ component of $\vec{L}$. Using least-squares regression, each plot has been fitted with a Gaussian of the form 
\begin{equation}
F_{\mathrm{p}}=Ae^{-\frac{1}{2}(\frac{x-\mu}{\sigma})^2} + k,
\end{equation} 
where $F_\mathrm{p}$ is the fraction of planets, $A$ is the amplitude of the curve (without offset), $x$ is either eccentricity or inclination, $\mu$ is the mean of the fitted distribution, $\sigma$ is the standard deviation of the fitted distribution and $k$ is the fitted offset constant. The fitted values are given in Table \ref{table:fitvalues}, and in the legend of each plot. The initial and final inclinations have been fitted with two distributions, the dotted blue line includes all points, and the solid red line does not include the most inclined point in each distribution, since there is a large gap between that point and the rest of the clumps, and a small sample size, it is unclear if this point is actually an outlier. We have provided these fits since current GI population synthesis models do not include  orbital eccentricity or inclination, and, despite our small sample size, this information may be useful in the future development of these models. Aside from a small decrease in standard deviation, there is little change between our initial and final eccentricity distribution; both peak at $e\sim 0.1$ and share an offset constant of $k=0.006$. However, the orbital inclination of our clumps is reduced by a factor of $\sim 100$ between the initial and final states, showing that clump orbital inclination, in our SPH simulations, is rapidly reduced after formation. Considering that many of our clumps undergo dynamical interactions that cause scattering and eccentricity pumping on short timescales, this high degree of coplanarity may be surprising, especially when considering that most exoplanets have mutual inclinations of a few degrees \citep{figueira2012,fangmargot2013}. However, it is consistent with our current understanding of highly inclined planet orbits relying on dynamical perturbations such as the Lidov-Kozai mechanism \citep{naozetal2013}. Our results may indicate that developing inclined orbits is difficult while a gas disc is present, even if substantial dynamical interactions between clumps take place in this time.  
\begin{table}
\centering
\begin{tabular}{lllll}
 \hline 
 Plot                      & $\mu$ & $\sigma$ & $A$ & $k$ \\
 \hline
 \hline
 Initial ecc.                & 0.094 & 0.095 & 0.070 & 0.006 \\
 Final ecc.                  & 0.107 & 0.071 & 0.079 & 0.006 \\
 Initial inc. (no outlier)   & 0.052 & 0.029 & 0.173 & 0.003 \\
 Initial inc. (with outlier) & 0.053 & 0.028 & 0.051 & 0.006 \\
 Final inc. (no outlier)     & 0.001 & 0.001 & 0.119 & 0.005 \\
 Final inc. (with outlier)   & 0.001 & 0.001 & 0.063 & 0.007 \\
 \hline
\end{tabular}
\caption{Parameter values of the Gaussian fits applied to the histograms in Figure \ref{fig:histogramecc}. From left to right, the columns are plot, mean, standard deviation, amplitude (without offset) and offset constant.\label{table:fitvalues}}
\end{table}
%
%\subsection{Spin properties}
%\label{subsec:spin}
%
%
%=========================
%\subsection{Spin properties}
%
%IMAGE WAS HERE CASS
\subsection{Spin properties}
\label{subsec:spin}
We analysed all of the fragments in our simulations, and found that several of them survive to the end of the simulation whilst undergoing retrograde rotation. This is shown in Figure \ref{fig:spinorbit}, which shows the relative alignments between the orbital angular momentum vector and the rotational angular momentum vector of the clumps. Both the top and bottom panel is split into two parts, showing significant disalignment at the top, marked in red crosses, and good-alignment at the bottom. The top panel shows the alignment as a function of mass, and the bottom panel shows the alignment as a function of semi-major axis.

This prompts the question - how did they get to be retro-rotating? Did they form like this, or were they perturbed in some way? Having checked all of the clumps with significant disalignment, we can see that all of them were perturbed by a close encounter with another fragment, which typically flung them rapidly further out into the disc. We give the most extreme example in Figure \ref{fig:encounter}, which shows the retrorotating clump 4 in simulation 4. In the leftmost panel, we see in the bottom left corner two clumps undergoing a close encounter. One of them then decreases its semi-major axis, whilst the other one is flung further out into the disc, to become retrorotating.
Figure \ref{fig:lspecific} shows the specific angular momentum profiles of a retro-rotating clump (top) and a prograde rotating clump (bottom). The blue line, at $T=2788$ years, is as soon is the clump is detected by our algorithm. We can see that the majority of the clump is in prograde rotation, with only the outer $\sim 1$ au in retrograde rotation. However, as time progresses and the clump continues to accrete material that is retro-rotating due to the encounter, this ultimately changes the rotation of the whole clump. For comparison, the bottom panel contains a prograde rotator of similar mass from simulation 5.

Figure \ref{fig:rotvel} shows the rotation velocity profiles of two clumps. Top shows the ultimately retro-rotating clump 4 in simulation 4, and the bottom, for comparison, is a clump of comparable mass that is always undergoing prograde rotation. Negative velocity is determined by the orbital angular momentum vector of the whole clump and the rotational angular momentum vector of the material being anti-aligned. Both panels show the clumps at four different times, and the dashed lines indicate the breakup velocity profile of each clump. Both clumps are rotating under their breakup velocity for radii below 1 au, and their velocity profiles are consistent with solid body rotation (i.e. $v\propto R$) at these radii. Much further out, the clumps become nebulous, but we have included information out as far as possible to show the interesting angular momentum exchange between material at $T=2788$ years and $T=3377$ years for the retro-rotator. For the prograde rotator, the ultimate configuration is a good approximation to a solid body rotation curve out to extended radii ($\sim$ 3 au). Interestingly, for the retro-rotator, the velocity profile at the outer part of the fragment ($\sim 2.2$ au to $\sim 4$ au) is consistent with Keplerian rotation (i.e. $v\propto\nicefrac{1}{\sqrt{R}}$). This suggests the presence of a disc, or a disc-like structure, around the clump. Unfortunately, it is not (at the time of writing) currently possible to self-consistently re-resolve such regions in SPH simulations, so we are unable to investigate this further.
%==================================================
% Prograde and retrograde radial profiles
%===================================================
\begin{figure}
\centering
  \includegraphics[width=0.5\textwidth]{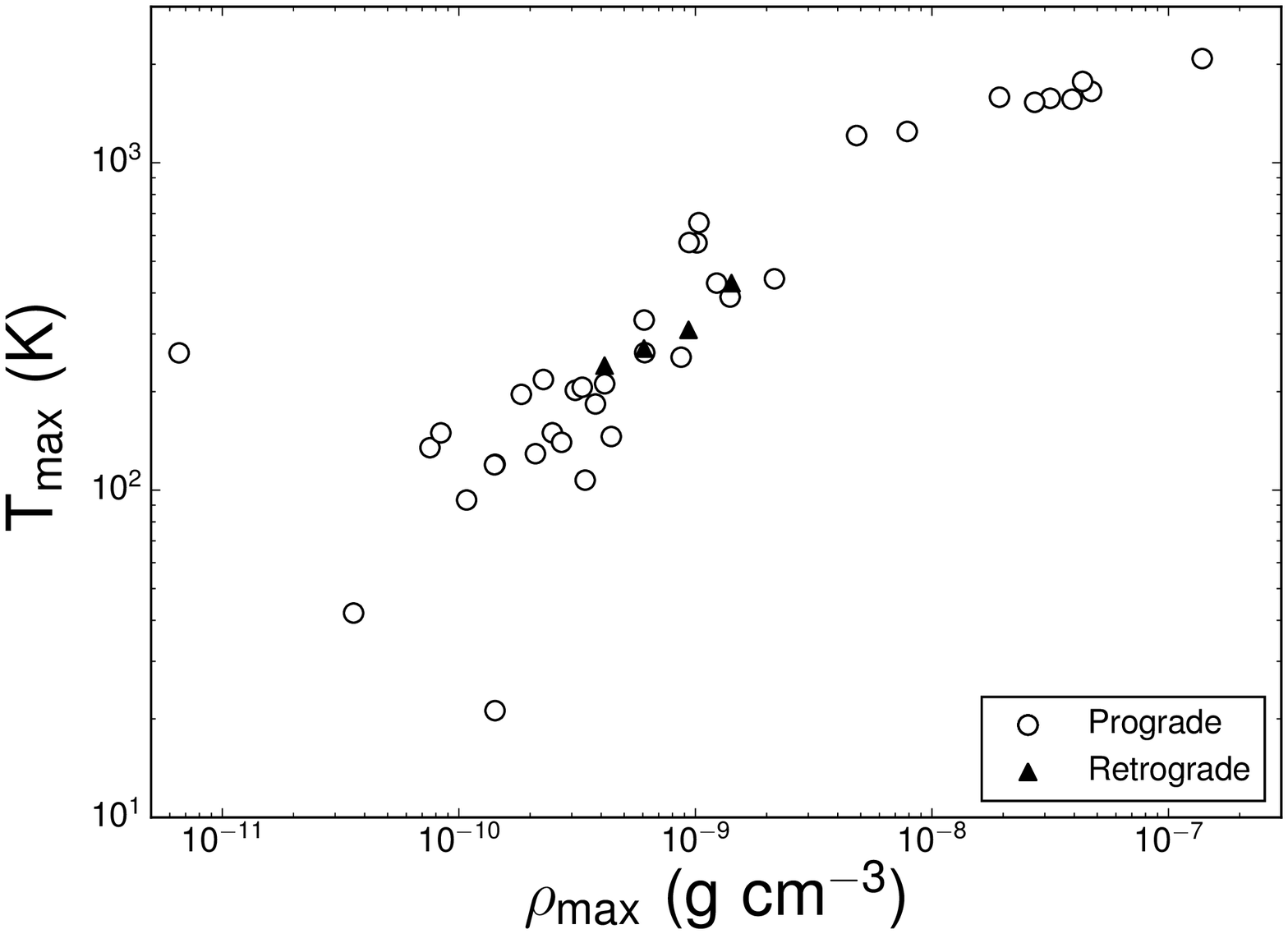}
\captionof{figure}{Final maximum temperature and density of clumps in all simulations. Open circles are prograde rotating clumps, closed triangles are retrograde rotating clumps.\label{fig:rhomaxTmax}}
\end{figure}

\begin{figure}
\centering
\includegraphics[width=1.0\linewidth]{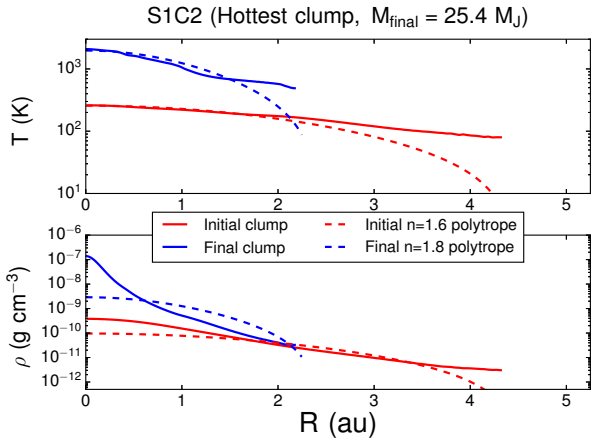}
 \caption{Radial temperature and density profile of the hottest clump we identified in our simulations, clump 2 in simulation 1. Red solid lines show initial temperature and density, blue solid lines show final temperature and density. For comparison, an $n=1.6$ polytrope is plotted for the initial clump, and an $n=1.8$ polytrope is plotted for the final clump. In both cases, the polytrope is a good fit to the temperature profile of the clump, however it is a poor fit to the density, particularly in the final state, where it underestimates central density by almost two orders of magnitude.\label{fig:hottestclumpprofile}}
\end{figure}

\begin{figure*}
\begin{center}
\begin{tabular}{cc}
\textbf{Prograde rotators} & \textbf{Retrograde rotators} \\
\includegraphics[width=0.5\linewidth]{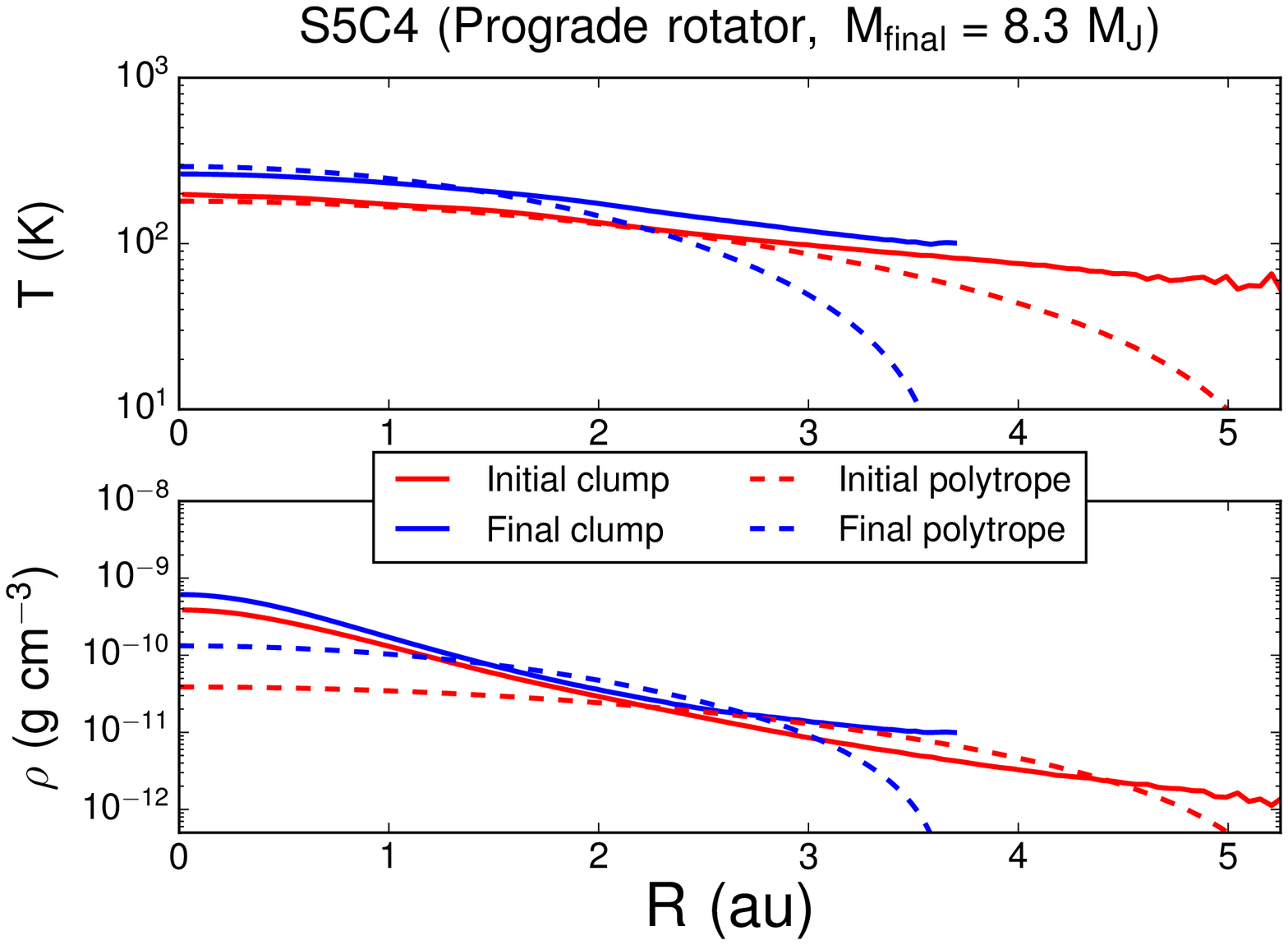} & \includegraphics[width=0.5\linewidth]{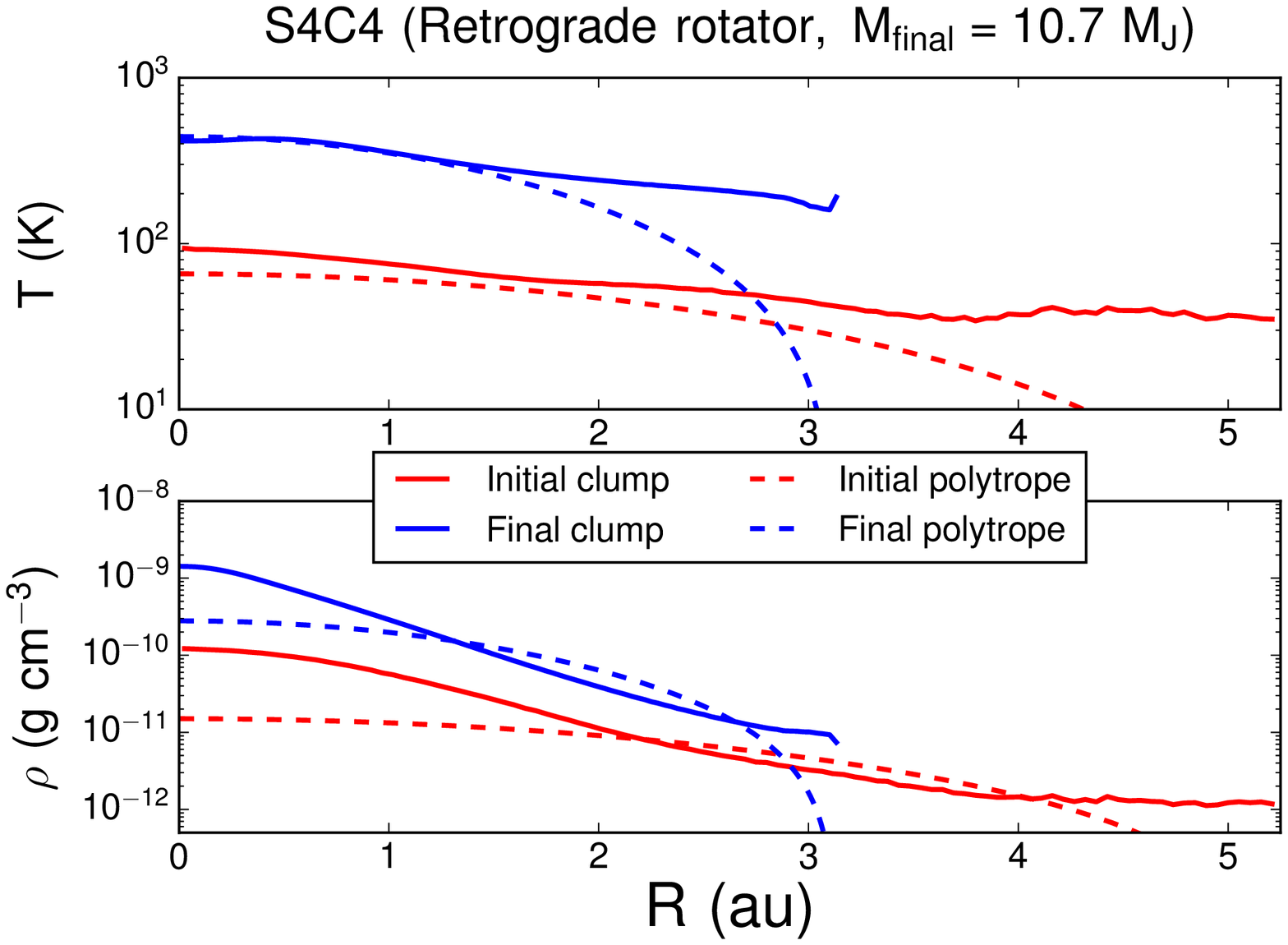}\\
\includegraphics[width=0.5\linewidth]{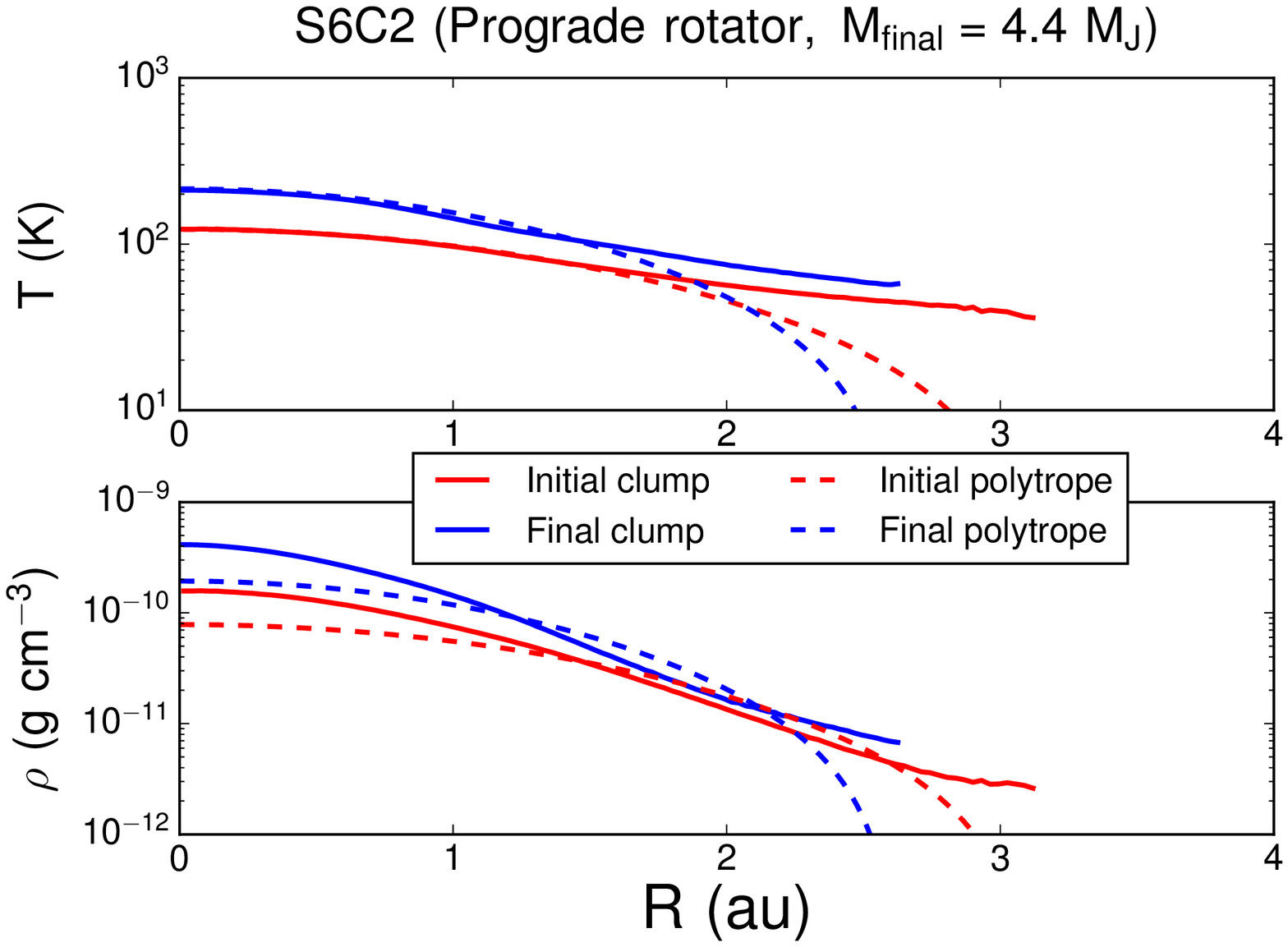} & \includegraphics[width=0.5\linewidth]{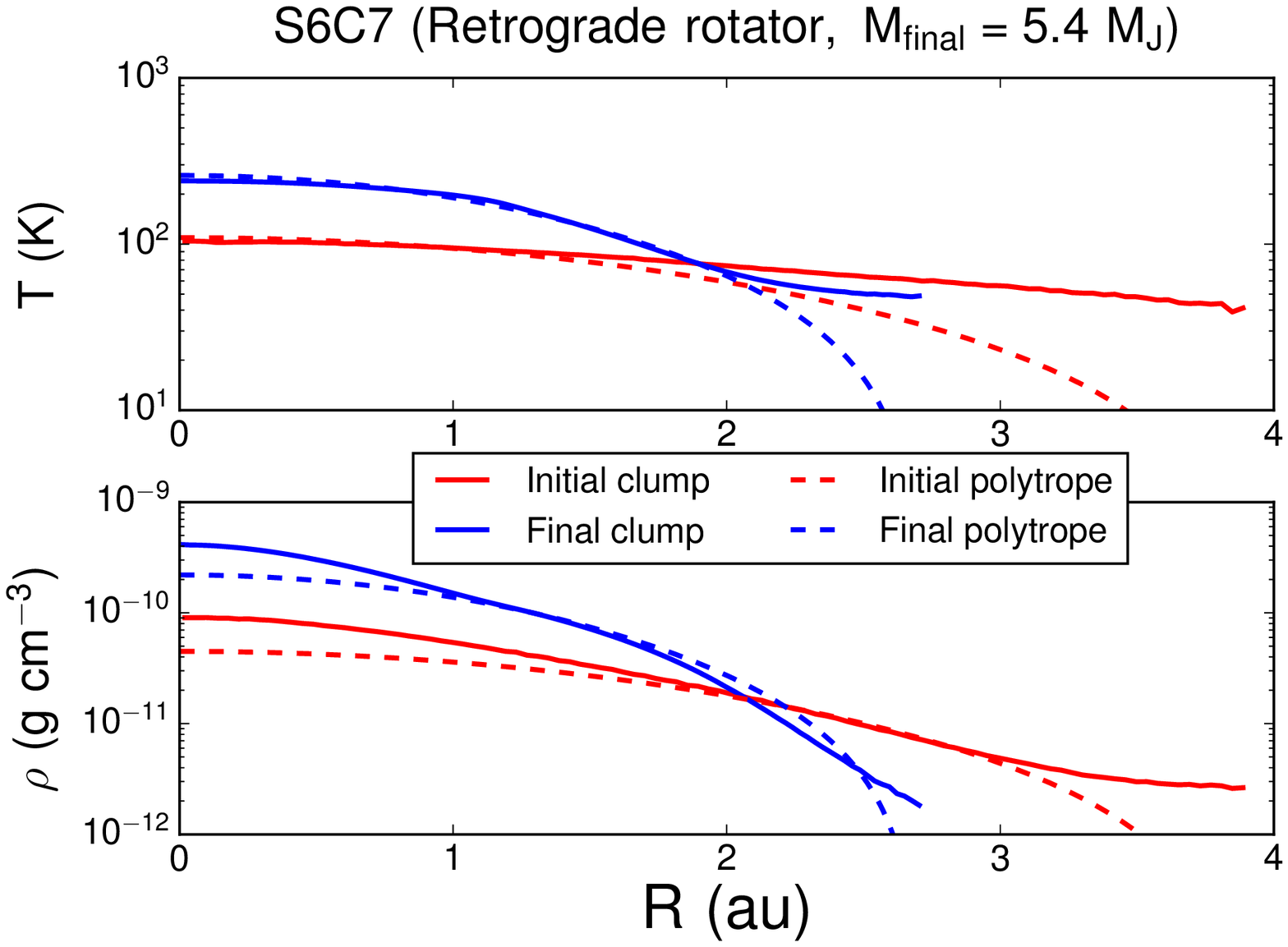}\\
\includegraphics[width=0.5\linewidth]{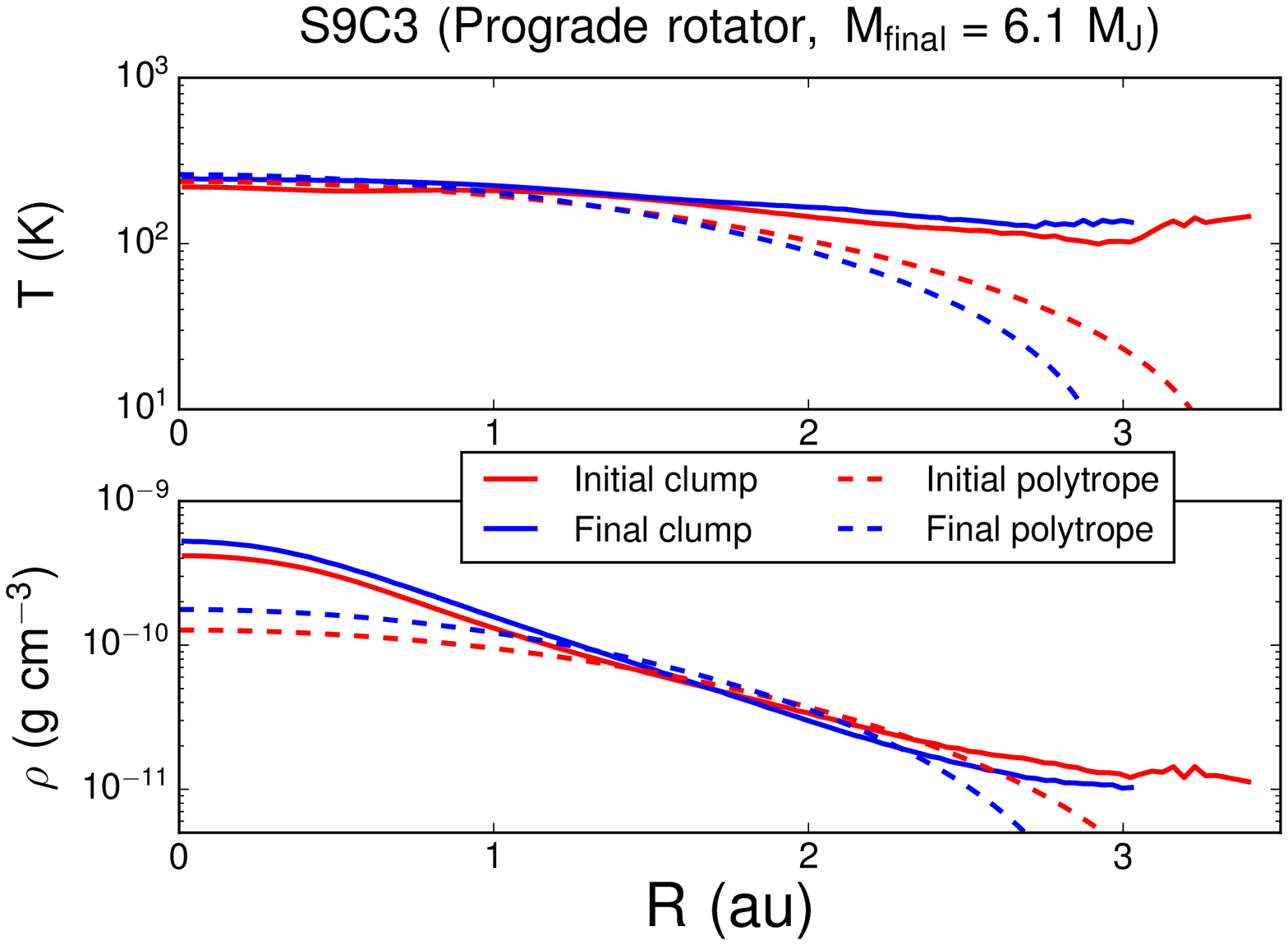} & \includegraphics[width=0.5\linewidth]{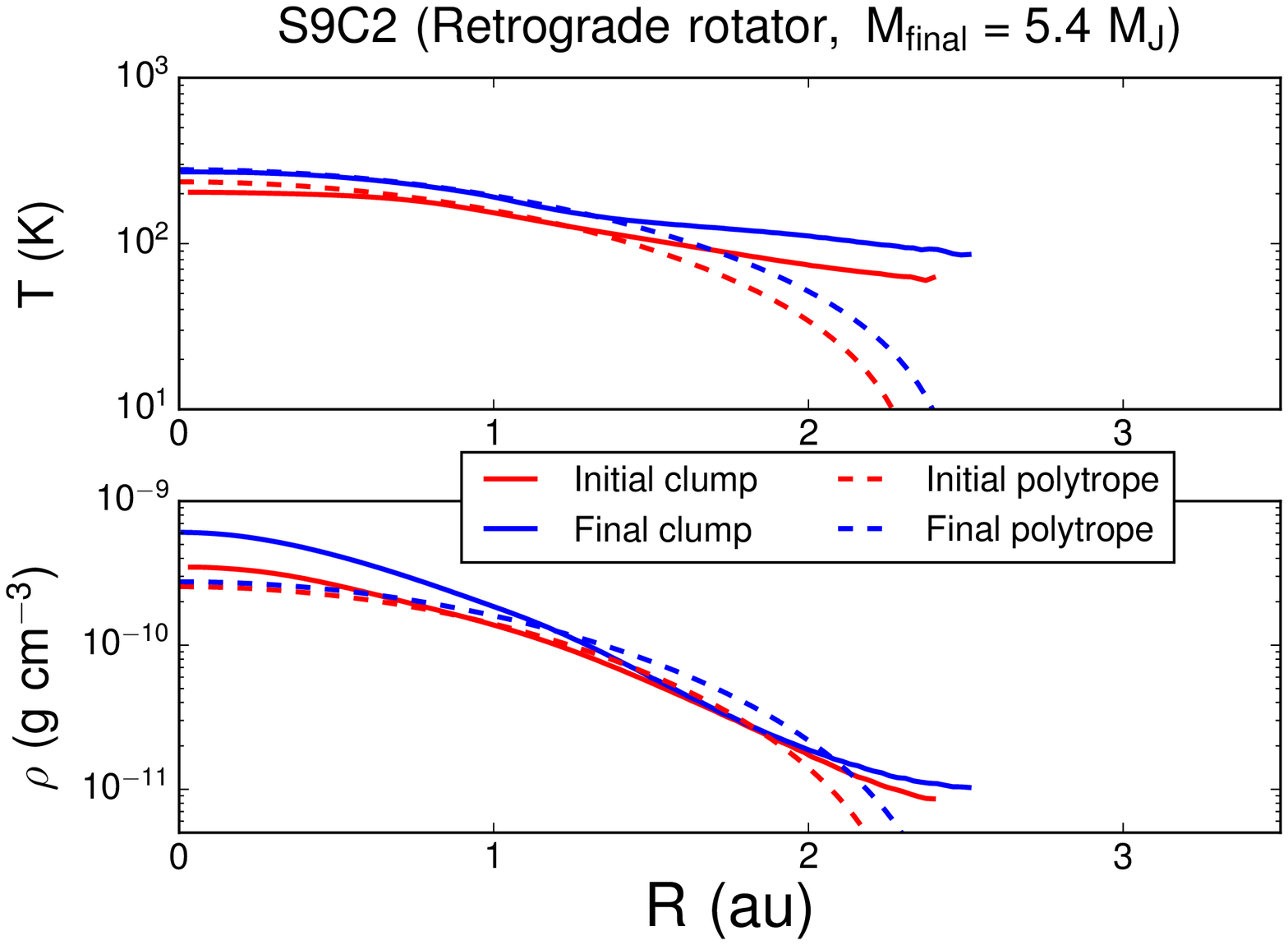}\\
%\caption{Left prograde right retrograde\label{fig:retroprofiles}}
\end{tabular}
\caption{Radial density and temperature profiles for the initial and final state of 6 clumps. Red solid lines show initial clump profile, blue solid lines show final clump profile. For comparison, all clumps have initial and final $n=1.5$ polytropic profiles plotted in initial and final states. The left column contains 3 clumps undergoing prograde rotation, and the right column shows 3 clumps undergoing retrograde rotation. In each row, the clumps are of comparable mass, and are intended to be compared, although bearing in mind that the clumps have had different evolutionary histories and their final masses are not identical. For most of the clumps, there is not much in their profiles that would mark them as a retro-rotating clump. The final inner and surface temperatures are similar, as is the shape of the density profile. Clump 4 in simulation 4 (top right panel) is somewhat the exception, with a large increase in both final inner density, inner temperature and surface temperature. Given the violent encounter it endured early in its history (shown in Figure \ref{fig:encounter}), this may be unsurprising, but it is interesting to note that high surface temperatures may indicate a violent encounter in the past. In all cases, a polytrope of index $n=1.5$ is a reasonable fit to the temperature profile, but consistently underestimates the inner density by around an order of magnitude.\label{fig:retroprofiles}}
\end{center}
\end{figure*}
\subsection{Density and temperature properties}
\label{subsec:profiles}

Maximum density and temperature for all clumps in our simulation is shown in Figure \ref{fig:rhomaxTmax}, open circles show prograde rotating clumps and dark triangles show retrograde rotating clumps. The maximum temperature we identify in a clump is 2081 K, which implies that none of our fragments have started to dissociate molecular hydrogen and can therefore be considered as the first hydrostatic cores. Additionally, a further 9 clumps have internal maximum temperatures above $\sim 1000$ K, which means they would begin to evaporate dust. Both of these results are in good agreement with previous studies of fragments, such as \citet{vorobyov2013}, who found similar results. Our measurements of clump maximum temperature are limited by our resolution, since running simulations at higher resolution would allow higher densities to be reached before becoming computationally infeasible to continue.  

As discussed in our introduction, the advantage of using a purely hydrodynamical simulation with no sink particles is that we can examine fragment internal structure during the simulation. With this in mind, we show the radial temperature and density profiles of 7 fragments, one of which  ultimately becomes the hottest fragment in our simulations, shown in Figure \ref{fig:hottestclumpprofile}, and the remaining 6 are three retro-rotators and three prograde rotators, of comparable mass and with similar simulation histories, shown in Figure \ref{fig:retroprofiles}. In both figures, each image is split into two panels, radial temperature profile on the top and radial density profile on the bottom. The initial clump profile is shown in solid red, and the final clump profile is shown in solid blue. For comparison, a polytrope is also plotted in each panel. 

Figure \ref{fig:hottestclumpprofile} shows the radial density and temperature profile of the hottest clump identified out of our 9 simulations. We can see that although the radial temperature profile is well described by a polytrope of index $n=1.6$, for the initial state, and $n=1.8$ for the final state, in both cases the actual density deviates from the polytropic density significantly. This is also true for the rest of the clumps, shown in Figure \ref{fig:retroprofiles}, where the left-hand column shows prograde rotators, and the right-hand column shows retrograde rotators. In all cases in Figure \ref{fig:retroprofiles}, we have plotted a polytrope of index $n=1.5$, which is appropriate for a fully convective star such as a brown dwarf. Since in a polytrope, pressure $P$ and density $\rho$ are related by
\begin{equation}
P = K \rho ^{\frac{n+1}{n}},
\end{equation}  
where $K$ is a constant and $n$ is the polytropic index, it implicitly assumes that pressure is a power law function of density which is constant throughout the star. For our clumps, it appears to be the case that a polytropic approximation may be too simplistic when estimating the internal structure of the clump. This may have some implications again for current GI population synthesis models, since orbital parameters are sensitive to the radial distribution of mass in the forming fragments. It is probably best to exercise caution when constructing clumps in GI population synthesis models, and it may be the case that a "follow the adiabats" approach, as used in \citet{nayakshinfletcher2015}, is more appropriate.

When we compare the final states of the retro-rotating clumps with their prograde rotating counterparts, there is not much that would mark them as retro-rotating, final density and final temperature profiles for both directions of rotation are similar, and are consistent with other simulations of fragmenting protostellar discs (see, e.g. \citealt{vorobyov2013}). There is one clump that is an exception when compared to the rest of the clumps, and is shown in the top right hand panel of Figure \ref{fig:retroprofiles}. This is S4C4. In its final state, the surface temperature of the clump is a factor of $\sim 3$ higher than we see in the rest of the clumps, and its central temperature is $\sim 425$ K, a factor of $\sim 2$ larger when compared to the rest of the clumps in Figure \ref{fig:retroprofiles}. This high temperature could possibly be explained by an encounter with another clump. Figure \ref{fig:encounter} shows this interaction, (clump 4 is marked by a green square). It is scattered by the more massive clump into the outer part of the disc, becoming shocked as it passes through a spiral arm. The encounter with the other clump causes the direction of rotation of clump 4 to change, but the large increase in temperature could be due to this motion through a region of increased density, entering perpendicular to the spiral arm where the density gradient is at its largest.
%===============================================
% Merger and tidal disruption plot
%===============================================
\begin{figure*}
\begin{center}
\begin{tabular}{ccc}
T = 2778 years & T = 2865 years & T = 2890 years\\

\begin{tikzpicture}
%\node [anchor=west] (note) at (-1,3) {\Large Note};
%\node [anchor=west] (water) at (-1,1) {\Large Water};
\begin{scope}[xshift=1.5cm]

    \node[anchor=south west,inner sep=0] (image) at (0,0) {\includegraphics[width=0.33\linewidth]{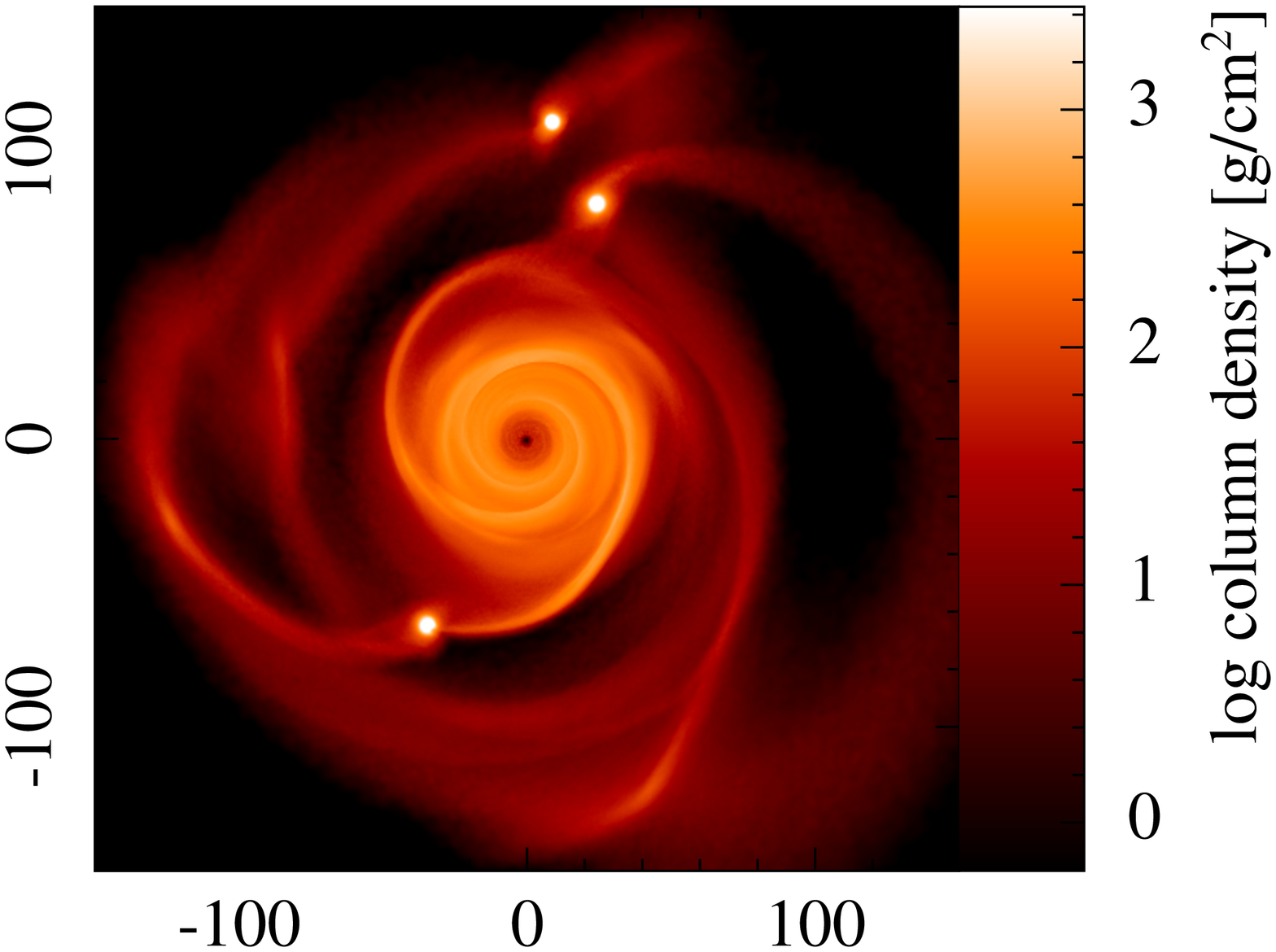}};
    \begin{scope}[x={(image.south east)},y={(image.north west)}]
    %{0.241,0.3057}{0.2766,0.3546}
   % {0.2489,0.3148}{0.2715,0.3437}
   %{0.4125,0.4152}{0.4538,0.4786}

       % \draw[green,thick] (0.4305,0.6904) rectangle(0.53,0.822);
        	\draw[green,thick](0.3977,0.7482)rectangle(0.509,0.9015);
        %\draw [-latex, ultra thick, red] (note) to[out=0, in=-120] (0.48,0.80);
        %\draw [-stealth, line width=5pt, cyan] (water) -- ++(0.4,0.0);
    \end{scope}
\end{scope}
\end{tikzpicture}% 

&
\begin{tikzpicture}
%\node [anchor=west] (note) at (-1,3) {\Large Note};
%\node [anchor=west] (water) at (-1,1) {\Large Water};
\begin{scope}[xshift=1.5cm]

    \node[anchor=south west,inner sep=0] (image) at (0,0) {\includegraphics[width=0.33\linewidth]{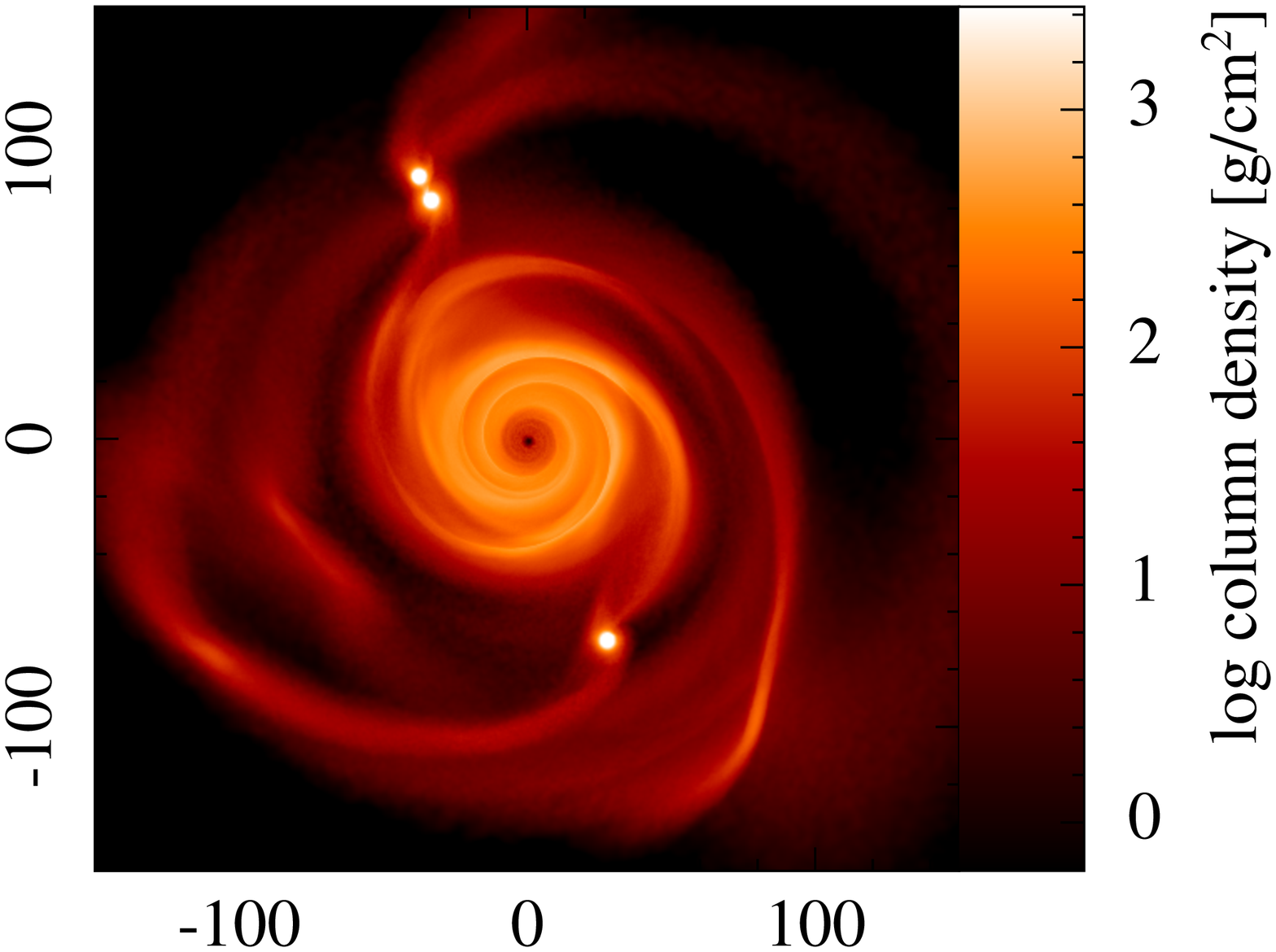}};
    \begin{scope}[x={(image.south east)},y={(image.north west)}]

        %\draw[green,thick] (0.3765,0.5077) rectangle(0.4178,0.5711);
        %\draw[green,thick] (0.3565,0.6648) rectangle(0.456,0.7964);
        	\draw[green,thick](0.2837,0.7173) rectangle (0.395,0.8706);
        %\draw [-latex, ultra thick, red] (note) to[out=0, in=-120] (0.48,0.80);
        %\draw [-stealth, line width=5pt, cyan] (water) -- ++(0.4,0.0);
    \end{scope}
\end{scope}
\end{tikzpicture}%
&
\begin{tikzpicture}
%\node [anchor=west] (note) at (-1,3) {\Large Note};
%\node [anchor=west] (water) at (-1,1) {\Large Water};
\begin{scope}[xshift=1.5cm]

    \node[anchor=south west,inner sep=0] (image) at (0,0) {\includegraphics[width=0.33\linewidth]{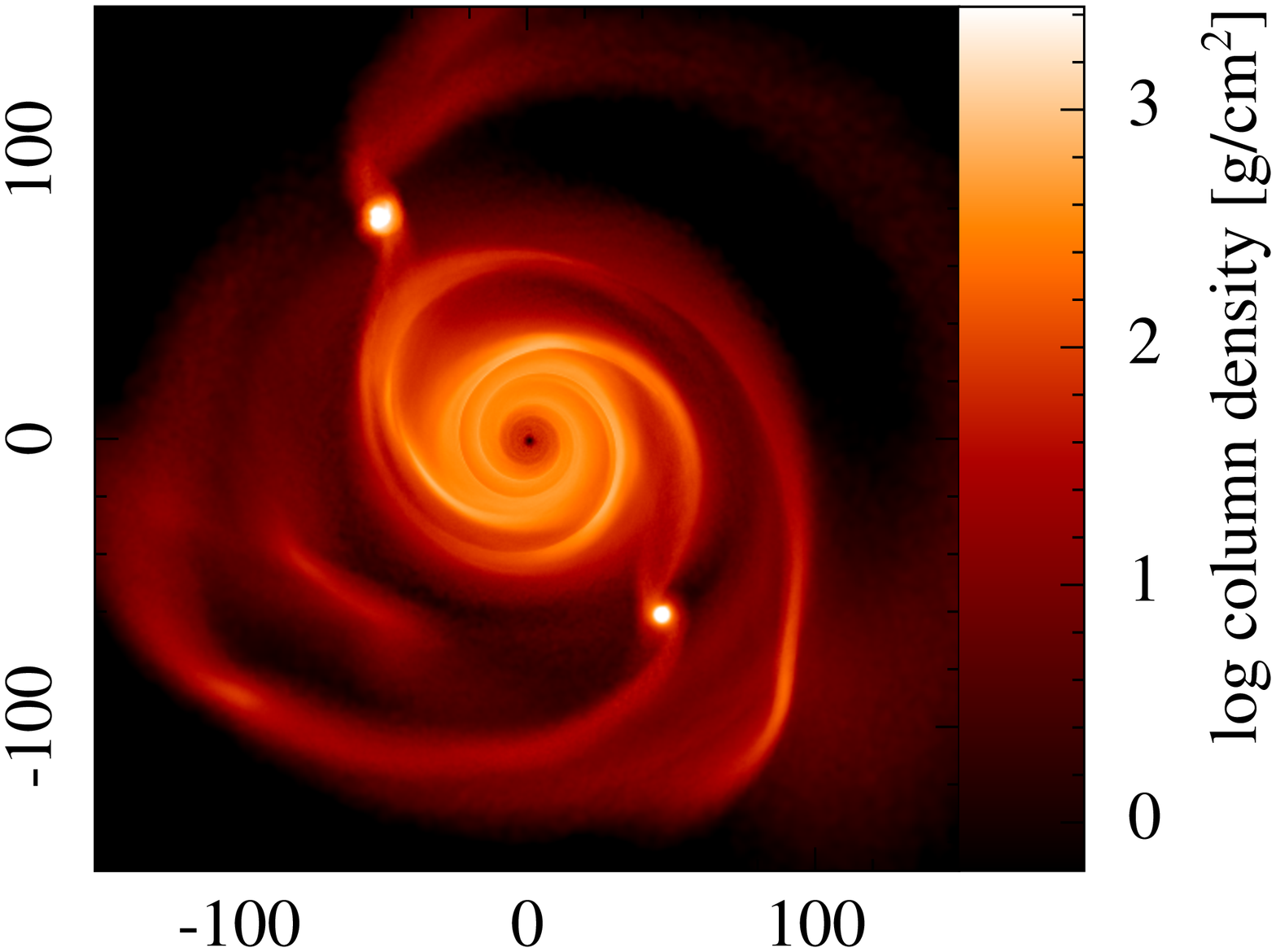}};
    \begin{scope}[x={(image.south east)},y={(image.north west)}]

       % \draw[green,thick] (0.3335,0.6463) rectangle(0.433,0.7779);
        \draw[green,thick] (0.2457,0.6985)rectangle(0.357,0.8518);
        %\draw [-latex, ultra thick, red] (note) to[out=0, in=-120] (0.48,0.80);
        %\draw [-stealth, line width=5pt, cyan] (water) -- ++(0.4,0.0);
    \end{scope}
\end{scope}
\end{tikzpicture}

\end{tabular}
\caption{Column density plots of simulation 3, where clumps 2 and 4 (highlighted in green) undergo a merger. \label{fig:merger}}
\end{center}
\end{figure*}
\begin{figure*}
\begin{center}
\begin{tabular}{ccc}
T = 3146 years & T = 3351 years & T = 3504 years\\
\begin{tikzpicture}
%\node [anchor=west] (note) at (-1,3) {\Large Note};
%\node [anchor=west] (water) at (-1,1) {\Large Water};
\begin{scope}[xshift=1.5cm]

    \node[anchor=south west,inner sep=0] (image) at (0,0) {\includegraphics[width=0.33\linewidth]{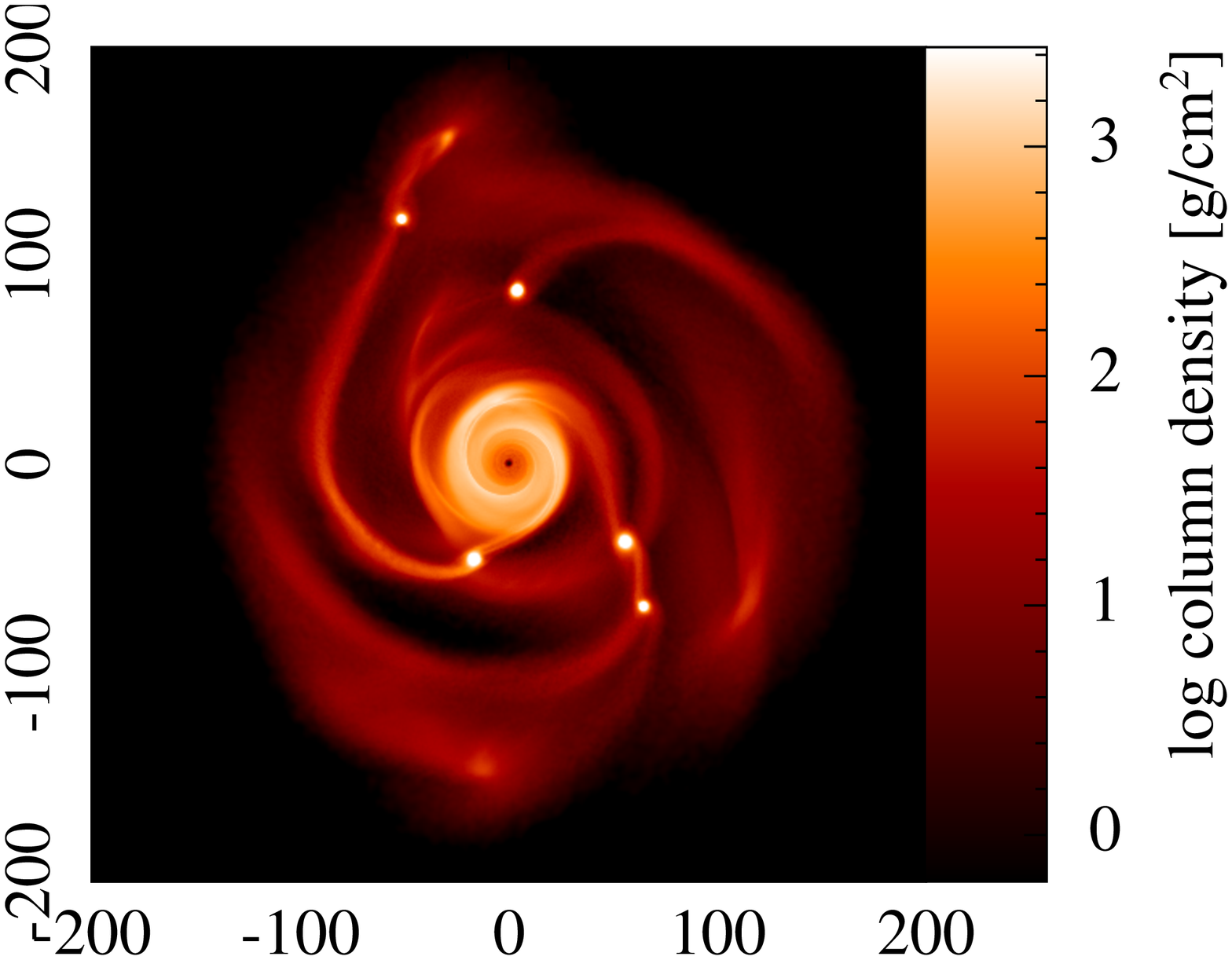}};
    \begin{scope}[x={(image.south east)},y={(image.north west)}]
    %{0.241,0.3057}{0.2766,0.3546}
   % {0.2489,0.3148}{0.2715,0.3437}
   %{0.4125,0.4152}{0.4538,0.4786}
        %\draw[green,thick] (0.4125,0.4152) rectangle(0.4538,0.4786);
        	\draw[green,thick](0.364,0.3893)rectangle(0.413,0.4512);
        %\draw [-latex, ultra thick, red] (note) to[out=0, in=-120] (0.48,0.80);
        %\draw [-stealth, line width=5pt, cyan] (water) -- ++(0.4,0.0);
    \end{scope}
\end{scope}
\end{tikzpicture}%
&
\begin{tikzpicture}
%\node [anchor=west] (note) at (-1,3) {\Large Note};
%\node [anchor=west] (water) at (-1,1) {\Large Water};
\begin{scope}[xshift=1.5cm]
    \node[anchor=south west,inner sep=0] (image) at (0,0) {\includegraphics[width=0.33\linewidth]{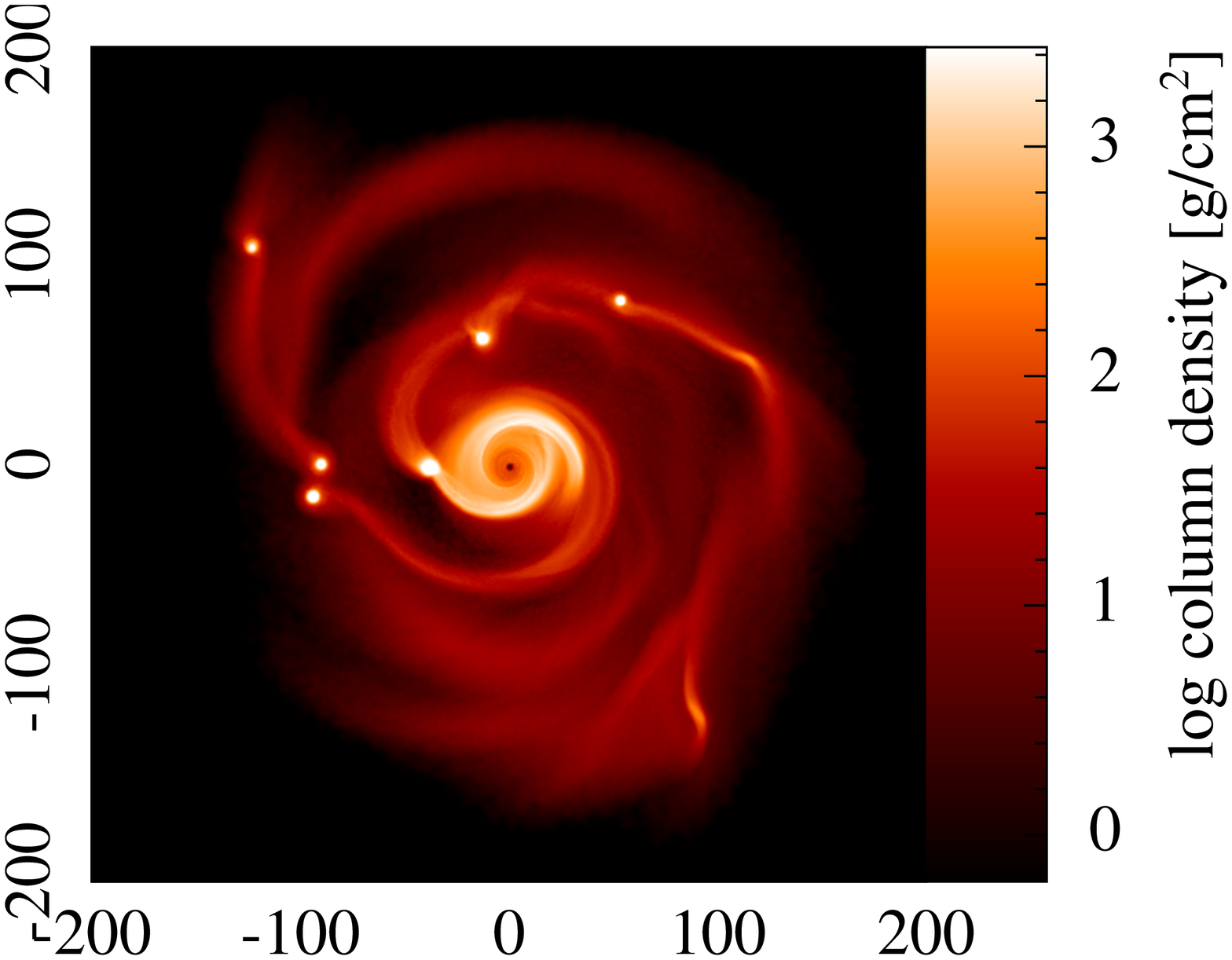}};
    \begin{scope}[x={(image.south east)},y={(image.north west)}]

        %\draw[green,thick] (0.3765,0.5077) rectangle(0.4178,0.5711);
        \draw[green,thick](0.327,0.4821)rectangle(0.376,0.544);
        %\draw [-latex, ultra thick, red] (note) to[out=0, in=-120] (0.48,0.80);
        %\draw [-stealth, line width=5pt, cyan] (water) -- ++(0.4,0.0);
    \end{scope}
\end{scope}
\end{tikzpicture}%
&
\begin{tikzpicture}
%\node [anchor=west] (note) at (-1,3) {\Large Note};
%\node [anchor=west] (water) at (-1,1) {\Large Water};
\begin{scope}[xshift=1.5cm]

    \node[anchor=south west,inner sep=0] (image) at (0,0) {\includegraphics[width=0.33\linewidth]{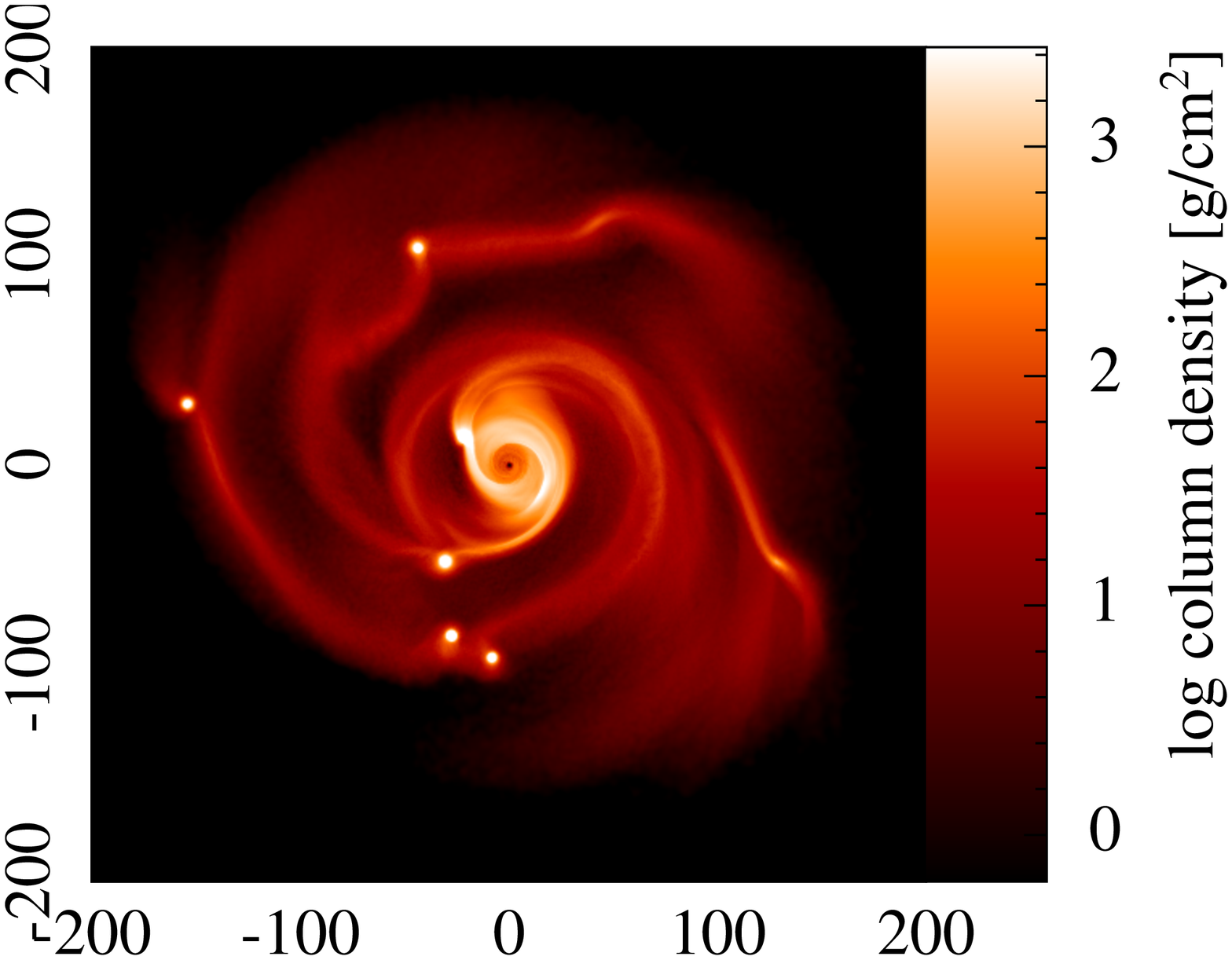}};
    \begin{scope}[x={(image.south east)},y={(image.north west)}]

        %\draw[green,thick] (0.4065,0.5404) rectangle(0.4478,0.6038);
        \draw[green,thick](0.353,0.5156)rectangle(0.402,0.5775);
        %\draw [-latex, ultra thick, red] (note) to[out=0, in=-120] (0.48,0.80);
        %\draw [-stealth, line width=5pt, cyan] (water) -- ++(0.4,0.0);
    \end{scope}
\end{scope}
\end{tikzpicture}\\

T = 3556 years & T = 3607 years & T = 3658 years\\
\includegraphics[width=0.33\linewidth]{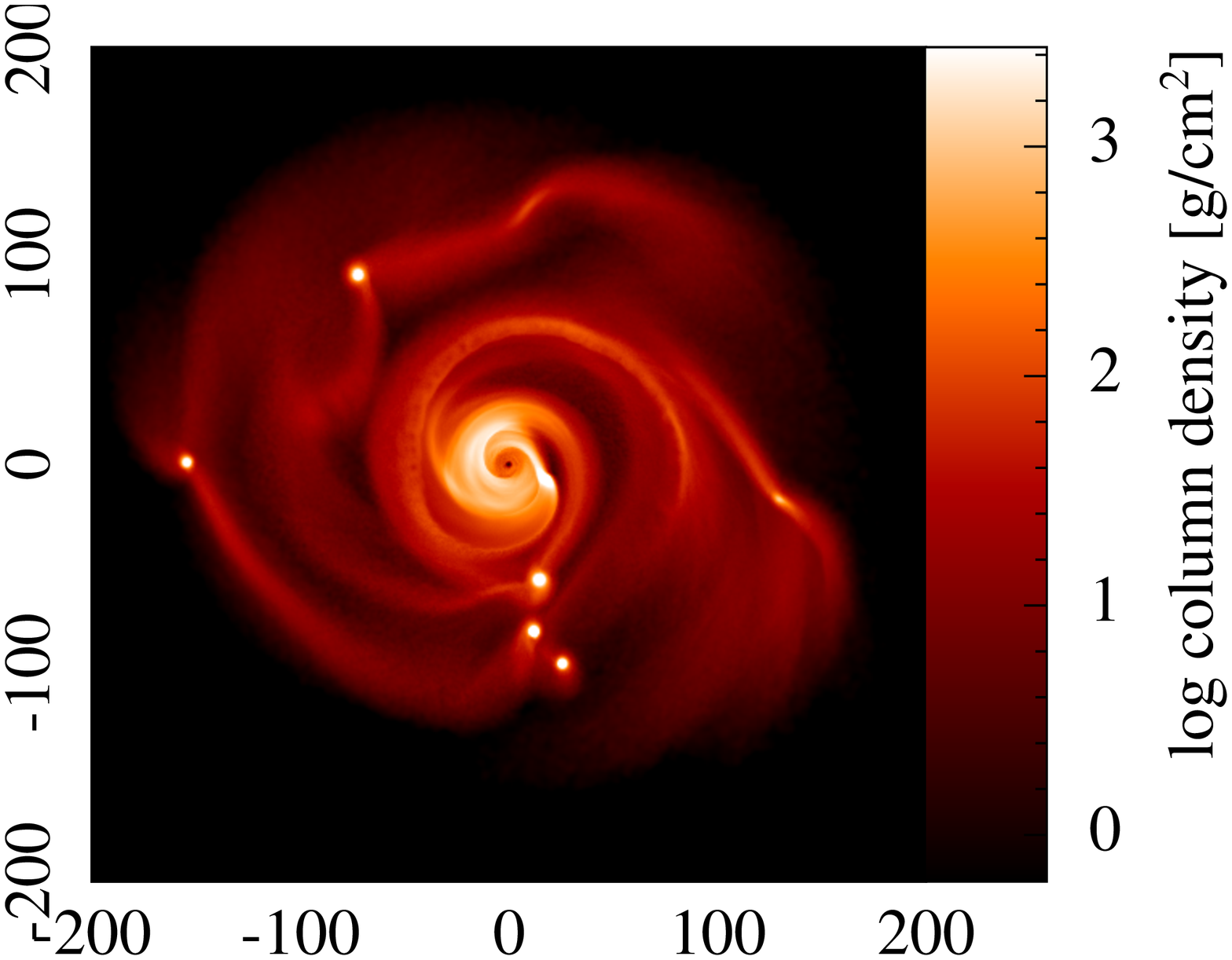} & \includegraphics[width=0.33\linewidth]{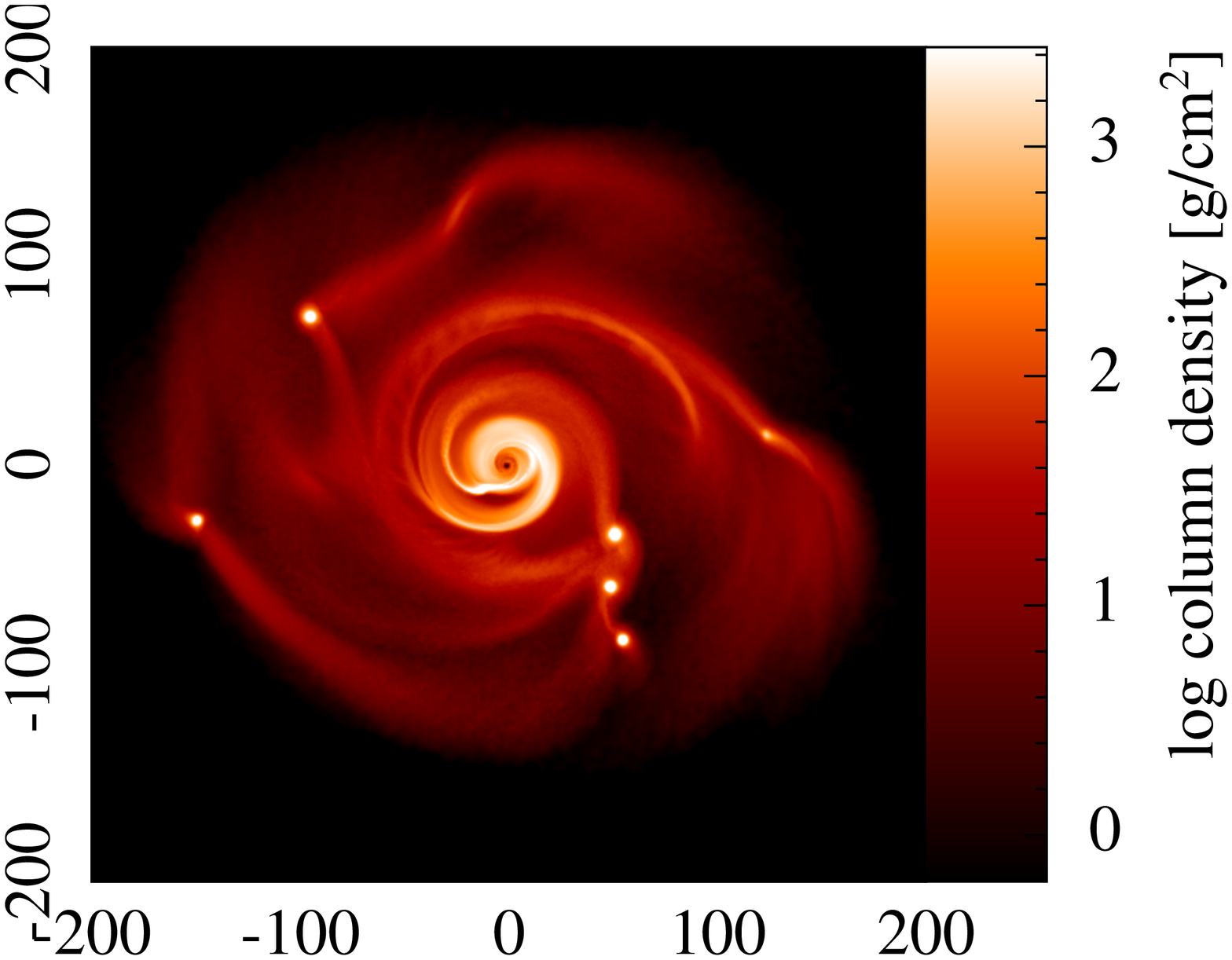} & \includegraphics[width=0.33\linewidth]{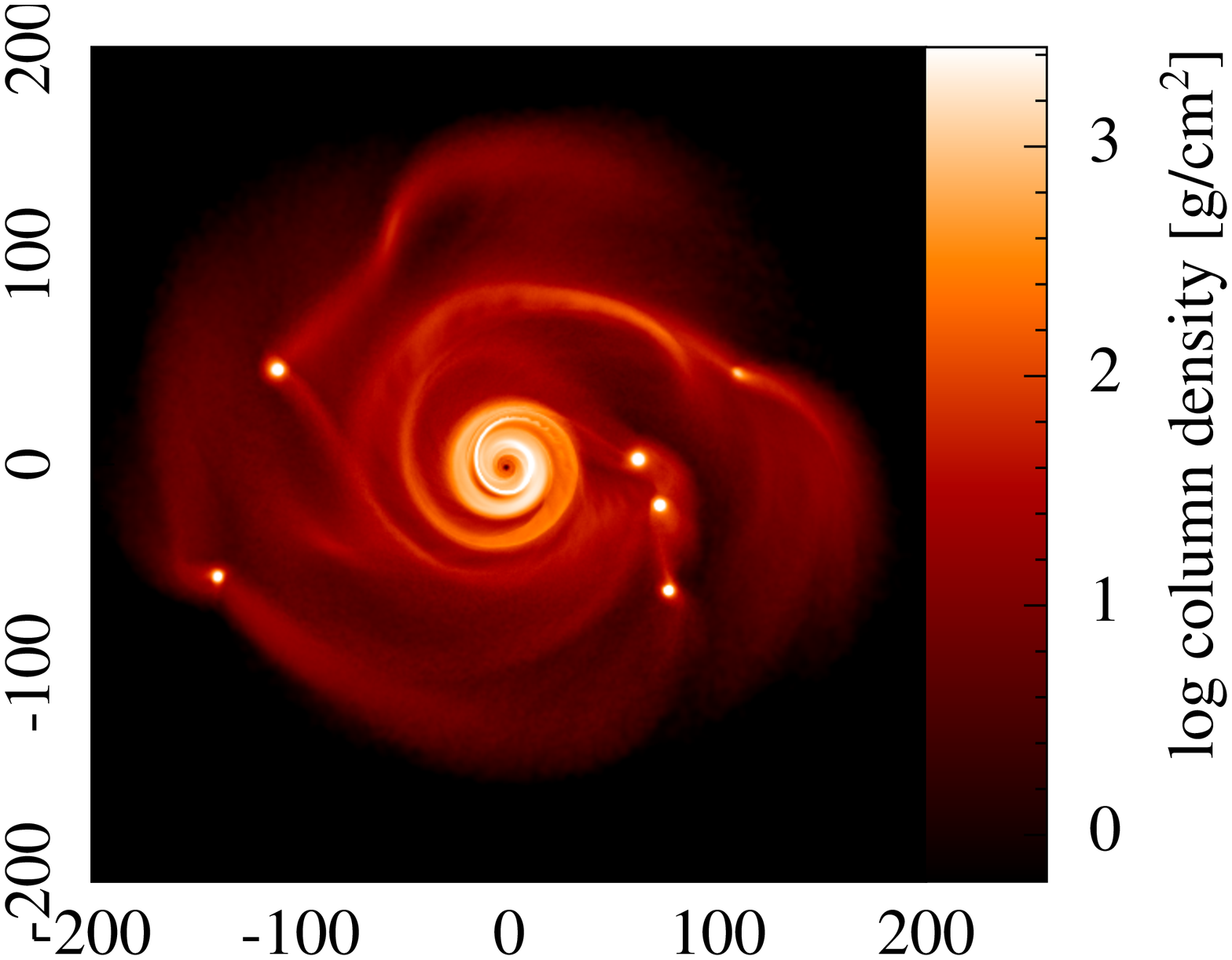}\\
\end{tabular}
\caption{Column density plots of simulation 7, increasing in time, showing the tidal disruption of clump 4, marked in green in the top three images before it begins to be tidally destroyed. \label{fig:tidal}}
\end{center}
\end{figure*}

We suggest, therefore, that clump-clump interactions may provide a mechanism for dramatic increase in temperature of forming clumps, either directly through the interaction, or their subsequent scattering through dense regions in the disc. This mechanism could present a problem for terrestrial planet formation through tidal downsizing. For a terrestrial planet to form in the tidal downsizing scenario, a clump must not accrete too much mass, and then become tidally disrupted after migrating too close to the host star \citep{nayakshin2010,boleyetal2010}. It then leaves behind a solid core, \textit{if} dust grain sedimentation was sufficiently rapid to form a core. Since we have tentatively shown that clump-clump interactions are common (given our small sample size), then it may be possible that clump temperatures are frequently too high, at too young an age, for dust sedimentation to have taken place inside clumps. For a solid core to form, clumps need to exist at a temperature below the dust sublimation temperature ($\sim 1200$ K) for the duration of the sedimentation process.
%We can see that our final surface temperatures for all clumps are in agreement with Vorobyov et al Fragmenting protostellar discs: properties and observational signatures (ads was down - update later), typically in the region of $T<100$ K. Although there are few retrorotating fragments, so it is difficult to draw a universal conclusion, it is interesting to note that fragments experiencing retro-rotation end up with the largest change in temperature profile, undergoing a dramatic, steep change when compared with the prograde rotators. WHY IS THIS. 
%Profile of largest clumps
%Profile of clumps in the process of merging
%Profile of clumps just before tidally destroyed
%Put in radial profiles of some of the most interesting clumps here, I guess the %retro-rotators and also the most massive clumps?
%We report on the properties of these simulations, splitting our results into two sections. We first discuss the simulations individually, and then draw more broad conclusions as we consider all clumps in all simulations. We run each simulation using only hydrodynamics, and no sink particles, until one or more fragments becomes too dense to continue.
%\subsection{Mergers and tidal disruptions}

\subsection{Tidal disruption and mergers}
\label{subsec:tidalmergers}
Of the 41 clumps detected by the DDS method in our simulations, 7 were tidally destroyed by the central star, and 4 clumps underwent mergers. Despite our small sample size, these results suggest that both tidal destruction and mergers are common amongst protostellar disc fragments. An example merger is shown in Figure \ref{fig:merger}, which shows simulation 3, clumps 2 and 4, merging together. An example tidal disruption is shown in Figure \ref{fig:tidal}, which shows simulation 7, clump 4 undergoing tidal disruption. Such tidal disruptions could potentially be an explanation for outburst type behaviour in young protostars, due to the rapid increase in accretion rate onto the central star (see, e.g.,\citealt{vorobyovbasu2005,boleyetal2010,nayakshinlodato2012}). We leave observational signatures of our tidal disruptions to future work.
\section{Discussion and conclusion}
\label{sec:conclusion}
In this paper, we have presented one original method for identifying fragments in a simulation, the density derivative search, and one method adapted from the \texttt{CLUMPFIND} \citep{clumpfind,smith2008,forganpricebonnel2016} algorithm. We ran 9 SPH simulations of a $0.25$ M$_{\odot}$ disc around a 1 M$_{\odot}$ star, each with a surface density profile of $\Sigma\propto r^{-1}$, an inner radius of 10 au and an outer radius of 100 au. Each simulation was run for as long as computationally feasible without converting dense regions to sink particles, since we wished to calculate orbital properties for our fragments, which are sensitive to their radial mass distribution.

Each disc fragmented to form at least 2 bound objects, and we analysed the fragments (which we call clumps, once they have been detected) using the density derivative search and the adapted \texttt{CLUMPFIND} method. We have shown that these two methods are complementary, as the density derivative search is able to detect low mass clumps, and clumps that are ultimately tidally destroyed, while the search using the adapted \texttt{CLUMPFIND} method filters out clumps which are unlikely to survive the simulation, but also has a relative insensitivity to low mass clumps.

We compare our sample of fragments to the population synthesis model of \cite{forganricepopsynth2013}, and find that our algorithm has some limitations at early times (i.e., during the initial period of fragment formation), which means that some radial migration has already happened before the algorithm detects the fragment. Despite this, it is fairly robust at late times (i.e., a few orbital periods after formation), and so our final mass and final semi-major axis functions, and the mass semi-major axis relationship, are representative of the final configurations (i.e., at $T\sim$ 4000 years) of our systems. Of course, the final mass and final semi-major axis distributions that we present here will differ from the \textit{ultimate} distributions of the systems. These will only be determined some $\sim 10^6$ years after formation, long after the disc has dispersed. What happens after disc dispersal is not considered in this work, but we refer the interested reader to \citet{forganparkerrice2015} and \citet{lietal2016}.

We examine the internal temperature and density structure of our fragments, and compare them with appropriately indexed polytropes. We find that the central density of our fragments are typically an order of magnitude denser than their polytropic equivalent, and since orbital parameters of a body are sensitive to its internal mass distribution, we recommend caution when using polytropes in GI population synthesis models to calculate orbital parameters. 

%review change begin
Furthermore, that the interiors of the clumps may not be well described by polytropes may raise concerns about the validity of the \citet{forgan2009} hybrid radiative transfer method, which uses the polytropic cooling formalism of \citet{stamatellos2007} to approximate radiative cooling. Essentially, the assumption is that each SPH particle is embedded in its own polytropic pseudo-cloud, of polytropic index $n$. Therefore, the cooling of any morphology that is not a polytrope of index $n$ may not be exactly described by the polytropic cooling formalism. However, so long as the geometry is \textit{approximately} spherical, which gravitationally bound clumps tend to be, then the polytropic cooling approximation is valid to within an order of unity for a variety of opacity laws \citep{wilkinsclarke2012}.

More generally, the polytropic cooling method of \citet{stamatellos2007} has been criticised when employed in planar geometry, such as in self-gravitating protostellar discs. The most alarming of these is the systematic under-estimation of cooling rates by a factor of $\sim$ 100 \citep{wilkinsclarke2012}, due to the use of local variables only in optical depth determination. This ignores the decreasing column density normal to the midplane of the disc, which may offer an easier escape path for photons.

However, this criticism is specific to the polytropic cooling method of \citet{stamatellos2007}, rather than the hybrid method of \citet{forgan2009}. By including flux limited diffusion, the problems outlined in \citet{wilkinsclarke2012} are addressed.
%By including flux-limited diffusion, as \citet{forgan2009} does, the heat diffusion between hotter and cooler regions can be accounted for.

More recently, the polytropic cooling approximation of \citet{stamatellos2007} has been updated to include radiative feedback from sink particles \citep{stamatellos2015}, which is not captured by the \citet{forgan2009} hybrid scheme. However, we do not use sink particles in this work, since we aim to characterise the radial properties of our fragments. By using only SPH particles, the flux limited diffuser in the \citet{forgan2009} algorithm naturally provides radiative feedback from the fragments, as heat is diffused from hotter to cooler regions. 

Ultimately, the hybrid method of \citet{forgan2009} is an approximation to a computationally much more expensive task, a full radiation hydrodynamical treatment of disc evolution. Therefore, results obtained using this approximation should be adopted with knowledge of its limitations in mind.

The radiative cooling of real protoplanetary discs depends strongly on opacity, which is not examined in this work. Dust opacities have a large degree of uncertainty; for temperatures below $\sim 1000$ K, dust grains dominate absorption, so this uncertainty is likely to matter in the region of parameter space considered in this work. For example, \citet{semenovetal2003} have shown that for temperatures $100\,{\mathrm{K}} < T < 1000{\,\,\mathrm{K}}$, opacity may differ by a factor of $\sim$ 10. Since it is known that initial fragment mass depends on opacity \citep{masunagainutsuka1999}, our initial fragment mass {distribution} may differ from what we state here.

 However, it is widely accepted that dust opacity is proportional to metallicity, and it has been shown that while low-metallicity fragments may be a factor of $\sim$ 3 more massive than solar metallicity fragments, the initial mass distribution of the fragments is, apart from the shift at low mass, very similar between the two metallicities \citep{bate2005}. We conclude, in light of this, that our fragment mass distributions are probably reasonable, but caution that the lower mass fragments may be larger if opacity is decreased. It is possible to estimate the mass dependence on opacity by appealing to the opacity limit for fragmentation, as follows.
 
 Usual derivations of the opacity limit for fragmentation include an \textit{efficicency factor}, $e$, which subsumes, amongst other unknowns, the opacity of the gas. Adjusting this efficiency factor allows us to examine a range of potential fragment masses. We begin with an expression for the power of a ball of gas collapsing in freefall:
\begin{equation}
\frac{|B|}{t_{\mathrm{ff}}} = \Bigg(\frac{3}{5\pi}\frac{\mathrm{G}^3 M^5}{R^5}\Bigg)^{\frac{1}{2}},
\end{equation}
where $B$ is the gravitational binding energy of the gas,  given by 
\begin{equation}
B=-\frac{3}{5}\frac{\mathrm{G}M^2}{R},
\end{equation}
and $t_{\mathrm{ff}}$ is the free-fall time of the gas,  given by
\begin{equation}
t_{\mathrm{ff}} = \sqrt{\frac{3\pi}{32\mathrm{G}\rho}}.
\end{equation}
For the collapse to continue isothermally, then
\begin{equation}
\label{eq:conditioncollapse}
\frac{|B|}{t_\mathrm{ff}} \lesssim L,
\end{equation}
where $L$ is the luminosity describing the radiation of the ball of gas, as a blackbody, given by
\begin{equation}
L=e4\pi R^2 \sigma T^{4},
\end{equation}
where $T$ is temperature, $\sigma$ is the Stefan-Boltzmann constant, and $e$ is the \textit{efficiency factor} of the radiation. Substituting our expression for power, $|B|/t_\mathrm{ff}$,  and luminosity, $L$ into equation \ref{eq:conditioncollapse}, we can rearrange equation \ref{eq:conditioncollapse} to arrive at an expression for the \textit{critical mass}:
\begin{equation}
M_{crit} = \Bigg( \frac{400}{9} \frac{e^2\pi^4 \sigma^4 T^8 R^9}{\mathrm{G}}  \Bigg)^{\frac{1}{5}}
\end{equation}
above which the collapse proceeds adiabatically. However, this mass must be above the local \textit{Jeans mass}, $M_\mathrm{J}$, for the collapse to continue, so we now have a condition
\begin{equation}
M_\mathrm{J} \lesssim M \lesssim M_\mathrm{crit},
\end{equation}
which must be satisfied. If $M_\mathrm{J}$ and $M_\mathrm{crit}$ are equal, we then arrive at the \textit{opacity limit for fragmentation}, which describes the minimum mass of a fragment that may form. Inserting constants, we now have an expression for the minimum fragment mass,
\begin{equation}
M_\mathrm{frag} \approx 6 \frac{T^{\frac{1}{4}}}{e^{\frac{1}{2}}} \,\, \mathrm{M}_{\mathrm{J}}.
\end{equation}
From this, we can see that for a fixed temperature, minimum fragment mass increases with a decrease in efficiency. If the efficiency of the radiation is 100\%,  i.e., the gas has a sufficiently low opacity such that it is effectively completely optically thin to escaping radiation, then for a temperature of $T\sim 10$ K a fragment mass of $\sim 10$ M$_{J}$ is expected.  For an efficiency of 20\%, then this rises to $\sim 23$ M$_{\mathrm{J}}$. Clearly, how strongly this efficiency depends on opacity will determine the relationship of proportionality, however, it is clear that opacity plays an important, if not dominant, role in determining initial fragment mass. In addition to determining initial fragment mass, the opacity will also play an important role in determining the final fragment mass through accretion rate onto the fragments, since the thermodynamics of the gas inside the Hill sphere of the fragment depends on this opacity \citep{stamatellos2015}.

%=\frac{3}{5\pi}\frac{\mathrm{G}^3 M^5}{R^5}^{\frac{1}{2}}
%ROBABLY WANT TO SWAP THE POSITION OF THIS WITH THE BATE 2005 ARGUMENT.
%5resulting in a min INCLUDE THE ARGUMENT IN MY THESIS.
% 
%Although it is certainly feasible that some of our fragment masses may \textit{increase} due to a reduction in opacity, we are naturally bound by the minimum Jeans mass, so if opacity is increased (by, say, a factor of 10), we are unlikely to see very small fragments as a result. The minimum Jeans mass for fragmentation, which dictates when fragmentation must stop, is proportional to $\kappa^{\frac{1}{7}}$, 
%We use the \citet{belllin1994} opacities, assuming identical opacities in all discs. 
%However, fragment mass, $M_\mathrm{F}$ has been shown to depend on opacity, $\kappa$, such that $M\propto \kappa ^{-1}$ \citep{masunagainutsuka1999}. However, this only makes physical sense for \textit{decreasing} opacity, otherwise there would be no minimum mass for fragmentation, and arbitrarily small fragments may form. TALK ABOUT MINIMUM FRAGMENT MASS THAT HAS EFFICIENCY FACTOR> ITS ON THESIS PAGE 60
%
%review change end
We find that fragment-fragment interactions play a substantial role in the ultimate fate of our systems. Low-mass fragments can be scattered out to large radii, (and therefore remain low mass, since there is less material to accrete), and are therefore unlikely to be tidally destroyed by the central star. Since current GI population synthesis models suggest that $\sim 40\%$ of initial fragments are ultimately tidally destroyed \citep{forganricepopsynth2013}, if a significant number of these fragments are actually scattered out to large separations by interactions with other fragments, this figure could potentially be much lower. We therefore recommend that fragment-fragment interactions in the gas phase of the disc be included in any new GI population synthesis models.

During their lifetime, we find that our fragments may be shocked as they pass through spiral arms, rapidly increasing the internal energy of the fragment. This could have implications for terrestrial planet formation through the tidal downsizing hypothesis, since solid core formation requires rapid dust sedimentation. If the interior of these fragments are hot enough to sublimate dust at very early stages in their lifetime, it may not be possible for them to form solid cores, a key assumption in the core-assisted gas-capture hypothesis of \citet{nayakshinhelledboley2014}.

In addition to encounters scattering fragments out to large radii, we also find a tentative relationship between the dominant azimuthal wavenumber in the disc, and the maximum semi-major axis of a clump in that disc, such that $a_{\mathrm{max}}\propto \nicefrac{1}{m}$. This seems to suggest that the spiral arms of protoplanetary discs play as large a role in the dynamical fate of the clumps as do the clump-clump scatterings. Although this relationship is preliminary, and requires more simulations in a range of disc-to-star mass ratios to confirm it (since $m\sim\nicefrac{1}{q}$ \citep{donghallrice2015}, where $q$ is disc-to-star mass ratio), it is unsurprising that such a relationship may exist, and we suspect that the relationship is one of inverse proportionality, rather than one of negative proportionality, due to the relationship between gravitational torque $\Gamma_{\mathrm{G}}$ and azimuthal wavenumber as follows (for a full derivation and comprehensive explanation, we refer the reader to \citealt{binneytremainedynamics}). The gravitational torque $\Gamma_{\mathrm{G}}$ exerted on material outside a radius $R_0$ in a disk is given by
\begin{equation}
\Gamma_{\mathrm{G}} = \mathrm{sgn}(k) \frac{\pi^2 m R_0 \mathrm{G}\Sigma_1^2}{k^2},
\end{equation}  
where $\Sigma_1$ is a gentle function of radius, $k$ is the radial wavenumber defined as 
\begin{equation}
k\equiv\frac{\partial(mf(R))}{\partial R},
\end{equation}
$mf(R)$ is the radial shape function of the spiral (for a simple example, see \citealt{halletal2016}), and $\mathrm{sgn}(k)=+1$ for trailing spirals (i.e, positive torque exerted outwards). At a given value of $R_0$, and the same function for $\Sigma_1$, we can see that 
\begin{equation}
\Gamma_{\mathrm{G}}\propto\frac{1}{m}.
\end{equation} 
To establish the relationship $a_\mathrm{max}\propto \nicefrac{1}{m}$, we have assumed that the amount of torque is directly proportional to the change in radial distance. Whilst this is almost certainly an over-simplification of matters, we can see from
\begin{align}
\Gamma_{\mathrm{G}} &= \frac{d|\vec{L}|}{dt} \nonumber \\
&= \frac{d}{dt} \bigg( m|\vec{v}\times\vec{r}| \bigg)  \nonumber \\
& = \frac{d}{dt} \bigg(m|\vec{v}||\vec{r}|\mathrm{sin}[\theta]\bigg),
\end{align}
that if we assume mass, velocity, and the angle between $\vec{v}$ and $\vec{r}$ stay fairly constant, then $\Gamma_{\mathrm{G}}\propto \nicefrac{dr}{dt}$, i.e. the distance we wish to move our fragment. Assuming that all fragments form at roughly the same $r$ in a given disc, (i.e. where that particular disc becomes susceptible to fragmentation), then we recover $a_{\mathrm{max}}\propto \nicefrac{1}{m}$.

Varying the surface density profile in a protostellar disc will alter disc torques and migration rates of planets (see, for example,\citealt{baruteauetal2011}). Since we simulated 9 protostellar discs that had initially identical surface density profiles, this is not something we have investigated, but it could potentially alter our results. For example, steeper surface density profiles correspond to more rapid migration rates. It is therefore feasible that a sharp cut-off at the disc outer edge, as we use in our initial conditions, may exacerbate migration torques in the disc. However, since the majority of the disc mass is contained well within the disc's numerical outer edge, it is probably reasonable to assume it is not the dominant effect when considering the radial migration of fragments.
%DELETED
%That $\Gamma_{\mathrm{G}}\propto\nicefrac{1}{m}$ could potentially impact the fragmentation radius in a given disc is unsurprising. For low $m$, large torques will be exerted on the disc, allowing mass to move inwards towards the central star. If the surface density profile of the disc changes significantly, then the radius at which the disc becomes susceptible to fragmentation will also change. With this in mind, we recommend that future GI population synthesis models have an extra layer of stochasticity to reflect this effect, since whatever $m$-mode is set up in the disc may affect the subsequent fragmentation radius.
%Since we performed 9 simulations where each disc had identical global properties, we did not examine the effect of varying surface density profile index on migration rates or torques in the disc (see, for example,\citealt{baruteauetal2011}). Since steeper surface density profiles correspond to more rapid migration rates, it is feasible that a sharp cut-off at the disc outer edge may exacerbate migration torques in the disc. However, since the majority of the disc mass is contained well within the disc's numerical outer edge, it is probably not the dominant effect.

Despite the relatively short timescales of our simulations (since we did not make use of sink particles), we can see that orbital properties of fragments are drastically altered by interactions with each other. Since disc fragmentation forms objects with initially low eccentricities ($e<0.1$), it is generally accepted that measurements of eccentricity as a function of orbital distance will constrain the formation mechanism of giant planets and brown dwarfs \citep{vorobyovwideorbits}, with high eccentricity being caused by dynamical scattering. 

However, our results in Figure \ref{fig:eccentricity} possibly suggest that these high eccentricity orbits could be formed at very early times, during the gas phase of the disc, and as such eccentricity measurements of brown dwarfs and giant planets may not, necessarily, constrain their formation mechanism. We have shown that the intial orbital inclination of our fragments is reduced by a factor of $\sim$ 100 over the duration of the simulation (Figure \ref{fig:histogramecc}), despite the significant dynamical interactions many of the fragments experience. This suggests that although dynamical interactions certainly can create highly inclined orbits, doing so while the gas disc is present may be much more difficult. 

On the other hand, our simulations are of discs in isolation. Inclination and eccentricity may be excited by environment, such as a stellar companion, or location within a cluster environment \citep{forganparkerrice2015}. Since current GI population synthesis models do not include eccentricity or inclination information, we have provided several Gaussian fits in Figure \ref{fig:histogramecc} from our SPH simulation data. Despite our small sample size, we hope these plots will be useful in further development of GI population synthesis models.

In this work, we do not consider how solid particles will effect the formation, evolution, and survival of the clumps, nor their effects on the behaviour of the gas. This is, quite possibly, the most important limitation of our work. As we have already discussed, opacity could heavily influence the initial mass distribution of our clumps. It is also feasible that altering the opacity in the clumps, due to movement of solids, could result in clumps cooling more (or less) rapidly, resulting in more (or fewer) clumps surviving \citep{nayakshin2010}. More clumps surviving because of local opacity changes could present a partial solution to the rapid inward migration and subsequent disruption of GI fragments found in so many other works (see, for example, \citealt{vorobyovbasu2005,baruteauetal2011,zhuetal2012GI}).  
%Quite possibly, the most important limitation of our investigation is how the inclusion of solid particles will effect the formation, evolution and survival of our clumps, which we do not consider. As we have discussed, opacity will most likely heavily dominate the initial mass distribution of our clumps. If, in addition to this, the clumps are able to cool more rapidly, it is, perhaps,  possible that an increase in magnitude of "boundness" will enhance their probability of survival, simply because more energy is required to tidally disrupt them. This mechanism, if it were to occur, could present a partial solution to the rapid inward migration and subsequent disruption of GI fragments found in so many other works (see, for example, \citealt{vorobyovbasu2005,baruteauetal2011,zhuetal2012GI}). This is not a new idea, and was discussed by \citet{nayakshin2010} in the original formulation of the "tidal downsizing" hypothesis.

%This could greatly increase the survival rate of objects that form through gravitational stability.
To date, no global 3D SPH simulations of dusty, self-gravitating, fragmenting protostellar discs have been conducted. This is a computationally expensive task, with complicated physics. Almost certainly, a full treatment of radiation hydrodynamics would be necessary to correctly capture the cooling of the system, and the spatial evolution of the dust within the gas. High resolution is needed in the core of these fragments to trace their compositional change. However, if we are considering the global evolution of the system, we must be able to capture the full dynamical range. This is probably not possible with current computational abilities, and is one of many multiphysics problems present in protostellar disc modelling \citep{haworthetal2016}.
%It does not appear to be possible with current computational abilities, although it may be possible to accept some limitations and run under-resolved simulations in the inner cores, with some analytical proxy for the behaviour in the central core. However, this may be the dominant effect, which could result in no clear solution becoming apparent.

How fragmentation proceeds when solid grains are included in the simulation is clearly an interesting question. Although not considered in 3D hydrodynamics simulations, this idea has recently been explored by \citet{nayakshinhelledboley2014}, in their so-called "core-assisted gas capture" (CAGC) paradigm. In this scenario, grain sedimentation inside a fragment forms a core of heavy elements. Upon reaching a critical core mass, the surrounding gaseous envelope collapses onto the core, analogous to the core accretion paradigm. The assumption in the work of CAGC is, of course, that these cores \textit{do} form. Gravitational instability tends to occur beyond $\sim 50$ au in protostellar discs \citep{rafikov2005}. For a significant core to form in GI clumps due to grain sedimentation, there must be a substantial local dust-to-gas ratio at the site of clump formation. Since dust grains tend to migrate rapidly inwards \citep{weidenschilling1977}, it may well be difficult to maintain a significant dust-to-gas ratio at the site of fragmentation, making core formation from grain sedimentation inside a fragment difficult. However, GI is also a rapid process, so as long as the fragmentation takes place rapidly, then the local change in pressure gradient may be sufficient to prevent inward migration of these grains.

If massive cores in GI clumps \textit{do} form, as assumed by CAGC, then we should probably expect differences in the final clump distributions to what we have presented here. An interesting feature of the (CAGC) paradigm is that if planets are formed by this method, then we should expect a positive metallicity correlation. 

Assuming, for now, that the collapse proceeds with no core mediation, then it is probably reasonable to expect no metallicity corellation; indeed, there is at least \textit{some} numerical evidence that we may expect a \textit{negative} metallicity corellation \citep{caietal2005}, since low metallicity corresponds to faster cooling (and therefore stronger spiral amplitudes, increasing the effective gravitational stress) which may result in fragmentation. This is because higher metallicity results in better cooling only in the optically thin regime, for example, at the tenuous surface of the disc. 

Conversely, higher metallicity results in slower cooling in the optically thick regime, i.e. at the disc midplane, which is the location of fragmentation. Therefore, gas-giant planets formed through GI should be preferentially found around \textit{low} metallicity stars. It seems to now be fairly clear that gravitational instability rarely forms planetary mass objects \citep{riceetal2015}, but if it does, these objects will be much larger than Jupiter. We should, then, find that planets more massive than Jupiter are more frequent around low-metal stars. However, it has been suggested that this is not the case, since planets more massive than $\sim 3$ M$_{\mathrm{J}}$ seem to be found \textit{less} often around low-metallicity stars \citep{thorngrenetal2015}.

Overall, the picture is unclear. While CAGC may result in a positive metallicity correlation, there seems to be enough evidence to suggest that fragmentation should preferentially occur in discs around low-metallicity stars, which, in turn, suggests that GI planets would be preferentially found around low-metallicity stars. In the CAGC paradigm, fragmentation must first take place, before the core can form through subsequent grain sedimentation. If this fragmentation happens less often around high-metallicity stars, this could still result in an over negative metallicity corellation. Ultimately, the ambiguity of what happens at low-metallicity does not remove the strong high-metallicity correlation with planet frequency. This is mostly suspected to be the work of core accretion, and is difficult to explain with any GI theory alone.
%Whilst this could be interpreted as evidence for CAGC, the CAGC paradigm stimulates that the gas/dust mixture must first fragment, before grain sedim

The work we have presented here outlines the first direct comparison between a GI population synthesis model and a suite of 3D, global hydrodynamics simulations of fragmenting protostellar discs. While the work of \citet{forganricepopsynth2013} presents the first attempt at a GI population synthesis model, it is not the only one (see, for example, \citealt{galvagnimayer2014,nayakshinfletcher2015}, and a recent extension to the \citealt{forganricepopsynth2013} model in \citealt{forganparkerrice2015}). As mentioned in the introduction to this work, our simulations are run for as long as is computationally feasible without the use of sink particles, since part of our aim was to characterise the internal density and temperature profile of these fragments. However, doing so limits us to very early times in the disc, typically around $\sim 4000$ years or so after the disc has initially formed.  For this reason, we did not compare our results to the models of \citet{forganparkerrice2015}, which consider the ultimate dynamical fate of the fragments after disc dispersal. 

Similarly, the work of \citet{forganricepopsynth2013} considers the fragmentation phase of a disc, unlike \citet{galvagnimayer2014}. Since our hydrodynamical simulations are analysed around the fragmentation phase, we wished to use a model that did not assume already that clumps exist in the disc, ruling out the \citet{galvagnimayer2014} models. Finally, we considered the \citet{nayakshinfletcher2015} models unsuitable for direct comparison to our hydrodynamics simulations because only one fragment per disc is simulated, unlike the population synthesis models of \citet{forganricepopsynth2013}, which places multiple fragments in a disc susceptible to fragmentation at separations of a few Hill radii. We stress here that we are not suggesting the superiority of the \citet{forganricepopsynth2013} models, simply that those particular models were best suited for direct comparison to our hydrodynamics simulations.
%We choose to compare the results of our hydrodynamics simulations to the population synthesis models of \citet{forganricepopsynth2013}, rather than other population synthesis models (see, for example, \citealt{galvagnimayer2014,})
%Despite this, even if gravitational instability rarely forms wide, giant planets, it may still be the dominant formation mechanism for such planets \citep{viganetal2017}.
%First, the formation of the gas clumps is from gas, not from a combined gas+solid. However, radiative cooling is effected by metallicity. Higher metallicity means better cooling in the optically thin regime (i.e. at the more tenuous disc surface). It also means worse cooling in the optically thick regime, at the disc midplane. Since collapse is required in the midplane, increased metallicity may be expected to produce a negative metallicity correlation for GI clumps. 
%However, Boss investigated this by modifying opacity in the disc midplane by several orders of magnitude, and found no effect on the behaviour of GI.
%It is difficult to explain the correlation at high metallicity with any GI-only prescription, leading to most people concluding this to be the work of core accretion. Additionally, planets much more massive than Jupiter are preferentially found around low metal starts (Rice et al.), which is is probably consistent with disc instability in the conventional sense, rather than the core mediated collapse.
%If this process was included in our simulations, this would most likely alter our results, resulting in far more fragme

Of the 41 clumps that are detected in these simulations, 7 were tidally destroyed ($\sim 20\%$), and 2 have orbits with eccentricity approaching unity ($e\sim 0.75$), which suggests that they are on their way to being ejected ($\sim 5\%$). If these clumps \textit{are} ultimately ejected, then gravitational instability could, perhaps, also contribute to the population of free-floating planets \citep{rice2003freefloating,forganparkerrice2015}. We have demonstrated that the orbital and structural evolution of neighbouring fragments are linked; we recommend, therefore, that any future population synthesis models are able to account for this.
\section{Acknowledgements}
We thank the anonymous referee for their constructive comments, which greatly improved the discussion in this paper. We would like to thank Daniel Price for his publicly available SPH plotting code \texttt{SPLASH} \citep{splash}, which we used to produce Figures \ref{fig:columndensity}, \ref{fig:encounter}, \ref{fig:merger} and \ref{fig:tidal}. We also thank Fabian Fischer for his online LaTex Overlay Generator (available here: \url{https://ff.cx/latex-overlay/}), which we used to annotate Figures \ref{fig:encounter}, \ref{fig:merger} and \ref{fig:tidal}. KR gratefully acknowledges support from STFC grant ST/M001229/1. DF gratefully acknowledges support from the ECOGAL project, grant agreement 291227, funded by the European Research Council under ERC-2011-ADG. The research leading to these results also received funding from the European Union Seventh Framework Programme (FP7/2007-2013) under grant agreement number 313014 (ETAEARTH). This project has received funding from the European Research Council (ERC) under the European Union's Horizon 2020 research and innovation programme (grant agreement No 681601).
%
%=====================================================
%\bibliographystyle{mn2e}
\bibliographystyle{mnras}
\bibliography{bib}{}

%\bsp

\label{lastpage}

\end{document}